\documentclass[11pt]{article}
\usepackage[a4paper, width = 150mm, top = 30mm, bottom = 30mm]{geometry}
\usepackage{jheppubmod}
\usepackage[usenames,dvipsnames]{xcolor}

\usepackage{epsfig,amsfonts,amssymb,amscd,amsmath,mathrsfs,xspace,mathrsfs,amsthm,dsfont,bbm}
\usepackage[small,bf,hang]{caption}
\usepackage{mathabx,amsmath}
\usepackage{slashed}

\usepackage{comment}
\usepackage{color}

\usepackage{bbold}
\usepackage{lipsum,framed}
\usepackage{cancel}
\usepackage{enumitem}

\usepackage{mathrsfs} 
\usepackage{braket} 
\usepackage{slashed} 

\usepackage{comment}
\usepackage{color}
\usepackage[most]{tcolorbox}
\usepackage{psfrag}
\usepackage{array}
\usepackage{amssymb}
\usepackage{amsmath}
\usepackage{amsthm}
\usepackage{mathtools}
\usepackage[T1]{fontenc} 
\usepackage{cancel}
\usepackage{cleveref}
\crefformat{section}{\S#2#1#3} 
\crefformat{subsection}{\S#2#1#3}
\crefformat{subsubsection}{\S#2#1#3}

\usepackage[labelsep=quad]{subcaption}
	
\usepackage{cite}
\usepackage{epsfig}
\usepackage{tikz}
\usetikzlibrary{arrows,plotmarks,positioning,calc}
\usepackage{comment}
\usepackage{IEEEtrantools}
\usepackage{graphicx}
\usepackage{physics}
\usepackage{float}

\theoremstyle{definition}
\newtheorem{definition}{Definition}[section]

\usepackage{setspace}
\usepackage{setspace}
\usepackage{amsmath}
\usepackage{amssymb}
\usepackage{slashed}
\usepackage{xcolor}
\usepackage{cancel}
\usepackage{mathtools}
\usepackage{tikz}
\usetikzlibrary{positioning,decorations.pathmorphing,decorations.markings,arrows}

\usepackage{empheq}
\usepackage{hyperref}
\hypersetup{
	pdfencoding=unicode,
	colorlinks=true,
	urlcolor=teal,
	linkcolor=RoyalBlue,
	citecolor=Maroon,
	pdfdisplaydoctitle=true,
	pdfstartview=FitH,
	linktocpage=true
}
\usepackage{pdfpages}
\usepackage{vmargin}
\usepackage{comment}
\setmarginsrb     { 2.5cm}  
{ 2.5cm}  
{ 2.5cm}  
{ 2cm}  
{ 0cm}  
{ 0cm}  
{ 0cm}  
{ 1.25cm}  

\numberwithin{equation}{section}

\newcommand{\p}{\partial}

\newcommand{\mxx}[1]{\text{max}\big(#1\big)}

\title{Positive Geometries, Corolla Polynomial and Gauge Theory Amplitudes}
\author[1]{Alok Laddha}\emailAdd{aladdha@cmi.ac.in}
\author[2,3]{Amit Suthar}\emailAdd{amitsuthar@imsc.res.in}
\affiliation[1]{Chennai Mathematical Institute, Siruseri, SIPCOT, , Chennai 603103}

\affiliation[2]{The Institute of Mathematical Sciences, IV Cross Road, CIT Campus, Taramani, Chennai, India 600113. }

\affiliation[3]{Homi Bhabha National Institute, Training School Complex, Anushakti Nagar, Mumbai, India 400085. }
\abstract{Arkani-Hamed, Bai, He and Yan (ABHY) discovered a convex realisation of the associahedron whose combinatorial and geometric structure generates tree-level amplitudes in bi-adjoint scalar theory. ABHY associahedron of dimension $k$ determines a unique meromorphic $k$-form in the kinematic space of Mandelstam invariants which is $k+3$ point tree-level amplitude in bi-adjoint $\phi^{3}$ theory. As ABHY further proved in \cite{nima1711}, while the color-ordered amplitude is a form in the kinematic space, the (double) \emph{color-dressed} amplitudes  are scalars obtained by the so called color-form duality. Color-form duality can be implemented by contracting a d-log form with a multi-vector field (MVF) whose coefficients are fixed by the color factors of the theory. We extend the latter result  to identify tree-level S-matrix of Yang Mills theory with a scalar obtained by contracting the canonical form of ABHY associahedron with a MVF in the kinematic space. Components of this MVF are also determined by the combinatorial structures that underlie associahedron and Corolla polynomial which was introduced by Kreimer, Sars and van Suijlekom (KSVS) in \cite{kreimer1208}. KSVS used the Corolla polynomial to obtain (at all orders in the loop expansion) the parametric representation of gauge theory Feynman integral from the corresponding Feynman integral in $\phi^{3}$ theory. The map from latter integral to the former was  generated by the Corolla polynomial. Using the full power of Corolla polynomial, we then extend these results to obtain Yang-Mills one loop planar integrand  
by contracting the Corolla generated MVF with the canonical form defined by $\hat{D}_{n}$ polytope discovered by Arkani-Hamed, Frost, Plamondon, Salvatori, Thomas recently. We also demonstrate that KSVS representation of Corolla graph differential in the parametric space can be readily extended to ``spin up" the curve integral formulae for $\textrm{Tr}(\phi^{3})$ amplitude discovered in \cite{nima2309, nima2311} and give an explicit construction of such formulae for tree-level and planar one loop gluon amplitudes.}
\begin{document}

\setcounter{tocdepth}{2}
\maketitle
\setcounter{page}{2}

\pagebreak

\makeatletter
\g@addto@macro\bfseries{\boldmath}
\makeatother

\section{Introduction}

One of the central questions in the analytic S matrix program is to identify the class of meromorphic functions in the Poincare invariant space of scattering data (also known as kinematic space in the literature) which qualify as amplitudes of unitary and local quantum field theories. Over last few decades, a striking answer to this question with an underlying universality has begun to emerge for a class of quantum field theories.  Namely, in the kinematic space, there is a class of combinatorial geometries that generate S matrix.  The strongest evidence in the favour of this radical proposal is ${\cal N}$ = 4  super Yang-Mills (SYM) theory in the planar limit. The integrand of planar amplitudes in this theory is a canonical form in the kinematic space parametrized by momentum twistors. This form is in fact completely fixed by a convex geometry known as amplituhedron \cite{The_Amplituhedron}. The ``amplituhedron program" has led to rather striking advances in our understanding of planar amplitudes in ${\cal N} = 4$ SYM theory. Seminal attempt at going beyond planarity can be found in \cite{Bern:2015ple}. 

In the world of scalar quantum field theories, the kinematic space corresponding to spin-less external particles is simply the Cartesian space of independent Mandelstam invariants .\footnote{For a fixed space-time dimension $D$, this space is in fact a algebraic variety as opposed to a Cartesian space  and is also parametrized by momentum twistors. However so long as number of external particles $n \leq\, D$, the kinematic space can also be co-ordinatized by the Mandelstam invariants modulo linear relations imposed by momentum conservation.}  A number of seminal developments in physics and mathematics has led to a discovery of rich class of convex polytopes in this kinematic space. The star of the show is a convex polytope known as associahedron (Stasheff polytope) whose rich combinatorial structure makes it amenable to a convex realisation inside the kinematic space associated to tree-level scattering,\cite{stasheff1963homotopy}.  This realisation was discovered by Arkani-Hamed, Bai, He and Yun (ABHY) and is known as ABHY associahedron,\cite{nima1711}.  In fact associahedron is a specific (and in many ways the most prominent!) member of a family of such combinatorial polytopes known as accordiohedra \cite{pilaud-maneville, chapoton, Pinaki-Alok-Prasanth:1811, Aneesh-Mrunmay-Nikhil:1906, Prashanth:1906, Mrunmay:1911}. All the accordiohedra are convexly realised in the kinematic space where the embedding is completely fixed by underlying combinatorial structures of these polytopes. 

Accordiohedra share a common property that it has boundaries of all co-dimensions and that every vertex is adjacent to number of co-dimension one faces which equals the dimension of the polytope. More generally, they belong to a specific family of geometries known as positive geometry \cite{nima1703}. 

Any positive geometry which is realised in the kinematic space, defines a meromorphic form that produce the tree-level S-matrices of a large class of colored scalar theories. For a sampling of the papers where the (ever-deepening) relationship between combinatorics, geometry and amplitudes is analysed, we refer the reader to the following papers and references there in,\cite{nima1711, nima1703, baziermatte, pppp1906, ferroreview, Herrmann_Trnka_review, Pinaki-Alok-Prasanth:1811, Prashanth:1906, mrunmay2104, mrunmay2206, nima2312:hiddenzero, Salvatori:stokes}. The one loop integrands of colored scalar theories also belong to this category where the geometries reside in enlarged kinematic space including loop momenta and the geometries arise from combinatorics of chords that dissect abtract polygon with genus $> 0$, \cite{ahst,afpst,mrunmay2007,giulio-halohedron}.  The geometry whose form generates  one loop planar integrands of bi-adjoint scalar theory is known as $\hat{D}_{n}$ polytope which has much the same combinatorial skeleton as Associahedron.\footnote{At higher loop orders, the corresponding polytopes are known as surfacohedra and are subject of an upcoming work  by Arkani-Hamed, Frost, Plamondon, Salvatori, Thomas, \cite{afpst}.} $\hat{D}_{n}$ polytope is also a positive geometry with boundary of all codimensions with a corresponding unique meromorphic form.  Analysing emergence of S-matrix via geometric and combinatorial structures is a radical attempt to penetrate the S-matrix program as some of the hard postulates such as unitarity and locality emerge thanks to the boundary structure of the positive geometries. 
 
Although the class of amplitudes (or more precisely amplitude integrands at loop level)  that are meromorphic forms associated to positive geometries is ever increasing,  (see e.g. \cite{gravitohedron, He:2023rou}), various garden variety QFTs such as QED, Yang-Mills (YM) theory has  been resistant to positive geometries program untill recently. 

During past couple of years however,  the real potential of positive geometry program is being realised. The ABHY associahedron (or $\hat{D}_{n}$ polytope e.g.) are a convex polytopes whose family of normals (called the ${\bf g}$-vectos) to the co-dimension one facets form a fan in ${\bf R}^{\vert {\cal K}_{n} \vert}$. This fan uniquely determines a set of piece-wise linear forms which are dual to the ${\bf g}$-vectors which are known as headlight functions, \cite{nima2309, pilaud-maneville}.\footnote{In representation theory literature, headlight functions are known as $c$-vectors, see e.g. \cite{pilaud-maneville}.}

In \cite{nima2309, nima2311} it was shown that the headlight functions along with co-ordinates of the kinematic space can be used to calculate the $\textrm{Tr}(\phi^{3})$ amplitude at arbitrary order in the loop expansion and the genus ($\frac{1}{N}$ for $SU(N)$) expansion in the parametric space of Schwinger parameters. This representation of amplitude is known as the curve integral formula and once we view the relationship between positive geometries and scalar amplitudes through the lens of parametric space, then a large class of amplitudes of theories such as non-linear sigma model (NLSM) theory and Yang-Mills theory arise as certain deformations (obtained via shift of kinematic variables) of the $\textrm{Tr}(\phi^{3})$ amplitudes,\cite{nima2312:hiddenzero, nima2401:scaffold}. In the seminal paper \cite{nima2401:scaffold}, the authors proved that a specific deformation of the (stringy) $\textrm{Tr}(\phi^{3})$ curve integral formula generates  YM amplitudes at all orders in perturbative expansion !

As striking as these results are, it is worth asking if there is a map from the canonical form defined by positive geometries in the kinematic space and S-matrix of gauge theories. \emph{This is the question we analyse in the current paper.} There have been several tantalising hints in the recent literature that there should exist a geometric operation that maps the meromorphic form in the kinematic space to the Yang-Mills amplitude. The most prominent among those is the well known color kinematics duality which has been explicitly established for tree-level and one loop amplitudes \cite{BCJ, Boels_CKdualityoneloop, colorkinreview}.  For tree-level amplitudes another evidence came from the beautiful paper of Cheung, Shen and Wen \cite{cheung:transmutation} in which an explicit map was constructed that maps the Yang-Mills amplitude to bi-adjoint scalar amplitude.  In \cite{mom_amplitu_meets_assoc}, authors derived a direct relationship between the canonical form of the ABHY associahedron and the canonical form of the (tree-level) momentum amplituhedron. But perhaps the strongest hint from our perspective, that such a map between associahedron form and Yang-mills amplitude may  exist in fact resides in one of the many results which were obtained in \cite{nima1711} for tree-level amplitudes. 

As reviewed above, the ABHY associahedron generates color ordered bi-adjoint scalar amplitude as the canonical form in the kinematic space. However, a perhaps less appreciated result of ABHY in \cite{nima1711} was their interpretation of the color dressed amplitude in bi-adjoint scalar theory. It was shown that the color dressed amplitude is in fact a scalar (as opposed to a meromorphic form) which is defined over an extension of the physical kinematic space, that ABHY termed big Kinematic space. Intuitively, the big kinematic space is the space of all (including the non-planar) generalised Mandelstam invariants generated by the on-shell momenta and momentum conservation.  We will review it's definition in section (\ref{kstlsec}).  ABHY proved that whereas the color-ordered amplitude in the bi-adjoint scalar theory  is the (pullback of the) canonical form fixed by the associahedron,  the color-dressed amplitude is a scalar in the big Kinematic space. Geometrically, this scalar can be obtained by considering the contraction of a certain multi-vector field (MVF) with the d-log form in the big Kinematic space. The multi-vector field is in fact a sum over all the vertices of the associahedron (which is in bijection with the set of all the tri-valent graphs) such that each term in the sum is determined by the orientation at each vertex of the graph, and structure constants associated to each of these tri-valent vertices.
 
 Although the re-writing of the color-dressed amplitude as a contraction of the form with an contravariant tensor in the kinematic space is rather a trivial observation, it  turns out to have some intriguing graph theoretic properties. These properties are also shared by an interesting mathematical object known as Corolla polynomial which was defined  in \cite{kreimer1208} by Kreimer, Sars and van Suijlekom.  At it's heart Corolla polynomial is a sum over monomials where each monomial is determined by the variables that label half edges of the graph and co-efficient of each monomial is either 0 or 1, \cite{kreimer-yeats}. 

In their seminal work, KSVS used the Corolla polynomial to construct a third graph polynomial $\hat{C}_{\gamma}$ (along with First and  Second Symanzik graph polynomials) in the parametric space of Schwinger parameters. $\hat{C}_{\gamma}$ was a differential operator valued meromorphic function in the parametric space which was used to spin up the parametric space integrand to obtain gauge theory integrand. KSVS proved the result at all loop orders as the Corolla differential naturally maps the $\phi^{3}$ vertex to a three gluon vertex  and maps a tri-valent vertex adjacent to a closed cycle to a ``gauge-invariant" three-gluon vertex such that the unphysical modes are removed, (see section \ref{correv} for details). As the differential is simply a representation of a polynomial, it has been used to construct parametric space representation of large class of gauge theory amplitudes on a fixed graph.  

This motivates us to consider contraction of the associahedron form (or in general canonical forms associated to positive geometries such as $\hat{D}_{n}$ polytope) with multi-vector field which is defined via Corolla polynomial. As we show in this paper, Kreimer's Corolla graph differential rather naturally leads to a map from each vertex of the positive geometry $v$ to a Corolla differential in the kinematic space $\hat{C}_{v}$. This leads to a Multi-vector field in the kinematic space whose contraction with the d log form generates color ordered YM amplitude at tree-level and the planar integrand at one loop. As we prove in section \ref{sec: transmutation},  the following diagram depicts the complete picture that combine our results with the seminal results obtained in \cite{cheung:transmutation}. 
\begin{align}
\begin{tikzpicture}[scale=0.5,baseline={([yshift=-.5ex]current bounding box.center)}]
\draw[-stealth] (-3,0.5) ..controls (0,2).. (3,0.5);
\draw[-stealth] (3,-0.5) ..controls (0,-2).. (-3,-0.5);
\node at (-8,-0.2) {${\cal A}_n^{\phi^3}  \ \xleftarrow[A_{n-3}^{\textrm{ABHY}}]{\text{     restricted to     }} \ \Omega_n(\{X_{ik}\})$};
\node at (3.9,0) {$\mathcal{A}_n^{\textrm{YM}}$};
\node at (0,0.65+0.5) {$P$};
\node at (0,1.75+0.5) {$\mathcal{A}_n = \langle P\,\Omega_n\rangle $};
\node at (0,-0.65-0.5) {$\Upsilon$};
\node at (0,-1.75-0.5) {$\Omega_n = \Upsilon \mathcal{A}_n$};
\end{tikzpicture}
\end{align}

Our representation of Corolla differential is in the kinematic space as opposed to the Corolla graph differential defined by KSVS which was in the parametric space. However ``fusion of the curve integral formula" with definition of Corolla graph differential follows immediately thanks to the existence of headlight functions and hence an off-shoot of our 
analysis is the curve integral formula for YM amplitude. We explicitly derive the formula for the integrand in the parametric space for both the tree-level amplitude and (planar) one loop amplitudes. 

This paper is organised as follows. In sections \ref{kst1l} and \ref{pgfsadnp}, we review the definition of kinematic space, positive geometries and d-log forms that these geometries define in the kinematic space. In section \ref{correv}, we review the  definition of Corolla polynomial.  In section \ref{sec: from triangulatiosn to MVF} we contract the Corolla induced MVF with the scattering form defined by the associahedron and show that it generates color-ordered YM amplitude at tree-level. In section \ref{sec: transmutation}, we present the inverse of the MVF, the differential form that \emph{transmutes} the YM amplitude into the canonical form $\Omega_n$ for associahedron. In section \ref{sec: one loop YM using POmega} we extend this analysis to $\hat{D}_{n}$ polytopes and use the Corolla polynomial associated to graphs with one closed cycle to define the MVF. Once again the planar one loop integrand for YM amplitude is realised as a contracted scalar. In section \ref{sec: curve integral}, we derive curve integral formula for gluon amplitudes. Although the power of curve integral formula resides in their universality with respect to loop expansion as well as genus expansion, we give explicit formula only for tree-level and planar one loop amplitudes. But as the Corolla polynomial is defined for a tri-valent graph with arbitrary number of cycles, the generalisation of our results to obtain curve integral formula for gluon at arbitrary order in perturbation theory including non planar contributions should be immediate. We finally end with a discussion of our results and several open issues.


\section{Kinematic Space for Tree-level and One loop amplitudes}\label{kst1l}
In this section we review the construction of the kinematic space ${\cal K}_{n}^{(I)}\, \vert\, I\, \in\, \{0,1\}$ when external states are scalars. The super-script $(I)$ indicates that the kinematic space is co-ordinatized by the kinematic variables whose zeros correspond to poles of tree-level amplitudes or one loop integrands respectively. We will also define an ambient space known as off-shell kinematic space which treats all the kinematic variables associated to internal or external edges as independent variables.

\subsection{Kinematic space for tree-level S matrix and it's enlarged cousins}\label{kstlsec}
The space of $n$ massless momenta modulo action of the Poincare group is known as the kinematic space ${\cal K}_{n}$. In $D\, >\, n$ dimensions, ${\cal K}_{n}$ is a $\frac{n(n-3)}{2}$ dimensional linear space and can be coordinatized in terms of so-called planar kinematic variables.
\begin{align}
X_{ij}\, =\, (p_{i}\, +\, \dots\, +\, p_{j-1})^{2}
\end{align}
In recent years, one has discovered that inspite of being a rather barren space with no dynamical information, ${\cal K}_{n}$ is the home of an infinite family of linear convex polytopes of a certain kind known as the positive geometry. Discovery of these positive geometries realised in ${\cal K}_{n}$ has geometrized the notion of S-matrix as amplitudes are naturally realised as canonical meromorphic forms on these positive geometries.  

In \cite{nima1711}, ABHY introduced a $2^{n-1} - n - 1$ dimensional space  ${\cal K}_{n}^{\star}$ which contains ${\cal K}_{n}$ as a linear subspace.  We review it's definition  essentially following \cite{nima1711}. 
Consider a family of variables $S_{I}$. Each $S_{I}$ is labelled by a subset  $I\, \subset\, \{1,\, \dots\, n\}$. These variables are abstract at this stage and are subjected to two conditions
\setlist{nolistsep}
\begin{enumerate}[noitemsep]
\item $ S_{I} = S_{\overline{I}}$
\item $ S_{I} = 0~,\quad \vert I\vert = \{\, 0, 1, n-1, n\}\label{sicon}$
\end{enumerate}
Thanks to these conditions, it can be seen that $S_{I}$ can be identified with all Mandelstam variables without imposing relations arising from momentum conservation at each vertex.\footnote{$S_{n-1} = S_{n} = 0$ can be identified with overall momentum conservation and on-shell constraints on external momenta.}
A simple example illustrates this point. 
\begin{align}\nonumber
{\cal K}_{4}^{\star}\, =\, \{\, S_{12},\, S_{23},\, S_{13}\, \}
\end{align}
It thus follows that ${\cal K}_{n}$ is a subspace of ${\cal K}_{n}^{\star}$ by imposing ``Seven term identity" \cite{nima1711} 
\begin{align}\label{seventerm}
S_{I_{1}I_{2}}\, +\, S_{I_{2}I_{3}}\, +\, S_{I_{1}I_{3}}\, =\, \sum_{i=1}^{4}\, S_{I_{i}}
\end{align}
${\cal K}_{n}^{\star}$ is the kinematic space on which color dressed amplitudes are well defined. We will now define a further enlargement of big kinematic space  and refer to it as ``off-shell Kinematic space", ${\cal K}_{n}^{\prime}$.

${\cal K}_{n}^{\prime}$ is defined by relaxing the second condition on $S_{I}$ variables, (condition (\ref{sicon})).    That is,  ${\cal K}_{n}^{\prime}$ is spanned by $S_{I}$ which are subjected to the following condition. 
\setlist{nolistsep}
\begin{enumerate}[noitemsep]
\item $ S_{I} = S_{\overline{I}}$
\item $ S_{I} = 0, \vert I\vert = \{\, 0, n\}$
\end{enumerate}
${\cal K}_{n}^{\prime}$ is $2^{n-1}-1$ dimensional. As an example, ${\cal K}_{4}^{\prime}$ is co-ordinatized by,
\begin{align}\nonumber
{\cal K}_{4}^{\prime}\, =\, \{\, S_{12},  S_{23},  S_{13}, S_{1}, S_{2}, S_{3}, S_{4}\, \}~.
\end{align}
For color ordered gauge theory amplitudes, the relevant space will in fact be a subspace ${\cal K}_{n}^{\textrm{os}}$ of ${\cal K}_{n}^{\prime}$ which is parametrized by treating all the chords and sides of an $n$-gon (whose vertices are labelled clockwise from $1\, \dots,\, n$) wih independent variables. As can be easily checked, ${\cal K}_{n}^{\textrm{os}}$ is $\frac{n(n-1)}{2}$ dimensional.

The projection, $\pi\, :\, {\cal K}_{n}^{\prime}\, \rightarrow\, {\cal K}_{n}^{\star}$ is via the following restriction. 
\begin{align}\label{starstartostar}
S_{I}\, =\, 0\, \forall\, \vert I\vert = \{1,n-1\}
\end{align}
As ${\cal K}_{n}^{\textrm{os}}\, :=\, {\bf R}^{\frac{n(n-1)}{2}}$, we can co-ordinatize it with orthogonal co-ordinates $ x_{i}, x_{ij}$. Upon mapping to the kinematic space of scattering data, these variables will become
\begin{equation}
\begin{aligned}
x_{i}\, &\rightarrow\, 0 ~,\qquad (1\leq i \leq n) \\
x_{ij}\, &\rightarrow\, (p_{i}\, +\, \dots\, p_{j-1})^{2}~,\qquad (1\leq i <j-1 < n) 
\end{aligned}
\end{equation}
We note that the $d\ln$ form on ${\cal K}_{n}$ can be trivially pulled back on ${\cal K}_{n}^{\textrm{os}}$ and without loss of generality we denote the pull-back as $\Omega_{n-3}$. 

We also introduce a family of four vectors $\{ \xi_{i}\, \vert 1\, \leq\, i\, \leq\, n\, \} \cup\, \{\, \xi_{ij}\, \vert\, (1\leq i <j-1 < n) \}$ such that
\begin{align}
x_{i} = \xi_{i}^{2}~,\quad  x_{ij}\, =\, \xi_{ij}^{2} ~,\quad (\xi_{ij} = -\xi_{ji})~. 
\end{align}
\subsection{Kinematic space with one loop planar variables}
We will denote the one loop kinematic space as ${\cal K}_{n}^{1-\textrm{l}}$.  It can be co-ordinatized by planar kinematic variables which are in bijection with set of all internal chords that can dissect a planar $n$-gon with a puncture. These variables are defined as, 
\begin{equation}
	\begin{aligned}
	X_{ij}\, &=\, (p_{i}\, +\, \dots\, p_{j-1})^{2}\\
	X_{i0}\, &=\, (p_{1}\, +\, \dots\, p_{i-1}\, +\, l)^{2}\\
	X_{ii}\, &=\, (p_{i+1}\, +\, \dots\, +\, p_{n}\, +\, p_{1}\, +\, \dots\, + p_{i-1})^{2}
	\end{aligned}
\end{equation}

One could enlarge ${\cal K}_{n}^{1-\textrm{l}}$ to the off-shell kinematic space ${\cal K}_{n}^{1-\textrm{l}\ \textrm{os}}$ exactly as for ${\cal K}_{n}$.  That is ${\cal K}_{n}^{1-\textrm{l}\, \textrm{os}}$ is a $\frac{n(n-3)}{2} + 3n$ dimensional linear space which is co-ordinatized by chordal lengths of internal chords as well as sides of the $n$-gon.  We denote the cartersian co-ordinates of  ${\cal K}_{n}^{1-\textrm{l}\, \textrm{os}}$ as $\{\, x_{ij}, x_{i0}, x_{ii}, x_{i}\, \}$ each of which can be written as a square of a 4-vector associated to the chords or boundaries of an $n$-gon with a puncture. 

\begin{align}
x_{ij} = \xi_{ij}^{2},\quad x_{ii}\, =\, \xi_{ii}^{2},\quad \xi_{i0}\, =\, \xi_{i0}^{2},\, x_{i}\, =\, \xi_{i}^{2} ~.
\end{align}
The projection from ${\cal K}_{n}^{1-\textrm{l}\, \textrm{os}}\, \rightarrow\, {\cal K}_{n}^{1-\textrm{l}}$ is obtained by 
\begin{align}
x_{i}&\rightarrow\, 0\nonumber\\
x_{ii}&\rightarrow\, (\sum_{j=1}^{n} p_{j})^{2}\nonumber\\
x_{i0} - x_{i-1,0}&=\, 2 p_{i} \cdot l + X_{1,i+1} - X_{1,i}
\end{align}

One can pull-back the d$\ln$ form on ${\cal K}_{n}^{1-l}$ to the big kinematic space. As we will now show, the contraction of this form with a multi-vector field defined by Corolla representation over the pseudo-triangulation generates color ordered one-loop Yang-Mills integrand.
\section{Positive geometries for scalars: Associahedron and $\hat{D}_{n}$ polytope.}\label{pgfsadnp}
In this section, we briefly review the definition of the (combinatorial) associahedron, $\hat{D}_{n}$ polytope and their convex realisation discovered by ABHY and AHST respectively, \cite{nima1711,ahst}. A more detailed discussion on these geometries and their embeddings inside the kinematic space (as well as their relationship with dissection quivers and so-called gentle algebras) can be found in \cite{nima1711,ferroreview, Herrmann_Trnka_review, baziermatte,pppp1906,mrunmay2206}.  We caution the reader that this section is not a self-contained introduction to the rich subject of positive geometries and their realisations via quiver representation theory.  One of our goals in  this section is to introduce the concept of ${\bf g}$ vector fan which forms the back-bone of many of the recent developments in emergence of scattering amplitudes from the positive geometry.

Associahedron $A_{n-3}$ is a simple polytope whose any two adjacent vertices are related by precisely one mutation.\footnote{A  polytope of dimension $D$ is called simple if every vertex is adjacent to $D$ co-dimension one faces.}
Although various convex realisations of $A_{n-3}$, all of which are diffeomorphic to the ABHY realisation generate $n$-point amplitudes of an infinite class of scalar field theories, we will not explicitly rely on this realisation in the kinematic space, but will only use certain data that $A_{n-3}$ generates.

The  positive geometry associated to one loop integrand of bi-adjoint scalar theory is known to be a close cousin of the cluster $D_{n}$ polytope which was discovered by Arkani-Hamed, Frost, Plamondon, Salavatori, and Thomas in \cite{afpst}. Their construction has been reviewed in \cite{mrunmay2206}.

\subsection{Combinatorial polytopes} 
Here, we review the construction of the \emph{combinatorial polytopes}, in terms of triangulations of the $n$-gons. 

\subsubsection{The associahedron} \label{sssec: combinatorial associahedron}
Associahedron (also known as Stasheff Polytope) is a a set of relations between triangulations of an $n$-gon. More in detail, this set has subsets indexed by $ k \in\, \{0, \dots, n-3\}$ where  $T_{k}$ is the set of all partial k-triangulations of an $n$-gon.  Thus ${\cal T}_{n-3}$ is the set of all complete triangulations of an $n$-gon. We refer to a complete triangulation as a ``vertex" and similarly all the elements in ${\cal T}^{k}$ as ``faces of co-dimension k". 

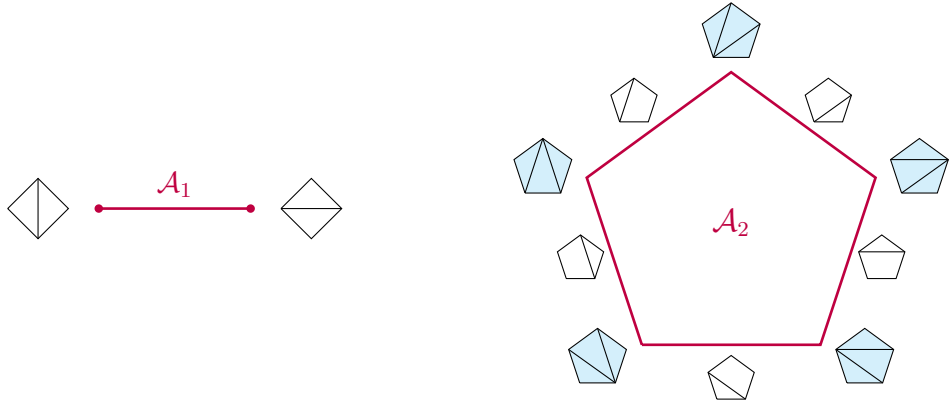
\begin{figure}[!tb]
    \centering
    \begin{tikzpicture}[scale=0.4,baseline={([yshift=-.5ex]current bounding box.center)}]
        \draw (-3,0) -- (-2,1) -- (-1,0) -- (-2,-1) -- (-3,0);
        \draw (-2,1) -- (-2,-1);
        \draw[line width=1pt, purple] (0,0) -- (5,0);
        \draw (6,0) -- (7,1) -- (8,0) -- (7,-1) -- (6,0);
        \draw (6,0) -- (8,0);
        \node[purple] at (2.5,0.8) {$\mathcal{A}_1$};
        \fill[purple] (0,0) circle [radius=0.14];
        \fill[purple] (5,0) circle [radius=0.14];
    \end{tikzpicture} 
    \hspace{2cm}
    \begin{tikzpicture}[scale=0.4,baseline={([yshift=-.5ex]current bounding box.center)}]
    \draw[line width=1pt, purple] (-5*0.588,-5*0.809) -- (-5*0.951,1.5) -- (0,5) -- (5*0.951,1.5) -- (5*0.588,-5*0.809) -- (-5*0.588,-5*0.809);
    \node[purple] at (0,0) {\large $\mathcal{A}_2$};
    \begin{scope}[scale=1, xshift=-6.2cm, yshift=1.8cm]
        \coordinate (a1) at (-0.588,-0.809);
        \coordinate (a2) at (-0.951,0.3);
        \coordinate (a3) at (0,1);
        \coordinate (a4) at (0.951,0.3);
        \coordinate (a5) at (0.588,-0.809);
        \fill[cyan!15] (a1) -- (a2) -- (a3) -- (a4) -- (a5);
        \draw (a1) -- (a2) -- (a3) -- (a4) -- (a5) -- (a1);
        \draw (a1) -- (a3);
        \draw (a3) -- (a5);
    \end{scope}
    \begin{scope}[scale=1, xshift=0cm, yshift=6.3cm]
        \coordinate (a1) at (-0.588,-0.809);
        \coordinate (a2) at (-0.951,0.3);
        \coordinate (a3) at (0,1);
        \coordinate (a4) at (0.951,0.3);
        \coordinate (a5) at (0.588,-0.809);
        \fill[cyan!15] (a1) -- (a2) -- (a3) -- (a4) -- (a5);
        \draw (a1) -- (a2) -- (a3) -- (a4) -- (a5) -- (a1);
        \draw (a1) -- (a3);
        \draw (a1) -- (a4);
    \end{scope}
    \begin{scope}[scale=1, xshift=6.2cm, yshift=1.8cm]
        \coordinate (a1) at (-0.588,-0.809);
        \coordinate (a2) at (-0.951,0.3);
        \coordinate (a3) at (0,1);
        \coordinate (a4) at (0.951,0.3);
        \coordinate (a5) at (0.588,-0.809);
        \fill[cyan!15] (a1) -- (a2) -- (a3) -- (a4) -- (a5);
        \draw (a1) -- (a2) -- (a3) -- (a4) -- (a5) -- (a1);
        \draw (a2) -- (a4);
        \draw (a1) -- (a4);
    \end{scope}
    \begin{scope}[scale=1, xshift=4.4cm, yshift=-4.5cm]
        \coordinate (a1) at (-0.588,-0.809);
        \coordinate (a2) at (-0.951,0.3);
        \coordinate (a3) at (0,1);
        \coordinate (a4) at (0.951,0.3);
        \coordinate (a5) at (0.588,-0.809);
        \fill[cyan!15] (a1) -- (a2) -- (a3) -- (a4) -- (a5);
        \draw (a1) -- (a2) -- (a3) -- (a4) -- (a5) -- (a1);
        \draw (a2) -- (a4);
        \draw (a2) -- (a5);
    \end{scope}
    \begin{scope}[scale=1, xshift=-4.4cm, yshift=-4.5cm]
        \coordinate (a1) at (-0.588,-0.809);
        \coordinate (a2) at (-0.951,0.3);
        \coordinate (a3) at (0,1);
        \coordinate (a4) at (0.951,0.3);
        \coordinate (a5) at (0.588,-0.809);
        \fill[cyan!15] (a1) -- (a2) -- (a3) -- (a4) -- (a5);
        \draw (a1) -- (a2) -- (a3) -- (a4) -- (a5) -- (a1);
        \draw (a5) -- (a3);
        \draw (a2) -- (a5);
    \end{scope}
    \begin{scope}[scale=0.8, xshift=-4cm, yshift=5cm]
        \coordinate (a1) at (-0.588,-0.809);
        \coordinate (a2) at (-0.951,0.3);
        \coordinate (a3) at (0,1);
        \coordinate (a4) at (0.951,0.3);
        \coordinate (a5) at (0.588,-0.809);
        \draw (a1) -- (a2) -- (a3) -- (a4) -- (a5) -- (a1);
        \draw (a1) -- (a3);
    \end{scope}
    \begin{scope}[scale=0.8, xshift=4cm, yshift=5cm]
        \coordinate (a1) at (-0.588,-0.809);
        \coordinate (a2) at (-0.951,0.3);
        \coordinate (a3) at (0,1);
        \coordinate (a4) at (0.951,0.3);
        \coordinate (a5) at (0.588,-0.809);
        \draw (a1) -- (a2) -- (a3) -- (a4) -- (a5) -- (a1);
        \draw (a1) -- (a4);
    \end{scope}
    \begin{scope}[scale=0.8, xshift=6.2cm, yshift=-1.5cm]
        \coordinate (a1) at (-0.588,-0.809);
        \coordinate (a2) at (-0.951,0.3);
        \coordinate (a3) at (0,1);
        \coordinate (a4) at (0.951,0.3);
        \coordinate (a5) at (0.588,-0.809);
        \draw (a1) -- (a2) -- (a3) -- (a4) -- (a5) -- (a1);
        \draw (a2) -- (a4);
    \end{scope}
    \begin{scope}[scale=0.8, xshift=0cm, yshift=-6.5cm]
        \coordinate (a1) at (-0.588,-0.809);
        \coordinate (a2) at (-0.951,0.3);
        \coordinate (a3) at (0,1);
        \coordinate (a4) at (0.951,0.3);
        \coordinate (a5) at (0.588,-0.809);
        \draw (a1) -- (a2) -- (a3) -- (a4) -- (a5) -- (a1);
        \draw (a2) -- (a5);
    \end{scope}
    \begin{scope}[scale=0.8, xshift=-6.2cm, yshift=-1.5cm]
        \coordinate (a1) at (-0.588,-0.809);
        \coordinate (a2) at (-0.951,0.3);
        \coordinate (a3) at (0,1);
        \coordinate (a4) at (0.951,0.3);
        \coordinate (a5) at (0.588,-0.809);
        \draw (a1) -- (a2) -- (a3) -- (a4) -- (a5) -- (a1);
        \draw (a3) -- (a5);
    \end{scope}
    \end{tikzpicture}
    \caption{Combinatorial associahedron: $\mathcal{A}_1$ and $\mathcal{A}_2$.}
    \label{fig:combinatorial associahedron}
\end{figure}

If 
\begin{align}
T_{1}^{n-3}, T_{2}^{n-3}\, \in\, {\cal T}_{n-3} ~,
\end{align}
such that
\begin{align}
T_{1}^{n-3}\, \cap\, T_{2}^{n-3}\, =\, T_{12}^{n-2} ~,
\end{align}
where $T_{12}^{n-2}\, \in\, {\cal T}^{n-2}$, then the two vertices are adjacent to each other through a one dimensional face. This defines a (hierarchy) of relations among partial triangulations which can be coherently put together as a positive geometry of dimension $n-3$ whose set of co-dimension $k$-faces is isomorphic to set of all $k$-partial triangulations. Refer to figure \ref{fig:combinatorial associahedron} for a picture of $A_1$ and $A_{2}$, the one and two dimensional associahedra. 

All the vertices of the associahedron (co-dimension $n$ facets) map to the complete triangulation of the $(n+3)$-gon. The definition implies that the two vertices are adjacent to each other, if and only if the two complete triangulations differ only by the flip of a single chord, keeping $(n-1)$ chords intact. The operation of changing precisely one chord is known as the mutation. Figure \ref{fig:combinatorial associahedron} show the simplest cases of associahedron: $\mathcal{A}_{1}$ and $\mathcal{A}_2$. 

Let $(ik)~, \ (1\leq i <k-1 < n)$ be the chord of $(n+3)$-gon connecting $i$ and $k$ vertices. Since the co-dimension 1 faces of $\mathcal{A}_n$ map to the triangulations using only one chord, we can label the faces by the chord present in partial triangulation bijective to the face: ($ik$). Counting the chords, the number of co-dimension 1 faces of the associahedron is $n(n-3)/2$. We denote the $k$ dimensional faces of $n$ dimensional associahedron as $\mathcal{A}_n^{[k]}$ so that we have: $|\mathcal{A}_n^{[n-1]}|=n(n-3)/2$. The number of vertices of the associahedron is as follows:
\begin{align} \label{number of vertices of associahedron}
    |\mathcal{A}_{n}^{[0]}| = C_{n+1} = \frac{(2n+2)!}{(n+2)!\,(n+1)!} ~,
\end{align}
$C_n$ being the Catalan number. For instance, we have $|\mathcal{A}_1^{[0]}|=2~, \quad |\mathcal{A}_2^{[0]}|=5~, \quad |\mathcal{A}_3^{[0]}|=14~,\quad\hdots$. 

The chord $(ik)$, of the $(n+3)$-gon decomposes the $(n+3)$-gon into an $(k-i+1)$-gon and another $(n-k+i+1)$-gon. So, we can conclude that the face of the associahedron is a product of smaller associahedra: 
\begin{align}
    \text{For } i\in [1,\hdots,n+1] ~, \quad k\in[2,\hdots,n+3-i]  ~:\qquad \mathcal{A}_{n} \xrightarrow{\p_{(i,i+k)}} \mathcal{A}_{k+1}\times \mathcal{A}_{n+1-k}
\end{align}

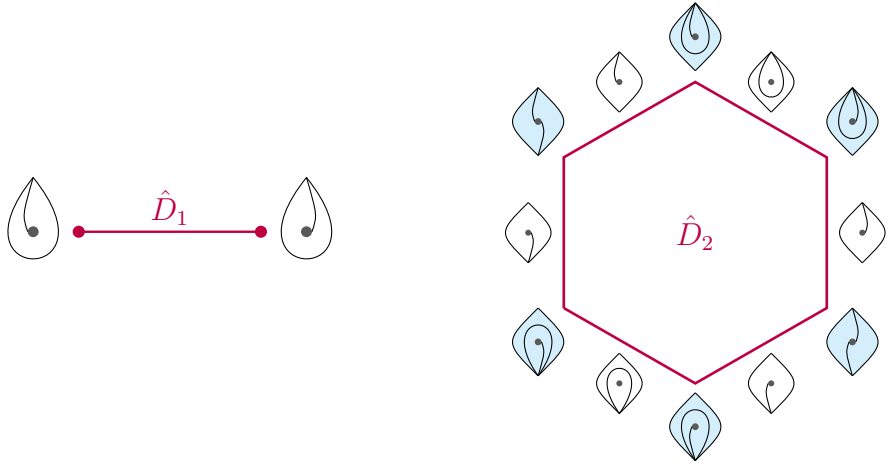
\begin{figure}[!ht]
    \centering
    \begin{tikzpicture}[scale=0.8,baseline={([yshift=-.5ex]current bounding box.center)}] 
        \draw[line width=0.9pt,purple] (-1.5,0) -- (1.5,0);
        \begin{scope}[scale=0.9,xshift=-2.5cm, yshift=0cm]
            \fill[gray!70!black] (0,0) circle [radius=0.1];
            \draw (0,1) .. controls (-1.6,-1) and (1.6,-1) .. (0,1);
            \draw (0,1) .. controls (-0.2,0.3) .. (-0.1,0);
            \end{scope}
        \begin{scope}[scale=0.9,xshift=2.5cm, yshift=0cm]
            \fill[gray!70!black] (0,0) circle [radius=0.1];
            \draw (0,1) .. controls (-1.6,-1) and (1.6,-1) .. (0,1);
            \draw (0,1) .. controls (0.2,0.3) .. (0.1,0);
            \end{scope}
        \node[purple] at (0,0.4) {\large $\hat{D}_1$};
        \fill[purple] (-1.5,0) circle [radius=0.1];
        \fill[purple] (1.5,0) circle [radius=0.1];
    \end{tikzpicture} \hspace{1cm}
    \begin{tikzpicture}[scale=0.5,baseline={([yshift=-.5ex]current bounding box.center)}]
        \draw[line width=1pt, purple] (-3.464,0) -- (0,-2) -- (3.464,0) -- (3.464,4) -- (0,6) -- (-3.464,4) -- (-3.464,0);
        \node[purple] at (0,2) {\large $\hat{D}_2$};
        \begin{scope}[scale=0.9,xshift=-4.6cm, yshift=-1cm]
        \fill[cyan!15] (0,1) .. controls (-1,0) .. (0,-1) .. controls (1,0) .. (0,1);
        \draw (0,1) .. controls (-1,0) .. (0,-1);
        \draw (0,1) .. controls (1,0) .. (0,-1);
        \fill[gray!70!black] (0,0) circle [radius=0.1];
        \draw (0,-1) .. controls (0.2,-0.3) .. (0.1,0);
        \draw (0,-1) .. controls (1.4,1) and (-1.4,1) .. (0,-1);
        \end{scope} 
        \begin{scope}[scale=0.8,xshift=-5.5cm, yshift=2.5cm]
        \draw (0,1) .. controls (-1,0) .. (0,-1);
        \draw (0,1) .. controls (1,0) .. (0,-1);
        \fill[gray!70!black] (0,0) circle [radius=0.1];
        \draw (0,-1) .. controls (0.2,-0.3) .. (0.1,0);
        \end{scope}
        \begin{scope}[scale=0.9,xshift=-4.6cm, yshift=5.5cm]
        \fill[cyan!15] (0,1) .. controls (-1,0) .. (0,-1) .. controls (1,0) .. (0,1);
        \draw (0,1) .. controls (-1,0) .. (0,-1);
        \draw (0,1) .. controls (1,0) .. (0,-1);
        \fill[gray!70!black] (0,0) circle [radius=0.1];
        \draw (0,-1) .. controls (0.2,-0.3) .. (0.1,0);
        \draw (0,1) .. controls (-0.2,0.3) .. (-0.1,0);
        \end{scope}
        \begin{scope}[scale=0.8,xshift=-2.5cm, yshift=7.5cm]
        \draw (0,1) .. controls (-1,0) .. (0,-1);
        \draw (0,1) .. controls (1,0) .. (0,-1);
        \fill[gray!70!black] (0,0) circle [radius=0.1];
        \draw (0,1) .. controls (-0.2,0.3) .. (-0.1,0);
        \end{scope}
        \begin{scope}[scale=0.9,xshift=0cm, yshift=8cm]
        \fill[cyan!15] (0,1) .. controls (-1,0) .. (0,-1) .. controls (1,0) .. (0,1);
        \draw (0,1) .. controls (-1,0) .. (0,-1);
        \draw (0,1) .. controls (1,0) .. (0,-1);
        \fill[gray!70!black] (0,0) circle [radius=0.1];
        \draw (0,1) .. controls (-0.2,0.3) .. (-0.1,0);
        \draw (0,1) .. controls (-1.4,-1) and (1.4,-1) .. (0,1);
        \end{scope}
        \begin{scope}[scale=0.8,xshift=2.5cm, yshift=7.5cm]
        \draw (0,1) .. controls (-1,0) .. (0,-1);
        \draw (0,1) .. controls (1,0) .. (0,-1);
        \fill[gray!70!black] (0,0) circle [radius=0.1];
        \draw (0,1) .. controls (-1.4,-1) and (1.4,-1) .. (0,1);
        \end{scope}
        \begin{scope}[scale=0.9,xshift=4.6cm, yshift=5.5cm]
        \fill[cyan!15] (0,1) .. controls (-1,0) .. (0,-1) .. controls (1,0) .. (0,1);
        \draw (0,1) .. controls (-1,0) .. (0,-1);
        \draw (0,1) .. controls (1,0) .. (0,-1);
        \fill[gray!70!black] (0,0) circle [radius=0.1];
        \draw (0,1) .. controls (-1.4,-1) and (1.4,-1) .. (0,1);
        \draw (0,1) .. controls (0.2,0.3) .. (0.1,0);
        \end{scope}
        
        \begin{scope}[scale=0.8,xshift=5.5cm, yshift=2.5cm]
        \draw (0,1) .. controls (-1,0) .. (0,-1);
        \draw (0,1) .. controls (1,0) .. (0,-1);
        \fill[gray!70!black] (0,0) circle [radius=0.1];
        \draw (0,1) .. controls (0.2,0.3) .. (0.1,0);
        \end{scope}
        \begin{scope}[scale=0.9,xshift=4.6cm, yshift=-1cm]
        \fill[cyan!15] (0,1) .. controls (-1,0) .. (0,-1) .. controls (1,0) .. (0,1);
        \draw (0,1) .. controls (-1,0) .. (0,-1);
        \draw (0,1) .. controls (1,0) .. (0,-1);
        \fill[gray!70!black] (0,0) circle [radius=0.1];
        \draw (0,1) .. controls (0.2,0.3) .. (0.1,0);
        \draw (0,-1) .. controls (-0.2,-0.3) .. (-0.1,0);
        \end{scope}
        \begin{scope}[scale=0.8,xshift=2.5cm, yshift=-2.5cm]
        \draw (0,1) .. controls (-1,0) .. (0,-1);
        \draw (0,1) .. controls (1,0) .. (0,-1);
        \fill[gray!70!black] (0,0) circle [radius=0.1];
        \draw (0,-1) .. controls (-0.2,-0.3) .. (-0.1,0);
        \end{scope}
        \begin{scope}[scale=0.9,xshift=0cm, yshift=-3.5cm]
        \fill[cyan!15] (0,1) .. controls (-1,0) .. (0,-1) .. controls (1,0) .. (0,1);
        \draw (0,1) .. controls (-1,0) .. (0,-1);
        \draw (0,1) .. controls (1,0) .. (0,-1);
        \fill[gray!70!black] (0,0) circle [radius=0.1];
        \draw (0,-1) .. controls (-0.2,-0.3) .. (-0.1,0);
        \draw (0,-1) .. controls (1.4,1) and (-1.4,1) .. (0,-1);
        \end{scope}
        \begin{scope}[scale=0.8,xshift=-2.5cm, yshift=-2.5cm]
        \draw (0,1) .. controls (-1,0) .. (0,-1);
        \draw (0,1) .. controls (1,0) .. (0,-1);
        \fill[gray!70!black] (0,0) circle [radius=0.1];
        \draw (0,-1) .. controls (1.4,1) and (-1.4,1) .. (0,-1);
        \end{scope}
    \end{tikzpicture}
    \caption{Combinatorial $\hat{D}_1$ and $\hat{D}_2$ polytopes.}
    \label{fig:comb Dhat 1 and 2 polytope}
\end{figure}

\paragraph{Bracketing}
A planar $n$-gon with clockwise orientation of vertices $\{\, 1,\, \dots,\, n\, \}$ a triangulation is a set of $(n-3)$ chords which can only intersect in their end points $\in\, \{1,\, \dots,\, n\, \}$. 
\begin{align}
T\, =\, \{\, (i_{1}, j_{1}),\, \dots,\, (i_{n-3}, j_{n-3})\, \} 
\end{align}
$T$ dissects the $n$-gon into $(n-2)$ triangles and there is a bijection between the set of all triangulation and placing $n-2$ parenthesis  on the sequence $\{\, 1,\, \dots,\, n\, \}$. 
\begin{align}
\{(13)\,,\,(14)\,,\hdots,\,(1,n-1)\} \equiv \big(\hdots (((12)3)4)\hdots \big)n
\end{align}
Mutation exchanges one of the brackets with a distinct one, keeping the rest of the sequence intact. For instance,
\begin{align}
\big(\hdots (((12)3)4)\hdots \big)n \ \ \longleftrightarrow \ \ \big(\hdots ((1(23))4)\hdots \big)n 
\end{align}

\subsubsection{$\hat{D}_{n}$ Polytope} \label{sssec: combinatorial Dhatn}
The $\hat{D}_n$ polytope, or the surfacehedron is the polytope that has a bijection between its co-dimension $k$ facets and the pseudo-triangulations of an $n$-gon with a puncture in its interior. In the presence of the puncture, the chords dissecting the $n$-gon can originate or terminate either at the vertices of the $n$-gon, or via spiralling around the puncture. As a result, there are two mutually incompatible chords joining a vertex of the $n$-gon with the puncture. To differentiate these two possible chords, we draw them curved, as shown in figure \ref{fig:comb Dhat 1 and 2 polytope}. Let us label the two possible chords joining the vertex $k$ of the $n$-gon with the puncture as $(k0)$ and $(k0')$. Note that the curve $(k0)$ is incompatible with $(i0')$, for all $i$ and $k$, i.e., the two can not appear together in a triangulation.

Along with the chords $(k0)$ and $(k0')$, and the chords connecting the vertices of the $n$-gon, here are all the possible chords (equivalently the faces of $\hat{D}_n$ polytope):
\begin{align}
    \big\{\,(i,i+k)\ \ \big|\ \  i\in [1,n] \, ,\ k\in [2,n]\,\big\} ~, \qquad \big\{ \big(k0\big) \ \ \big| \ \ k\in [1,n] \big\} ~, \qquad \big\{ \big(k0'\big) \ \ \big| \ \ k\in [1,n] \big\}
\end{align}
Note that the chords $(ik)$ and $(ki)$ are distinct, corresponding to two ways of going around the puncture. We refer to the curve going \emph{left} around the puncture, while going from $i$ to $(i+k)$, as $(i,i+k)$. For instance, look at \eqref{fig: example of chords ik ki}.

The $\hat{D}_n$ polytope has $n(n+1)$ faces. The faces are themselves smaller $\hat{D}_n$ polytopes, or the associahedra. Let us consider the face $(k0)$. With a $(k0)$ chord present, the $n$-gon with vertices $[1,\hdots,n]$ with a puncture effectively become a $(n+2)$-gon with vertices $[1,\hdots,k-1,k,0,k,k+1,\hdots,n]$ without any puncture. So, we have
\begin{align}
    \hat{D}_n \xrightarrow{\p_{(k0)}} \mathcal{A}_{n-1}~, \qquad \quad     \hat{D}_n \xrightarrow{\p_{(k0')}} \mathcal{A}_{n-1}~.
\end{align}
Let us consider the face $(i,i+k)$. This chord splits the $n$-gon into $(k+1)$-gon without the puncture, and an $(n-k+1)$-gon with the puncture, so we have:
\begin{align}
    \hat{D}_n \xrightarrow{\p_{(i,i+k)}} \mathcal{A}_{k-2} \times \hat{D}_{n-k+1}~.
\end{align}
For example, let us look at the faces $(25)$ and $(52)$ of $\hat{D}_5$:
\begin{align} \label{fig: example of chords ik ki}
    \begin{tikzpicture}[scale=0.9,baseline={([yshift=-.5ex]current bounding box.center)}]
        \coordinate (a1) at (-0.588,-0.809);
        \coordinate (a2) at (-0.951,0.3);
        \coordinate (a3) at (0,1);
        \coordinate (a4) at (0.951,0.3);
        \coordinate (a5) at (0.588,-0.809);
        \draw (a1) -- (a2) -- (a3) -- (a4) -- (a5) -- (a1);
        \fill[gray!70] (0,0) circle [radius=0.1];
        \draw (a2) ..controls (0.4,0.5).. (a5);
        \draw (a2) ..controls (-0.2,-0.3).. (a5);
        \node at (-0.9,-0.9) {1};
        \node at (-1.3,0.3) {2};
        \node at (0,1.35) {3};
        \node at (1.3,0.3) {4};
        \node at (0.9,-0.9) {5};
        \node at (-0.4,-0.5) {\small $(52)$};
        \node at (0,0.65) {\small $(25)$};
    \end{tikzpicture}\hspace{1cm}
    \mathcal{A}_1 \times \hat{D}_3 \xleftarrow{\p_{(25)}} \hat{D}_5 \xrightarrow{\p_{(52)}} \mathcal{A}_0 \times \hat{D}_4 ~.
\end{align}

While discussing the convex realization of the $\hat{D}_n$ polytope, we shall count the number of vertices. Here, we mention the result:
\begin{align}
    |\hat{D}_n^{[0]}| = 2(2n-1)C_{n-1} = \frac{2(2n-1)!}{n!\,(n-1)!}
\end{align}

\subsection{The g-vector fan, and the convex realizations}\label{gvfacr}
In this section, we review the computation of $g$-vectors which are uniquely determined by the combinatorics of dissections of an $n$-gon. $g$-vectors are one of the most crucial ingredients in linking combinatorial polytopes such as accordiohedra with S-matrix in quantum field theory. We will only analyse the $g$-vectors which are generated by combinatorics of triangulation of an $n$-gon and pseudo-triangulation of $n$-gon with a puncture in the center, \cite{pppp1906,mrunmay2206}. Recent advancements in the quiver representation theory lead to following results.

Let $\{{\cal D}^{I}_{ij}\}$ be the set of all the dissections of an n-gon with $I$ punctures where $I\, \in\, \{0,1\}$. Then, 
\begin{align}
\big\{ {\cal D}_{ij}^{I}\big\}\, \xleftrightarrow{\text{one to one}}\, \left\{{\bf g}_{ij}^{I}\, \in\, \left(\mathcal{E}\equiv{\bf R}^{\frac{n(n-3)}{2} + I (\frac{n^{2}}{2} + \frac{3n}{2})}\right) \right\}
\end{align}
That is, each dissection corresponds to a unique vector in a cartesian space of dimension $\frac{n(n-3)}{2}$ in the case of triangulation and $n^{2}$ in the case of pseudo-triangulations.  The map depends on a reference (pseudo)-triangulation which fixes the labels of the axis of embedding space. Set of all $g$ vectors in both of these cases form an essential and complete fan, which means the following:
\begin{itemize}
\item Given a set of dissections that triangulate an n-gon, or pseudo-triangulate an $n$-gon with a puncture, the positive span of the corresponding set of ${\bf g}$ vectors forms a cone such that set of all such cones only intersect in the origin and their union span the entire ${\cal E}$.  For example, look at figures \ref{fig:Dhat 2 polytope} and \ref{fig:cones for n=3}.
\item There is a unique convex realisation of the associahedron and $\hat{D}_{n}$ polytope such that the set of normal vectors to all the co-dimension one faces is in bi-jection with the set of all the ${\bf g}$-vectors. 
\item These realisations are precisely, the ABHY associahedron and convex realisation of $\hat{D}_{n}$ polytope discovered by Arkani-Hamed, Frost, Plamondon, Salvatori and Thomas.
\end{itemize}
We start with a reference triangulation or a pseudo-triangulation  (with a puncture in case of loop-level graphs) and drawing triangular paths in each of the cells of $T_0$. Let us mark the triangular paths clockwise in each cell. Refer to figures \ref{fig:paths and curves for n=5}, \ref{fig:nonspiral curves}, and \ref{fig:spiral curves} for examples. Now, we look for paths starting from an external edge (particle) $i$. We associate a unique vector $g_C$ in $\mathcal{E}$ to a path $C$. Note that there are clockwise `directions' on each segment of the path. We obtain the `mountainscape' for a path, with the chords of $T_0$ as intermediate nodes, by imagining a `forward' (in the direction of path) direction as going up, and a `backward' direction as coming down. Let $\{l_i\}_{i=[1,\hdots,\text{dim}(\mathcal{E})]}$ be the edges of the graph dual to $T_0$. With a peak (pinch in directions) in the mountainscape at the edge $l_i$, we include $+1$ in the $i$th entry of $\vec{g}_C$ and a $-1$ for a dip in the mountainscape (both directions pointing away from the chord):
\begin{align}
    \big(\vec{g}_C\big){}_m = \sum_{\{k |\ \text{peaks at } l_k\}} \delta_{m,k} - \sum_{\{k |\ \text{dips at } l_k\}} \delta_{m,k} 
\end{align}

Another way to characterize all the $g$-vectors is due to a recent work \cite{nima2309}, where the authors choose a particular triangulation $T_0$ and draw all the non-self-intersecting curves on the ribbon graph for $T_0$.  The two methods are completely equivalent, and the \emph{curves} are precisely the \emph{paths} described earlier. We shall treat \emph{curves}, \emph{paths} and \emph{chords} as synonymes in this article. As an extra structure, one can associate the following momentum with a path/curve $C$:
\begin{align}
    P_C = p_{\text{initial}} + \sum_{\text{right turns}} p_{\text{(incoming from left)}} ~.
\end{align}
Analogous to the pseudo-triangulation, we can obtain the `mountainscape' for a curve, with the edges of the graph dual to $T_0$ as intermediate nodes, by imagining a left turn as going up, and a right turn as coming down. Once we have the mountainscape, we can read off the $g$-vectors for curves.

\subsubsection{g-vector fan for tree-level graphs} 
\begin{figure}[!hb]
    \centering
    \includegraphics[scale=0.2]{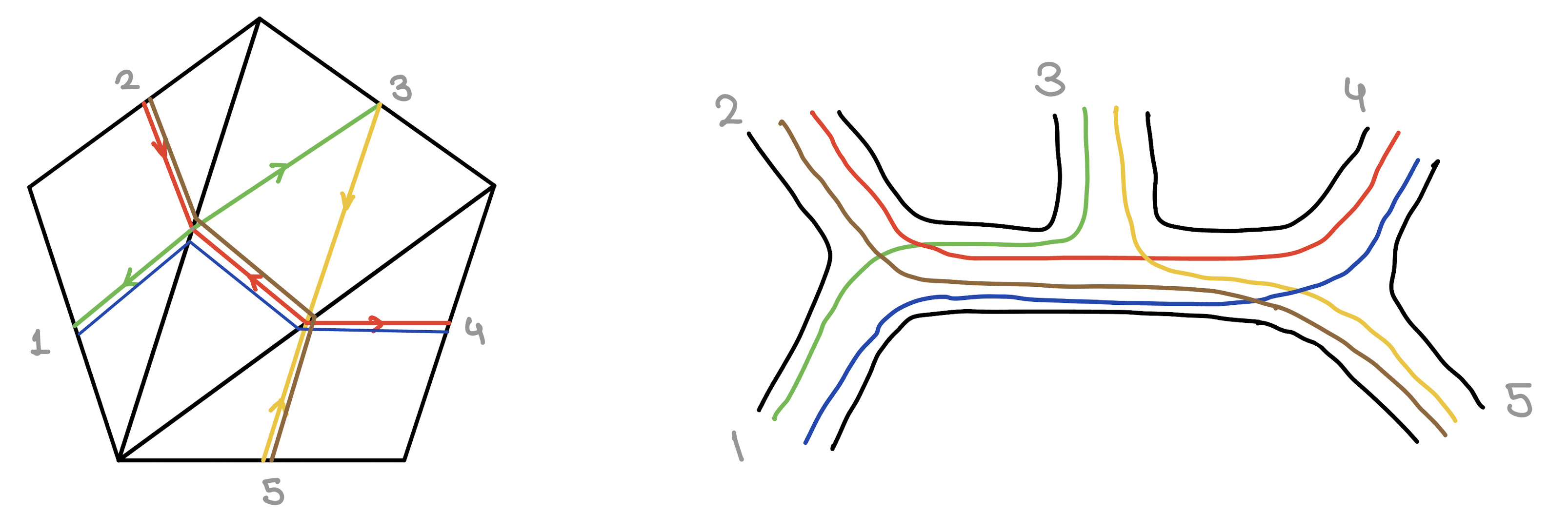}
    \caption{Triangulated paths and curves on the ribbon graph for tree level $n=5$.}
    \label{fig:paths and curves for n=5}
\end{figure}

Let us choose the ray-like triangulation as the reference triangulation $T_0$. We draw the paths on the triangulated $n$-gon or equivalently the curves on the ribbon graph for $T_0$. Let us denote the paths joining the leg $i$ to the leg $j$ by $C_{ij}$. These are precisely the chords (or faces) $(ij)$ discussed in the section \ref{sssec: combinatorial associahedron}. There are $n(n-3)/2$ such paths. Note that $C_{ij}$ and $C_{ji}$ are equivalent and have the momenta $P_{C_{ij}} = \big(p_i+\hdots + p_{j-1}\big)$ associated to 
these. For example, figure \ref{fig:paths and curves for n=5} depicts all the curves and equivalent paths for $n=5$. 

For arbitrary $n$, the following are the general $g$-vectors: 
\begin{equation}
\begin{aligned}
    \text{For } k\in [3,\hdots,n-1]~, &\qquad \big(\vec{g}_{1k}\big){}_m = \delta_{(k-2)m} \\
    \text{For } k\in [2,\hdots,n-2]~, &\qquad \big(\vec{g}_{kn}\big){}_m = -\delta_{(k-1)m} \\
    \text{For } i\in [2,\hdots,n-3]~, k\in[i+2,\hdots,n-1]~, &\qquad \big(\vec{g}_{ik}\big){}_m = -\delta_{(i-1)m} + \delta_{(k-2~)m} 
\end{aligned}
\end{equation}
The set of all these vectors gives us the \emph{g-vector fan}. 

\subsubsection{g-vector fan for planar one-loop}
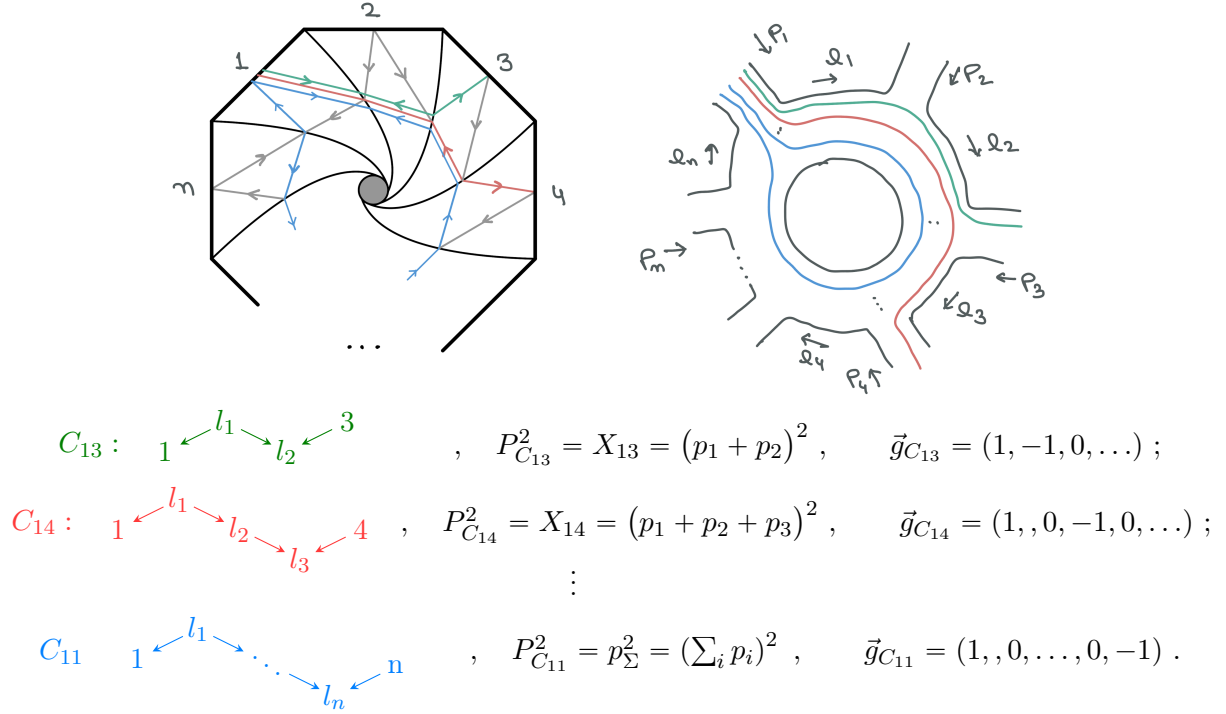
\begin{figure}[!ht]
    \centering
    \includegraphics[scale=0.2]{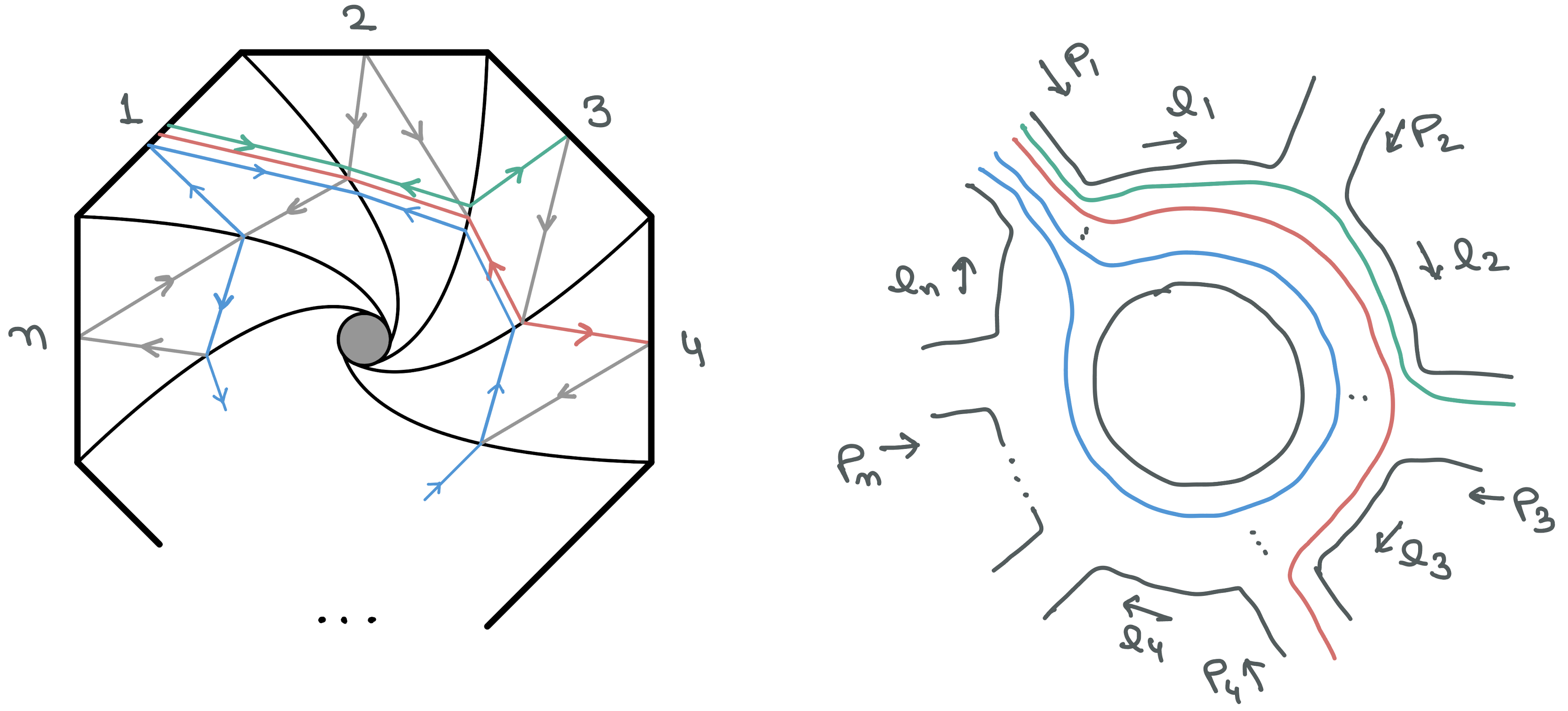}
    \begin{tikzpicture}[scale=0.2,baseline={([yshift=-.5ex]current bounding box.center)}]
    \begin{scope}[green!50!black]
        \node at (-5,1) {$C_{13}: $};
        \node at (0,0.5) {1};
        \draw[-stealth] (3,2) -- (1,1);
        \node at (4,2.5) {$l_1$};
        \draw[-stealth] (5,2) -- (7,1);
        \node at (8,0.5) {$l_2$};
        \draw[-stealth] (11,2) -- (9,1);
        \node at (12,2.5) {3};
    \end{scope}
    \node at (42,1) {$~, \quad P_{C_{13}}^2 = X_{13} = \big(p_1+p_2\big)^2 ~, \qquad \vec{g}_{C_{13}} = (1,-1,0,\hdots)~;$};
    \end{tikzpicture} \\
    \begin{tikzpicture}[scale=0.2,baseline={([yshift=-.5ex]current bounding box.center)}]
    \begin{scope}[red!70!pink]
        \node at (-5,1) {$C_{14}: $};
        \node at (0,0.5) {1};
        \draw[-stealth] (3,2) -- (1,1);
        \node at (4,2.5) {$l_1$};
        \draw[-stealth] (5,2) -- (7,1);
        \node at (8,0.5) {$l_2$};
        \draw[-stealth] (9,0) -- (11,-1);
        \node at (12,-1.5) {$l_3$};
        \draw[-stealth] (15,0) -- (13,-1);
        \node at (16,0.5) {4};
    \end{scope}
    \node at (45,1) {$ ~, \quad P_{C_{14}}^2 = X_{14} = \big(p_1+p_2+p_3\big)^2 ~, \qquad \vec{g}_{C_{14}} = (1,,0,-1,0,\hdots)~;$};
    \node at (30,-2.5) {$\vdots$};
    \end{tikzpicture}\\
    \begin{tikzpicture}[scale=0.2,baseline={([yshift=-.5ex]current bounding box.center)}]
    \begin{scope}[blue!50!cyan]
        \node at (-5,1) {$C_{11}$};
        \node at (0,0.5) {1};
        \draw[-stealth] (3,2) -- (1,1);
        \node at (4,2.5) {$l_1$};
        \draw[-stealth] (5,2) -- (7,1);
        \node at (8.5,0.6) {$\ddots$};
        \draw[-stealth] (10,-0.5) -- (12,-1.5);
        \node at (13,-2) {$l_n$};
        \draw[-stealth] (16,-0.5) -- (14,-1.5);
        \node at (17,0) {n};
    \end{scope}
    \node at (45,1) {$    ~, \quad P_{C_{11}}^2 = p_{\Sigma}^2 = \left(\sum_ip_i\right)^2 ~, \qquad \vec{g}_{C_{11}} = (1,,0,\hdots,0,-1)~.$};    
    \end{tikzpicture}
    \caption{Paths (curves) with their g-vectors lying on the $\Delta$ plane. Attributable to the cyclic symmetry, there are $n$ copies of such curves, obtained by shifting the starting point $1\to 2\to \hdots \to n$.}
    \label{fig:nonspiral curves}
\end{figure}

Let us choose $\{(k0)\ | \ k\in[1,n]\}$ as the preferred triangulation $T_0$, dual to the $n$-gon loop graph. Let $\{l_i\}_{1,\hdots,n}$ denote the chords. We draw all the paths on the $n$-gon with a puncture with $\{(k0)\}$ chords, and equivalently the paths on the corresponding ribbon graph. Thanks to the cyclic symmetry of the problem, all the paths come in $n$ copies. In figure \ref{fig:nonspiral curves}, we enlist all the \emph{non-spiral} paths, momenta associated with them, and the corresponding $g$-vectors. These are bijective to the $(ik)$ chords (or faces) discussed in the section \ref{sssec: combinatorial Dhatn}.

Let us denote the curve joining $i$ to $j$ by $C_{ij}$, such that it takes immediate left turn after $i$:  
\begin{align}
    \text{For }\quad i\in [1,n] ~,\quad j\in [i+2,i+n] ~,\qquad \quad  C_{ij} = i\,L\,l{i}\,R\,l_{i+1} \,R\hdots R\,t_{j-1}Lj ~,\nonumber\\
    P^2_{C_{ij}} =X_{ij}= (p_i+p_{i+1}+\hdots  + p_{j-1})^2 ~, \qquad (\vec{g}_{ij}){}_m = \delta_{im} - \delta_{(j-1)m} \label{general gvectors}
\end{align}

A total of $n(n-1)$ number of such $C_{ij}$ curves exist. Given $\{t_i\}_{i=1}^n$ being the coordinates on the $\mathcal{E}$, we can see that the g-vectors corresponding to these $C_{ij}$ curves lie on a co-dimension 1 hyperplane: $\sum_{i=1}^nt_i = 0$, called the $\Delta$-plane. One can check this by the explicit form of $g$-vectors from \eqref{general gvectors}. 
 
\begin{figure}[!ht]
    \centering
    \includegraphics[scale=0.2]{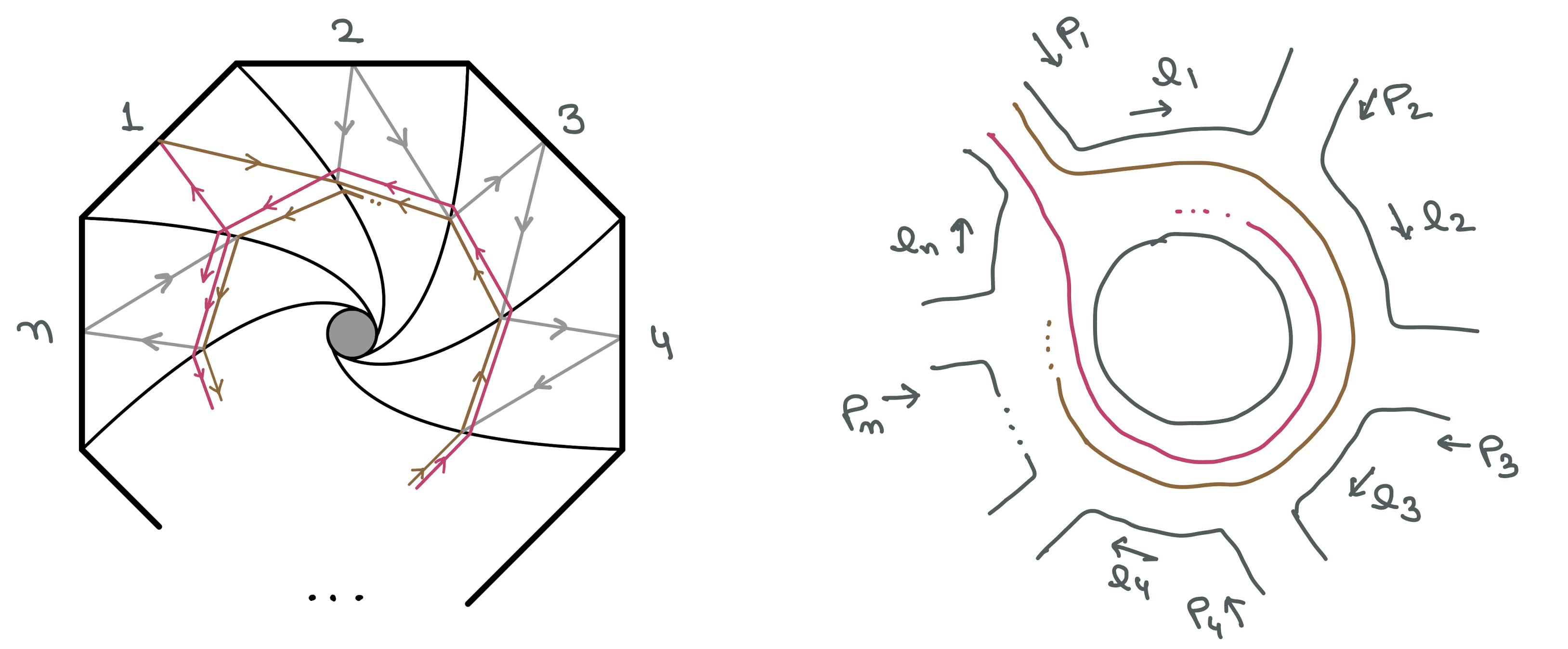}
        \begin{tikzpicture}[scale=0.2,baseline={([yshift=-.5ex]current bounding box.center)}]
        \begin{scope}[brown!70!black]
            \node at (-5,1) {$S_1~ :$};
            \node at (0.5,0.5) {1};
            \draw[-stealth] (3.5,2) -- (1.5,1);
            \node at (4.5,1) {$\bigg($};
            \node at (6,2.5) {$l_1$};
            \draw[-stealth] (7,2) -- (9,1);
            \node at (10.5,0.6) {$\ddots$};
            \draw[-stealth] (12,-0.5) -- (14,-1.5);
            \node at (15,-2) {$l_n$};
            \draw[-stealth] (16,-2.5) -- (18,-3.5);
            \node at (19.5,-2) {$\bigg)^{\infty}$};
        \end{scope}
        \node at (37,1) {$    ~, \quad P_{S_1} = l_{n} ~, \qquad \vec{g}_{S_1} = \big(1,0,\hdots,0\big)~;$};    
    \end{tikzpicture}
    \begin{tikzpicture}[scale=0.2,baseline={([yshift=-.5ex]current bounding box.center)}]
    \begin{scope}[pink!30!purple]
        \node at (-5,1) {$S_1'~ :$};
        \node at (0.5,-0.5) {1};
        \draw[-stealth] (1.5,-1) -- (3.5,-2);
        \node at (4.5,-1) {$\bigg($};
        \node at (6,-2.5) {$l_n$};
        \draw[-stealth] (9,-1) -- (7,-2);
        \node at (10.5,-0.5) {$\hdots$};
        \draw[-stealth] (14,1.5) -- (12,0.5);
        \node at (15,2) {$l_1$};
        \draw[-stealth] (18,3.5) -- (16,2.5);
        \node at (19.5,2) {$\bigg)^{\infty}$};
    \end{scope}
    \node at (37,1) {$    ~, \quad P_{S_1'} = -l_{n} ~, \qquad \vec{g}_{S_1'} = \big(0,\hdots,0,-1\big)~.$};    
    \end{tikzpicture}
    \caption{The spiraling paths (curves): $S_1$ and $S_1'$. There are $n$ copies of such curves, obtained by shifting the starting point $1\to 2\to \hdots \to n$.}
    \label{fig:spiral curves}
\end{figure}
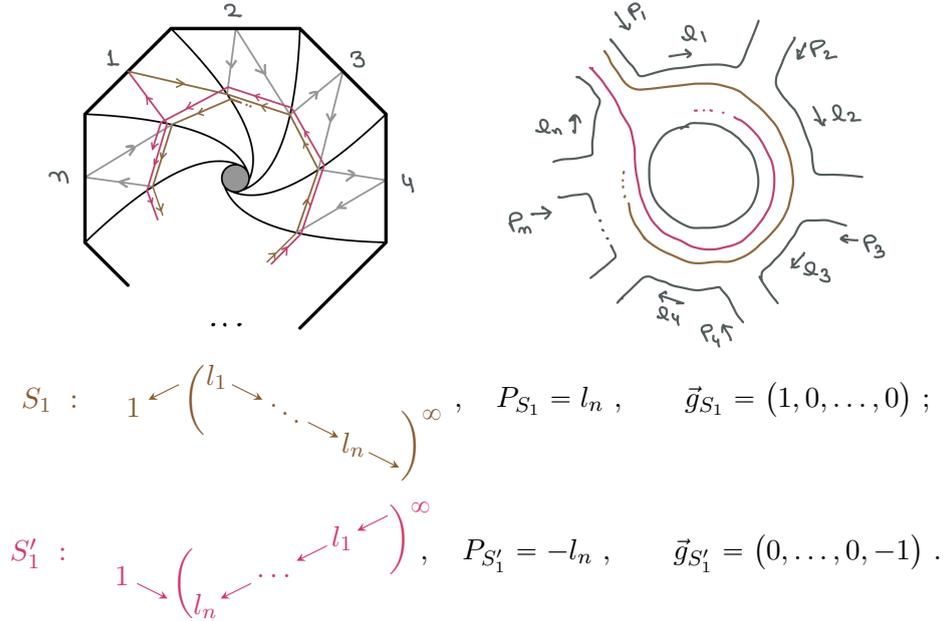

Along with the paths mentioned above, there are spiral paths depicted in figure \ref{fig:spiral curves}, which are termed as $S_i$ and $S_i'$:
\begin{align}
    \text{For }\quad i\in [1,n],\qquad \quad S_i &= iL\,\big(l_iR\,l_{i+1} R\,l_{i+2}R\hdots R\,l_{i+n-1}\,R\big)^{\infty}~,\nonumber \\
    &\hspace{2cm} (\vec{g}_{S_i})_m = \delta_{im} ~, \qquad\quad   P_{S_i} = l_{i-1}~.\\
    \text{For }\quad i\in [1,n],\qquad \quad S_i' &= iR\,\big(l_{i-1}L\,l_{i-2} L\,l_{i-3}L\hdots L \,l_{i-n}\,L\big)^{\infty} ~,\nonumber \\
    &\hspace{2cm} (\vec{g}_{S_i'})_m = -\delta_{(i-1)m} ~, \qquad \quad P_{S_i'} = -l_{i-1}~.
\end{align}
These are bijective to the $(k0)$ and $(k0')$ chords or faces discussed in combinatorial polytope.

There is a bijection between the pseudo-triangulations of $n$-gon with a puncture and the one-loop planar graphs. The two curves $(k0)$ and $(k0')$ joining vertex $k$ with the puncture correspond to two directions of loop momenta. So, there exists a mutation that flips the loop momentum. For instance, refer to figure \ref{fig:Dhat 1 polytope} and \ref{fig:Dhat 2 polytope}.
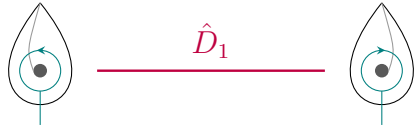
\begin{figure}[!ht]
    \centering
    \begin{tikzpicture}[baseline={([yshift=-.5ex]current bounding box.center)}] 
        \draw[line width=0.9pt,purple] (-1.5,0) -- (1.5,0);
        \begin{scope}[scale=0.9,xshift=-2.5cm, yshift=0cm]
            \fill[gray!70!black] (0,0) circle [radius=0.1];
            \draw (0,1) .. controls (-1.6,-1) and (1.6,-1) .. (0,1);
            \draw[gray!80] (0,1) .. controls (-0.2,0.3) .. (-0.1,0);
            \draw[teal] (0,-0.8) -- (0,-0.3);
            \draw[teal] (0,0) circle [radius=0.3];
            \draw[teal,-stealth] (0.0,0.3) -- (-0.05,0.3);
        \end{scope}
        \begin{scope}[scale=0.9,xshift=2.5cm, yshift=0cm]
            \fill[gray!70!black] (0,0) circle [radius=0.1];
            \draw (0,1) .. controls (-1.6,-1) and (1.6,-1) .. (0,1);
            \draw[gray!80] (0,1) .. controls (0.2,0.3) .. (0.1,0);
            \draw[teal] (0,-0.8) -- (0,-0.3);
            \draw[teal] (0,0) circle [radius=0.3];
            \draw[teal,-stealth] (0.0,0.3) -- (0.05,0.3);
        \end{scope}
        \node[purple] at (0,0.4) {\large $\hat{D}_1$};
    \end{tikzpicture}
    \caption{$\hat{D}_1$ polytope: a line segment.}
    \label{fig:Dhat 1 polytope}
\end{figure}
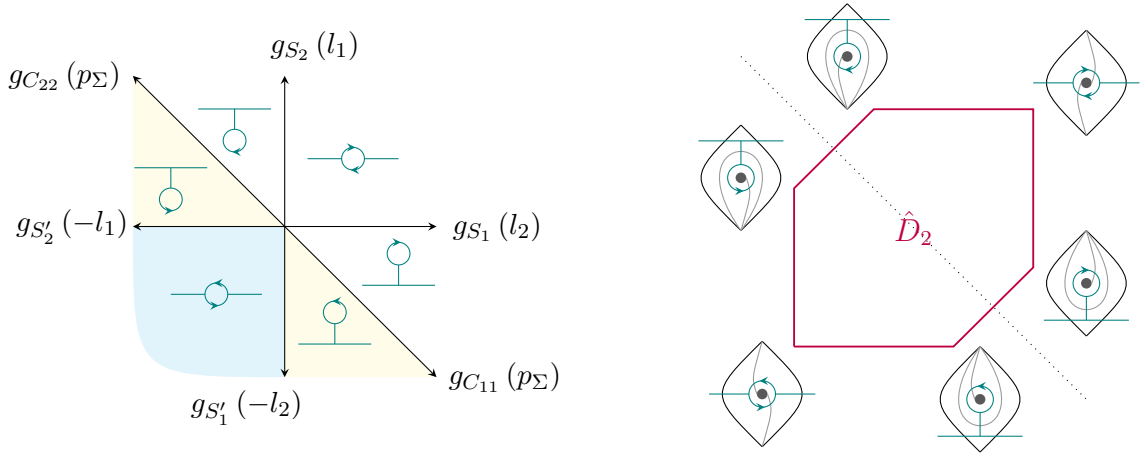
\begin{figure}[!hb]
    \centering
    \begin{tikzpicture}[baseline={([yshift=-.5ex]current bounding box.center)}]
        \fill[cyan!10] (-2,0) .. controls (-2,-2) .. (0,-2) -- (0,0) -- (-2,0);
        \fill[yellow!10] (-2,0) -- (-2,2) -- (0,0);
        \fill[yellow!10] (0,-2) -- (2,-2) -- (0,0);
        \draw[-stealth] (0,0) -- (0,2);
        \draw[-stealth] (0,0) -- (0,-2);
        \draw[-stealth] (0,0) -- (2,0);
        \draw[-stealth] (0,0) -- (-2,0);
        \draw[-stealth] (0,0) -- (2,-2);
        \draw[-stealth] (0,0) -- (-2,2);
        \node at (2.8,0) {$g_{S_1} \, (l_2)$};
        \node at (2.9,-2) {$g_{C_{11}} \, (p_{\Sigma})$};
        \node at (-2.9,2) {$g_{C_{22}} \, (p_{\Sigma})$};
        \node at (-2.8,0) {$g_{S_2'} \, (-l_1)$};
        \node at (0.4,2.4) {$g_{S_2} \, (l_1)$};
        \node at (-0.5,-2.4) {$g_{S_1'} \, (-l_2)$};
        \begin{scope}[scale=0.6,xshift=-1.1cm,yshift=1.9cm]
            \draw[teal] (-0.8,0.7) -- (0.8,0.7);
            \draw[teal] (0,0.7) -- (0,0.25);
            \draw[teal] (0,0) circle [radius=0.25];
            \draw[-stealth,teal] (-0.01,-0.25) -- (-0.05,-0.25);
        \end{scope}
        \begin{scope}[scale=0.6,xshift=-2.5cm,yshift=0.6cm]
            \draw[teal] (-0.8,0.7) -- (0.8,0.7);
            \draw[teal] (0,0.7) -- (0,0.25);
            \draw[teal] (0,0) circle [radius=0.25];
            \draw[-stealth,teal] (0.01,-0.25) -- (0.05,-0.25);
        \end{scope}
        \begin{scope}[scale=0.6,xshift=-1.5cm,yshift=-1.5cm]
            \draw[teal] (-1,0) -- (-0.25,0);
            \draw[teal] (1,0) -- (0.25,0);
            \draw[teal] (0,0) circle [radius=0.25];
            \draw[-stealth,teal] (-0.01,0.25) -- (-0.05,0.25);
            \draw[-stealth,teal] (0.01,-0.25) -- (0.05,-0.25);
        \end{scope}
        \begin{scope}[scale=0.6,xshift=1.1cm,yshift=-1.9cm]
            \draw[teal] (-0.8,-0.7) -- (0.8,-0.7);
            \draw[teal] (0,-0.7) -- (0,-0.25);
            \draw[teal] (0,0) circle [radius=0.25];
            \draw[-stealth,teal] (-0.01,0.25) -- (-0.05,0.25);
        \end{scope}
        \begin{scope}[scale=0.6,xshift=2.5cm,yshift=-0.6cm]
            \draw[teal] (-0.8,-0.7) -- (0.8,-0.7);
            \draw[teal] (0,-0.7) -- (0,-0.25);
            \draw[teal] (0,0) circle [radius=0.25];
            \draw[-stealth,teal] (0.01,0.25) -- (0.05,0.25);
        \end{scope}
        \begin{scope}[scale=0.6,xshift=1.5cm, yshift=1.5cm]
            \draw[teal] (-1,0) -- (-0.25,0);
            \draw[teal] (1,0) -- (0.25,0);
            \draw[teal] (0,0) circle [radius=0.25];
            \draw[-stealth,teal] (-0.01,-0.25) -- (-0.05,-0.25);
            \draw[-stealth,teal] (0.01,0.25) -- (0.05,0.25);
        \end{scope}
    \end{tikzpicture}
    \hspace{1cm}
    \begin{tikzpicture}[scale=0.7,baseline={([yshift=-.5ex]current bounding box.center)}]
    \draw[line width=0.7pt, purple] (0,0) -- (0,3) -- (1.5,4.5) -- (4.5,4.5) -- (4.5,1.5) -- (3,0) -- (0,0);
    \draw[dotted] (-1,5.5) -- (5.5,-1);
    \node[font = {\large}, purple] at (2.25,2.25) { $\hat{D}_2$};
    \begin{scope}[xshift=3.5cm, yshift=-1cm]
        \draw (0,1) .. controls (-1,0) .. (0,-1);
        \draw (0,1) .. controls (1,0) .. (0,-1);
        \fill[gray!70!black] (0,0) circle [radius=0.1];
        \draw[gray!80] (0,1) .. controls (-1.4,-1) and (1.4,-1) .. (0,1);
        \draw[gray!80] (0,1) .. controls (-0.2,0.3) .. (-0.1,0);
        \draw[teal] (-0.8,-0.7) -- (0.8,-0.7);
        \draw[teal] (0,-0.7) -- (0,-0.25);
        \draw[teal] (0,0) circle [radius=0.25];
        \draw[-stealth,teal] (-0.01,0.25) -- (-0.05,0.25);
    \end{scope}
    \begin{scope}[xshift=5.5cm, yshift=1.2cm]
        \draw (0,1) .. controls (-1,0) .. (0,-1);
        \draw (0,1) .. controls (1,0) .. (0,-1);
        \fill[gray!70!black] (0,0) circle [radius=0.1];
        \draw[gray!80] (0,1) .. controls (-1.4,-1) and (1.4,-1) .. (0,1);
        \draw[gray!80] (0,1) .. controls (0.2,0.3) .. (0.1,0);
        \draw[teal] (-0.8,-0.7) -- (0.8,-0.7);
        \draw[teal] (0,-0.7) -- (0,-0.25);
        \draw[teal] (0,0) circle [radius=0.25];
        \draw[-stealth,teal] (0.01,0.25) -- (0.05,0.25);
    \end{scope}
    \begin{scope}[xshift=5.5cm, yshift=5cm]
        \draw (0,1) .. controls (-1,0) .. (0,-1);
        \draw (0,1) .. controls (1,0) .. (0,-1);
        \fill[gray!70!black] (0,0) circle [radius=0.1];
        \draw[gray!80] (0,-1) .. controls (-0.2,-0.3) .. (-0.1,0);
        \draw[gray!80] (0,1) .. controls (0.2,0.3) .. (0.1,0);
        \draw[teal] (-1,0) -- (-0.25,0);
        \draw[teal] (1,0) -- (0.25,0);
        \draw[teal] (0,0) circle [radius=0.25];
        \draw[-stealth,teal] (0.01,0.25) -- (0.05,0.25);
        \draw[-stealth,teal] (-0.01,-0.25) -- (-0.05,-0.25);
    \end{scope}
    \begin{scope}[xshift=1cm, yshift=5.5cm]
        \draw (0,1) .. controls (-1,0) .. (0,-1);
        \draw (0,1) .. controls (1,0) .. (0,-1);
        \fill[gray!70!black] (0,0) circle [radius=0.1];
        \draw[gray!80] (0,-1) .. controls (1.4,1) and (-1.4,1) .. (0,-1);
        \draw[gray!80] (0,-1) .. controls (-0.2,-0.3) .. (-0.1,0);
        \draw[teal] (-0.8,0.7) -- (0.8,0.7);
        \draw[teal] (0,0.7) -- (0,0.25);
        \draw[teal] (0,0) circle [radius=0.25];
        \draw[-stealth,teal] (-0.01,-0.25) -- (-0.05,-0.25);
    \end{scope}
    \begin{scope}[xshift=-1cm, yshift=3.2cm]
        \draw (0,1) .. controls (-1,0) .. (0,-1);
        \draw (0,1) .. controls (1,0) .. (0,-1);
        \fill[gray!70!black] (0,0) circle [radius=0.1];
        \draw[gray!80] (0,-1) .. controls (1.4,1) and (-1.4,1) .. (0,-1);
        \draw[gray!80] (0,-1) .. controls (0.2,-0.3) .. (0.1,0);
        \draw[teal] (-0.8,0.7) -- (0.8,0.7);
        \draw[teal] (0,0.7) -- (0,0.25);
        \draw[teal] (0,0) circle [radius=0.25];
        \draw[-stealth,teal] (0.01,-0.25) -- (0.05,-0.25);
    \end{scope}
    \begin{scope}[xshift=-0.6cm, yshift=-0.9cm]
        \draw (0,1) .. controls (-1,0) .. (0,-1);
        \draw (0,1) .. controls (1,0) .. (0,-1);
        \fill[gray!70!black] (0,0) circle [radius=0.1];
        \draw[gray!80] (0,-1) .. controls (0.2,-0.3) .. (0.1,0);
        \draw[gray!80] (0,1) .. controls (-0.2,0.3) .. (-0.1,0);
        \draw[teal] (-1,0) -- (-0.25,0);
        \draw[teal] (1,0) -- (0.25,0);
        \draw[teal] (0,0) circle [radius=0.25];
        \draw[-stealth,teal] (-0.01,0.25) -- (-0.05,0.25);
        \draw[-stealth,teal] (0.01,-0.25) -- (0.05,-0.25);
    \end{scope}
\end{tikzpicture}
    \caption{The Feynman fan for planar one-loop $n=2$, and the $\hat{D}_2$ polytope: a hexagon.}
    \label{fig:Dhat 2 polytope}
\end{figure}
Note that all the graphs are doubled in the $\hat{D}_n$ polytope, on top of the doubling of the tadpoles due to the `switch' mutation. The co-dimension 1 hypersurface called $\Delta$-plane, defined by $\sum t_i=0$ divides the $\hat{D}_n$ polytope into two halves.

\paragraph{Revisiting the decomposition of $\hat{D}_n$ on its faces}
Let us revisit how the $\hat{D}_n$ polytope decomposes into smaller polytopes as we approach a boundary, thinking of trivalent graphs. Consider the face labeled by $S_i$ ($S_i'$), conjugate to the loop momenta $l_{i-1}$. Approaching the face is equivalent to performing a single cut on the loop, resulting in a `forward limit' $n+2$ tree-level diagram. So, we have:
\begin{align}
    \hat{D}_n \xrightarrow{\partial_{S_i}} \mathcal{A}_{n-1} ~, \qquad \quad     \hat{D}_n \xrightarrow{\partial_{S_i'}} \mathcal{A}_{n-1} ~.
\end{align}
Approaching the $C_{ij}$ face, the graphs decompose into a tree graph and a loop graph. Precisely, we have:
\begin{align}
    \text{For } i\in [1,n] ~, \qquad k\in [2,n]~, \qquad \hat{D}_n \xrightarrow{\partial C_{i,i+k}} \mathcal{A}_{k-2} \times \hat{D}_{n-k+1}
\end{align}
For instance, let us rewrite \eqref{fig: example of chords ik ki} in terms of graphs:
\begin{align}
    \begin{tikzpicture}[scale=0.5,baseline={([yshift=-.5ex]current bounding box.center)}]
        \draw (-1,1) -- (0,0) -- (-1,-1);
        \draw (0,0) -- (3,0) -- (4,-1);
        \draw (3,0) -- (4,1);
        \draw (1,0) -- (1,0.8);
        \draw (1,1.3) circle [radius=0.5];
        \draw (2,0) -- (2,-1);
        \node at (0.5,-0.3) {\tiny $(52)$};
        \node at (1.5,0.3) {\tiny $(25)$};
        \node at (-1.3,-1) {5};
        \node at (-1.3,1) {1};
        \node at (4.3,1) {2};
        \node at (4.3,-1) {3};
        \node at (2.2,-1) {4};
    \end{tikzpicture} \hspace{1cm}
    \mathcal{A}_1 \times \hat{D}_3 \xleftarrow{\p_{C_{25}}} \hat{D}_5 \xrightarrow{\p_{C_{52}}} \mathcal{A}_0 \times \hat{D}_4 ~.
\end{align}

Let us consider the the boundary along $C_{ii}$: $\hat{D}_n \xrightarrow{\partial C_{i,i+n}} \mathcal{A}_{n-2} \times \hat{D}_{1}$~. This is `cutting' the `zero' momentum propagator $p_{\Sigma}$. There are $n$ such faces, $C_{ii}$, in the $\hat{D}_n$ polytope, and all the tadpole graphs necessarily lie on these faces. Thus all the tadpoles on one side of the $\Delta$-plane, are arranged neatly into $n$ copies of associahedra $\mathcal{A}_{n-2}$. The remaining $\hat{D}_1$ mutates the tadpoles across the $\Delta$-plane. 

All the tadpole graphs belonging to a particular $\mathcal{A}_{n-2}$ associahedron share a common loop propagator. On the upper side of $\Delta$-plane, $\sum t_i>0$, all the vertices on the face $C_{ii}$ contains the the loop momenta $l_{i-1}$ corresponding to $S_i$. Figure \ref{fig:cones for n=3} depicts this for $n=3$. On the other side of the $\Delta$-plane, $\sum t_i<0$, the tadpole cones dual to the face $C_{ii}$ have the loop momenta $-l_{i-1}$ corresponding to $S'_i$.

\subsubsection*{Counting the number of vertices}
On one side of the $\Delta$ plane, the number of graphs (top dimensional cones in $g$-vector space, or the vertices of $\hat{D}_n$ polytope) is as follows:  
\begin{align*}
    &\text{Number of tadpole graphs}: 2(2n-3) |\mathcal{A}_{n-3}^{[0]}| \\
    &\text{Number of bubble graphs}: (2n-3) |\mathcal{A}_{n-3}^{[0]}| \\
    &\text{Number of triangle graphs}: (n-2) |\mathcal{A}_{n-3}^{[0]}| \\
    & \qquad \hdots \\
    &\text{Number of } (n-1)\text{ -gon graphs}: n  \\
    &\text{Number of }n\text{-gon graphs}: 1  
\end{align*}
$|\mathcal{A}_{n-3}^{[0]}|$ is the number of vertices of the $n-3$ dimensional associahedron \eqref{number of vertices of associahedron}.

Note that the number of $g$-vectors in $\Delta$-plane for $n$ particles is the same as the total number of $g$-vectors for $n-1$ particles: $n(n-1)$. Thus, the total number of top dimensional cones for $n-1$ particles on both sides of the $\Delta$-plane is the same as the ($n-1$ dim) cones inside the $\Delta$-plane for $n$ particles. The latter is identical to the number of tadpole graphs on one side of the $\Delta$-plane for $n$ particles, which is $2(2n-3)|\mathcal{A}_{n-3}^{[0]}|$. Thus, on both sides of the $\Delta$-plane, \textbf{the total number of vertices of the $\hat{D}_{n-1}$ polytope} is $2(2n-3)|\mathcal{A}_{n-3}^{[0]}| $. We write down the explicit counting for some simple cases in table \ref{tab:number of vertices of Dhat polytope}.
\begin{table}[!ht]
    \centering
    \begin{tabular}{c l l}
        $(n=1)\qquad$  & $2\times\big(1\big)$ &$=2$ \\ 
        $(n=2)\qquad$  & $2\times\big(1+2\big)$  &$=6$ \\
        $(n=3)\qquad$  & $2\times\big(1+3+6\big)$  &$=20$ \\
        $(n=4)\qquad$  & $2\times\big(1+4+10+20\big)$  &$=70$ \\
        $(n=5)\qquad$  & $2\times\big(1+5+15+35+70\big)$  &$=252$ \\
        $(n=6)\qquad$  & $2\times\big(1+6+21+56+126+252\big)$  &$=924$ \\
        & $\qquad \vdots$ & \\
        $n\qquad $ & $2\times \big(|n$-gons$| + |(n-1)$gons$| + \hdots+ |$bubbles$| + |$tadpoles$|$ $\big)$ & $= 2(2n-1)|\mathcal{A}_{n-2}^{[0]}|$
    \end{tabular}
    \caption{Number of vertices of the $\hat{D}_n$ polytope.}
    \label{tab:number of vertices of Dhat polytope}
\end{table}

We have thus counted the total number of trivalent planar one-loop diagrams (accounting for the doubling of tadpole graphs):
\begin{align}
    \text{Total planar one-loop graphs for } n \text{ particles} = \frac{1}{2}|\hat{D}_n{}^{[0]}| = (2n-1)\,C_{n-1} = \frac{(2n-1)!}{n!\,(n-1)!} ~.
\end{align}

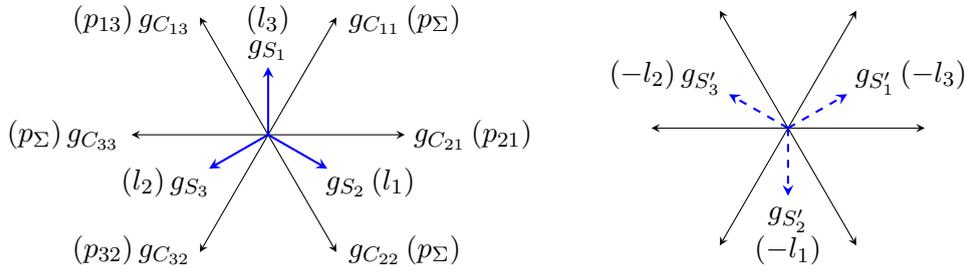
\begin{figure}[!ht]
    \centering
    \begin{tikzpicture}[scale=0.9]
    \draw[-stealth] (0,0) -- (2,0);
    \draw[-stealth] (0,0) -- (-2,0);
    \draw[-stealth] (0,0) -- (1,1.732);
    \draw[-stealth] (0,0) -- (1,-1.732);
    \draw[-stealth] (0,0) -- (-1,1.732);
    \draw[-stealth] (0,0) -- (-1,-1.732);
    \draw[-stealth,thick,blue] (0,0) -- (0,1);
    \draw[-stealth,thick,blue] (0,0) -- (0.866,-0.5);
    \draw[-stealth,thick,blue] (0,0) -- (-0.866,-0.5);
    \node at (2,1.7) {$g_{C_{11}}\,(p_\Sigma)$};
    \node at (2,-1.7) {$g_{C_{22}}\,(p_\Sigma)$};
    \node at (-3,0) {$(p_\Sigma)\,g_{C_{33}}$};
    \node at (3,0) {$g_{C_{21}}\,(p_{21})$};
    \node at (-2,1.7) {$(p_{13})\,g_{C_{13}}$};
    \node at (-2,-1.7) {$(p_{32})\,g_{C_{32}}$};
    \node at (0,1.25) {$g_{S_1}$};
    \node at (0,1.7) {$(l_3)$};
    \node at (1.5,-0.7) {$g_{S_2}\,(l_1)$};
    \node at (-1.5,-0.7) {$(l_2)\,g_{S_3}$};
\end{tikzpicture}
\hspace{0.5cm}
\begin{tikzpicture}[scale=0.9]
    \draw[-stealth] (0,0) -- (2,0);
    \draw[-stealth] (0,0) -- (-2,0);
    \draw[-stealth] (0,0) -- (1,1.732);
    \draw[-stealth] (0,0) -- (1,-1.732);
    \draw[-stealth] (0,0) -- (-1,1.732);
    \draw[-stealth] (0,0) -- (-1,-1.732);
    \draw[-stealth,dashed,thick,blue] (0,0) -- (0,-1);
    \draw[-stealth,dashed,thick,blue] (0,0) -- (-0.866,0.5);
    \draw[-stealth,dashed,thick,blue] (0,0) -- (0.866,0.5);
    \node at (0,-1.35) {$g_{S'_2}$};
    \node at (0,-1.8) {$(-l_1)$};
    \node at (1.8,0.75) {$g_{S_1'}\,(-l_3)$};
    \node at (-1.8,0.75) {$(-l_2)\,g_{S_3'}$};
\end{tikzpicture}
    \caption{The $g$-vectors for $n=3$, planar one-loop. The solid (dashed) blue vectors are pointing out (in) of the plane. The six black vectors lie on the $\Delta$-plane: $\sum t_i =0$.}
    \label{fig:fan for n=3}
\end{figure}

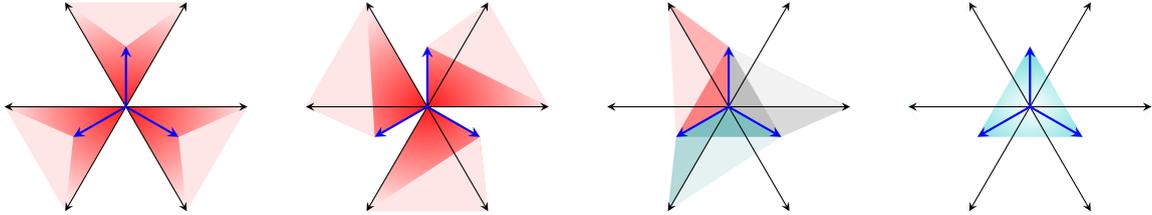
\begin{figure}[!bh]
    \centering
    \begin{tikzpicture}[scale=0.8,baseline={([yshift=-.5ex]current bounding box.center)}]
    \fill[pink!40] (0,0) -- (-1,1.732) -- (1,1.732) -- (0,0) -- (0,0) -- (-1,-1.732) -- (-2,0) -- (0,0) -- (0,0) -- (2,0) -- (1,-1.732) -- (0,0);
    \shade[inner color=red!90,outer color=pink!40](0,0) -- (1,1.732) -- (0,1) -- (-1,1.732) -- (0,0) -- (-2,0) -- (-0.866,-0.5) -- (-1,-1.732) -- (0,0) -- (1,-1.732) -- (0.866,-0.5) -- (2,0) -- (0,0);
    \draw[-stealth] (0,0) -- (2,0);
    \draw[-stealth] (0,0) -- (-2,0);
    \draw[-stealth] (0,0) -- (1,1.732);
    \draw[-stealth] (0,0) -- (1,-1.732);
    \draw[-stealth] (0,0) -- (-1,1.732);
    \draw[-stealth] (0,0) -- (-1,-1.732);
    \draw[-stealth,thick,blue] (0,0) -- (0,1);
    \draw[-stealth,thick,blue] (0,0) -- (0.866,-0.5);
    \draw[-stealth,thick,blue] (0,0) -- (-0.866,-0.5);
\end{tikzpicture}
\hspace{0.5cm}
\begin{tikzpicture}[scale=0.8,baseline={([yshift=-.5ex]current bounding box.center)}]
    \shade[inner color=red!90,outer color=pink!40] (0,0) -- (0,1) -- (1,1.732) -- (2,0) -- (0,0) -- (0.866,-0.5) -- (1,-1.732) -- (-1,-1.732) -- (0,0) -- (-0.866,-0.5) -- (-2,0) -- (-1,1.732) -- (0,0); 
    \fill[pink!40] (-1,1.732) -- (-2,0) -- (-0.866,-0.5) -- (-1,1.732) -- (0,0) -- (0,1) -- (1,1.732) -- (2,0) -- (0,1) -- (0,0) -- (0.866,-0.5) -- (1,-1.732) -- (-1,-1.732) -- (0.866,-0.5) -- (0,0);
    \draw[-stealth] (0,0) -- (2,0);
    \draw[-stealth] (0,0) -- (-2,0);
    \draw[-stealth] (0,0) -- (1,1.732);
    \draw[-stealth] (0,0) -- (1,-1.732);
    \draw[-stealth] (0,0) -- (-1,1.732);
    \draw[-stealth] (0,0) -- (-1,-1.732);
    \draw[-stealth,thick,blue] (0,0) -- (0,1);
    \draw[-stealth,thick,blue] (0,0) -- (0.866,-0.5);
    \draw[-stealth,thick,blue] (0,0) -- (-0.866,-0.5);
\end{tikzpicture}
\hspace{0.5cm}
\begin{tikzpicture}[scale=0.8,baseline={([yshift=-.5ex]current bounding box.center)}]
    \fill[gray!10] (0,1) -- (2,0) -- (0,0);
    \fill[gray!30] (0,0) -- (0.866,-0.5) -- (2,0);
    \fill[gray!50] (0,0) -- (0,1) -- (0.866,-0.5);
    \fill[teal!10] (0.866,-0.5) -- (-1,-1.732) -- (0,0);
    \fill[teal!30] (-0.866,-0.5) -- (-1,-1.732) -- (0,0);
    \fill[teal!50] (-0.866,-0.5) -- (0.866,-0.5) -- (0,0);
    \fill[red!10] (0,0) -- (-1,1.732) -- (-0.866,-0.5);
    \fill[red!30] (0,0) -- (-1,1.732) -- (0,1);
    \fill[red!50] (0,0) -- (0,1) -- (-0.866,-0.5);
    %
    \draw[-stealth] (0,0) -- (2,0);
    \draw[-stealth] (0,0) -- (-2,0);
    \draw[-stealth] (0,0) -- (1,1.732);
    \draw[-stealth] (0,0) -- (1,-1.732);
    \draw[-stealth] (0,0) -- (-1,1.732);
    \draw[-stealth] (0,0) -- (-1,-1.732);
    \draw[-stealth,thick,blue] (0,0) -- (0,1);
    \draw[-stealth,thick,blue] (0,0) -- (0.866,-0.5);
    \draw[-stealth,thick,blue] (0,0) -- (-0.866,-0.5);
\end{tikzpicture}
\hspace{0.5cm}
\begin{tikzpicture}[scale=0.8,baseline={([yshift=-.5ex]current bounding box.center)}]
    \shade[outer color=teal!50!cyan, inner color=purple!0] (2,0) -- (1.45,0.866) -- (1,1.732) -- (0,1.68) -- (-1,1.732) -- (-1.45,0.866) -- (-2,0) -- (-1.45,-0.866) -- (-1,-1.732) -- (0,-1.68) -- (1,-1.732) -- (1.45,-0.866); 
    \fill[white] (0,1) -- (2,0) -- (1,1.732)-- (-1,1.732) -- (-2,0) -- (-1,-1.732) -- (1,-1.732) -- (2,0) -- (0,1) -- (0.866,-0.5) -- (-0.866,-0.5) -- (0,1); 
    \draw[-stealth] (0,0) -- (2,0);
    \draw[-stealth] (0,0) -- (-2,0);
    \draw[-stealth] (0,0) -- (1,1.732);
    \draw[-stealth] (0,0) -- (1,-1.732);
    \draw[-stealth] (0,0) -- (-1,1.732);
    \draw[-stealth] (0,0) -- (-1,-1.732);
    \draw[-stealth,thick,blue] (0,0) -- (0,1);
    \draw[-stealth,thick,blue] (0,0) -- (0.866,-0.5);
    \draw[-stealth,thick,blue] (0,0) -- (-0.866,-0.5);
\end{tikzpicture}
    \caption{Cones for $n=3$ on the positive side $\sum t_i>0$ of $\Delta$-plane. Two left ones show total of six tadpoles, then three bubbles, and one triangle graph.}
    \label{fig:cones for n=3}
\end{figure}

In the $\Delta$ plane, there are $|\mathcal{A}_{n-3}^{[0]}|\times 2(2n-3)$ number of $n-1$ dim cones. These lead to an equal number of top-dimensional cones dual to the tadpole graphs on one side of the $\Delta$-plane, sharing a co-dimension 1 space with it. For example, figure \ref{fig:fan for n=3} and \ref{fig:cones for n=3} depicts the situation for $n=3$. The plane of the paper is the two-dimensional $\Delta$-plane. The six cones in the first two diagrams in figure \ref{fig:cones for n=3} correspond to tadpoles, sharing a 2d face with $\Delta$ plane. The cones corresponding to bubbles on the propagators share co-dimension 2 space with the $\Delta$ plane. In figure \ref{fig:cones for n=3}, the three cones in the third diagram correspond to bubbles and share a one-dimensional edge with the $\Delta$ plane. Note that the g-vector with `zero 'momentum: $p_{\Sigma} \equiv \sum_1^np_i$ is adjacent only to the cones dual to tadpoles. The final cone in figure \ref{fig:cones for n=3} depicts the `triangle' graph.

\subsection{From ABHY associahedron in ${\cal K}_{n}$ to Scattering form in ${\cal K}_{n}^{\star}$}\label{cdfdln}
As a positive geometry that sits inside ${\cal K}_{n}$ (or more precisely, in the positive octant of ${\cal K}_{n}$), it defines a unique $n-3$ form $\Omega_{n-3}$ on ${\cal K}_{n}$. $\Omega_{n-3}$ has simple poles on all the hyperplanes that contain any boundary of the ABHY associahedron  such that the residue on a hyper-plane $H$ containing a co-dimension $k$ boundary $\partial_{k}A_{n-3}$ equals the canonical form on the (product) of associahedron $\partial_{k}A_{n-3}$. 

More explicitly, given \emph{any} vertex $v_{0}$ of $A_{n-3}$, let $\sigma(v, v_{0})$ be the number of mutations required to flip the triangulation associated to $v_{0}$ to the triangulation associated to $v$. Then the planar scattering form is defined as, 
\begin{align}
\Omega_{n-3}\, =\, \sum_{v\, \in\, A_{n-3}}\, (-1)^{\sigma(v, v_{0})}\, \bigwedge_{(i,j)\, \in\, v}\, d\ln X_{ij}
\end{align}
This form can be pulled back to ${\cal K}_{n}^{\star}$ as
\begin{align}
\pi^{\star}\Omega_{n-3}\, =\, \sum_{v\, \in\, A_{n-3}}\, (-1)^{\sigma(v, v_{0})}\, \bigwedge_{(i,j)\, \in\, v}\, d\ln S_{i,j}
\end{align}
where
\begin{align}
S_{i,j} =: S_{I}\, \vert I = \{i,\dots,j-1\}
\end{align}
A remarkable and by now well known property of $\Omega_{n-3}$ is that the form is projective under $X_{ij}\, \rightarrow\, f(\{X_{mn}\})\, X_{ij}$. In fact it was proved in \cite{nima1711} that projective invariance uniquely fixes $\Omega_{n-3}$ (upto an overall sign which would change the S-matrix only by a phase $\pm 1$.) 

It was proved in \cite{nima1711} that although the invariance under of projective transformation uniquely fixes the planar scattering form in ${\cal K}_{n}$,there exists a class of scattering forms in the big kinematic space ${\cal K}_{n}^{\star}$ which satisfy this property. Of course $\Omega_{n-3}$ can be pulled back to ${\cal K}_{n}^{\star},\, {\cal K}_{n}^{\textrm{os}}$ and defines one such projective invariant $n-3$ form, but there are many other interesting examples, \cite{nima1711} which are relevant for study of scattering amplitudes.  One such form leads to the color dressed amplitudes of biadjoint scalar theory. We review this construction below as it will be central to us going forward. 

\newcommand{\bigslant}[2]{{\raisebox{.2em}{$#1$}\left/\raisebox{-.2em}{$#2$}\right.}}
Let $\alpha\, \in\, \bigslant{S_{n}}{Z_{n}}$ and fix a  planar 3-valent graph $\gamma_{\alpha}\, \forall\, \alpha\, \in\, \bigslant{S_{n}}{Z_{n}}$ which is compatible with the $\alpha$ ordering. Let the (internal) edges of the graph be denoted as $e_{a}^{\alpha}\, \vert\, 1\, \leq\, a\, \leq\, n-3$.  Let the variable of ${\cal K}_{n}^{\star}$ corresponding to $e_{a}^{\alpha}$ be denoted as $x_{e_{a}^{\alpha}}$.
We denote,
\setlist{nolistsep}
\begin{enumerate}[noitemsep]
\item The standard clock-wise ordering $(1,\, \dots,\, n)$ as ${\bf I}$ and the corresponding chosen graph as $\gamma_{{\bf I}}$, 
\item The number of ( flips + mutations ) required to map a graph $\gamma_{{\bf I}}$ to $\gamma_{\alpha}$ as $\textrm{flip}(\alpha, {\bf I})$. 
\end{enumerate}
Consider the $n-3$ form in  ${\cal K}_{n}^{\star}$, 
\begin{align}
\omega_{n-3}\, =\, \sum_{\alpha\, \in\,  \bigslant{S_{n}}{Z_{n}}}\, c(\gamma_{\alpha})\, w(\gamma_{\alpha})\, \prod_{a=1}^{n-3}\, \frac{1}{x_{e_{a}^{\alpha}}}
\end{align}
where
\begin{align}
w(\gamma_{\alpha})\, =\, (-1)^{\textrm{flip}(\alpha,{\bf I})}\, \bigwedge_{a=1}^{n-3}\, d x_{e_{a}^{\alpha}}
\end{align}
Using the Seven term identity of eqn. (\ref{seventerm}) it was shown in \cite{nima1711} that $\omega_{n-3}$  is projective and that $\omega_{n-3}$ can be used to generate a scalar in ${\cal K}_{n}^{\star}$  which is the color-dressed amplitude of bi-adjoint scalar theory. We recast the ABHY derivation in a slightly different manner which will then be generalised to obtain tree-level \emph{color ordered} Yang-Mills amplitude from $\Omega_{n-3}$.  Consider a three-regular graph $\gamma$, where each vertex is equipped with a clockwise orientation.\footnote{For graphs embedded in a plane, such an ordering can be inherited from the orientation of the underlying manifold, but even for algebraic graphs, we can simply choose an ordering at each vertex.} We color each (internal as well as external) edge of $\gamma$ with a color label. If the three edges $e_{1}(v), e_{2}(v), e_{3}(v)$ are incident on $v$, we color them with $a_{v}, b_{v}, c_{v}$ respectively. We can now associate a color tensor $\tilde{c}(\gamma)$ to the graph as, 
\begin{align}
\tilde{c}(\gamma)\, :=\, \prod_{v\, \in\, \gamma}\, f_{v}^{a_{v}b_{v}c_{v}}
\end{align}
where $f^{abc}$ is the structure constant of a compact group and the ordering of color indices at each vertex is determined by the orientation. 

In ${\cal K}_{n}^{\star}$ we can now define a multi-vector field (MVF)\footnote{A multi-vector field of rank $k\, \leq\, n$ on an ${\bf R}^{n}$ is a section of the exterior algebra bundle $\Lambda^{k}\, TM$. Contraction of such a field with a $k$ form is defined via
\begin{align}
\langle\, \partial_{x_{1}^{i_{1}}}\, \wedge\, \dots\, \wedge\, \partial_{x_{k}^{i_{k}}},\, dx^{j_{1}}\, \wedge\, \dots\, \wedge\, d x^{j_{k}}\, \rangle\, =\, \delta^{i_{1}\dots i_{k}}_{[j_{1}\, \dots\, j_{k}]}
\end{align}}
\begin{align}\label{mvfcf}
{\cal P}^{\textrm{color}}_{n}\, =\, \sum_{\alpha\, \in\,  \bigslant{S_{n}}{Z_{n}}}\, \tilde{c}(\gamma_{\alpha})\, (-1)^{\textrm{flip}(\alpha,{\bf I})}\, \bigwedge_{a=1}^{n-3}\, \frac{\partial}{\partial s_{e_{a}^{\alpha}}}
\end{align}
Then,
\begin{align}
\langle\, {\cal P}^{\textrm{color}}_{n}, \pi^{\star}\omega_{n}\, \rangle\, =\, \sum_{\alpha\, \in\,  \bigslant{S_{n}}{Z_{n}}}\, \tilde{c}(\gamma_{\alpha})\, c(\gamma_{\alpha})\, \prod_{a=1}^{n-3}\, \frac{1}{s_{e^{\alpha}_{a}}}
\end{align}
which is the color-dressed bi-adjoint amplitude $\sum_{\alpha}\, {\cal M}_{n}(\alpha\, \vert\, \alpha)$ with color factors denoted as $c, \tilde{c}$ respectively.   There are two important aspects of this interpretation of the color-dressed bi-adjoint amplitude from our perspective. 
\setlist{nolistsep}
\begin{itemize}[noitemsep]
\item As emphasized above, unlike the color ordered amplitude which is the volume of the simplex dual to the ABHY associahedron in ${\cal K}_{n}$, the color dressed amplitude is a scalar function in the big kinematic space.
\item This scalar can be understood as a contraction of the planar scattering form with multi-vector field. 
\item ${\cal P}^{\textrm{color}}_{n}$ is a weighted sum over set of all graphs obtained from $\gamma_{I}$ by mutation and flips. It's valuation on each graph is a polynomial (in color space) determined by the orientation  at each vertex of the graph and the set of half-edges incident at that vertex.\footnote{An half-edge is simply a pair $(e,v)$ of an edge along with one of it's end points. For a closed loop incident on a single vertex, half edge is same as the edge itself.}
\end{itemize}


\section{A Brief Review of Corolla Polynomial}\label{correv}
In \cite{kreimer1208}, D. Kreimer, M. Sars and W. D. van Suijlekom  defined a new graph polynomial  $C_{\gamma}$ for any 3-regular graph $\gamma$ which is uniquely fixed by the  
\begin{enumerate}
\item The genus  of the graph
\item The set of  half-edges of the graph and 
\item A clockwise orientation at each vertex of the graph.
\end{enumerate}
As such Corolla polynomial is simply a counting polynomial (see \cite{kreimer-yeats}) which enumerates the set ${\cal T}$ of all the subsets of the half-edges $T$ such that $\frac{\gamma}{T}$ has no cycle,\cite{kreimer-yeats}. An intriguing mathematical object in it's own right, the real interest for us in the Corolla polynomial is due to seminal results proved in \cite{kreimer1208}, where a specific representation of the Corolla polynomial (which we review below) was used to generate a large class of gauge theory correlators from correlators in massless $\phi^{3}$  theory. 

More in detail, it was shown in \cite{kreimer1208} that the Corolla polynomial can be represented  as a differential operator which can then  ``spin up" an $n$-particle, $L$ loop  Feynman diagram in  $\phi^{3}$ theory in Schwinger parametrisation, for all $n,\, L$.  The result is a  the Schwinger integrand of Yang-Mills theory where the underlying graph is built out of (1) 3-gluon vertex, (2) 4-gluon vertices and (3) closed ghost loops.   In our opinion, this is a rather  striking result where the manifest gauge invariance and locality arises from a remarkable fact that the action of the Corolla polynomial can also be viewed as generating elements of the graph homology, \cite{vogtmann, berghoff2008}.  Although, the complete ramfications of the rich structure of the Corolla polynomial still remains to be appreciated and explored, it has already been used to define amplitude for a class of gauge theories including (1) spontaneously broken gauge theories \cite{prinz} and (2) gauge theories coupled to charged matter fields such as QED \cite{Golz:2017aoa}. 

We review this construction below. Many of the beautiful properties of the Corolla polynomial have been derived in \cite{kreimer-yeats, berghoff2008}. Our review will only scratch the surface of these results, as our primary goal is to define a new representation of the Corolla polynomial  as differential operators in ${\cal K}_{n}^{\textrm{os}}$.\footnote{Kreimer's introduction of the Corolla polynomial as a third graph polynomial which along with First and Second symanzik polynomials generate gauge theory amplitudes. We will review the  Corolla graph differential in the space of Schwinger parameters  in section (\ref{cgdksvs}) and extend it to define a ``Corolla curve differential". This will lead us to the curve integral formulae for gluons as derivatives of the formulae discovered in \cite{nima2309}.} Interested reader is encouraged to peruse the references cited above for more details about Corolla Polynomial.
\subsection{Half edges, Graph Orientation and Edge ordering}
In this section we review the graph theoretic data required to define the Corolla polynomial on tri-valent (cubic) graphs with a fixed ordering of external particles. Throughout the paper, we will only encounter three-valent graphs which can be embedded on a disc with $n$ maked points on the boundary and at most one puncture.

Given an edge $e$ incident on a vertex $v$, by half edge we mean the pair $(e,v)$. In a cubic graph, every vertex $v$ has three half edges associated to it denoted by $h_{v}^{I}\, \vert\, I = \{1,2,3\}$. We denote the set of tri-valent vertices of a graph $\gamma_{n}$ with $n$ external vertices as $V(\gamma_{n})$.  
\begin{definition}
A corolla $C_{v}$ is a vertex $v\, \in\, V(\gamma_{n})$ along with the set of all the half edges incident on $v$. 
\end{definition}
To define the Corolla polynomial, a bare graph is not enough but we need to assign some additional data to it. 
\begin{enumerate}
\item All the external edges are outgoing.
\item We assign a clockwise orientation to each vertex in $V(\gamma_{n})$ once and for all. It generates a cyclic ordering on the half edges, $h_{v}^{1} < h_{v}^{2} < h_{v}^{3}$.  
\item Each half edge $h_{e}$ is assigned a variable, $a_{h_{e}}$. 
\item Every internal edge is directed. 
\item To every vertex of the associahedron, we assign a unique ``ordering label" which orders the set of (internal) edges of the corresponding 3-regular planar graph. 
\end{enumerate}
The last two points requires some deliberation. For a fixed graph, we can assign a ordering to the set of  vertices as well as set of internal edges once and for all, but as $A_{n-3}$ ``glues" all 3-regular planar graphs together, the corresponding ordering on the set $E_{\textrm{int}}(\gamma_{1})$ and $E_{\textrm{int}}(\gamma_{2})$ also has to commute with mutations. We claim that indeed, such an ordering assignment exists.

Let us consider a given 3-regular graph with ordered external vertices as (following J. Stasheff) parenthesizing $(1, 2, \dots, n)$ with  $n-3$ parenthesis which are either non intersecting or nested, \cite{stasheff1963homotopy}.  In the case of $n = 4$ the two possible triangulations can hence be represented as following parenthetical expressions.
\begin{align}
T_{1}\, =\,  (1 2), 3, 4\nonumber\\
T_{2}\, =\, 1, (2 3), 4 
\end{align}
We note that given a parenthetical expression that corresponds to a fixed triangulation, one can always add a redundant parenthesis without changing the triangulation. We color the redundant parenthesis in blue.
\begin{align}
T_{1}\, =\, (12), \textcolor{blue}{(}34\textcolor{blue}{)}\nonumber\\
T_{2} =\, \textcolor{blue}{(}1, (23), 4\, \textcolor{blue}{)}
\end{align}
We also note that in $n=4$ case, given a triangulation $T$ and the associated parenthesised expression, every pair of parenthesis then corresponds  to a unique tri-valent vertex in $\gamma_{T}$. \begin{align}
(12)\, \leftrightarrow\, v_{1,2,(12)}\nonumber\\
(34)\, \leftrightarrow\, v_{3,4,(34)}
\end{align}
We now provide a natural ordering on the set parenthetisized expressions via
\begin{align}\label{criteria}
(i,j) < (k,l)\, \textrm{if}\,  (i,j)\, \cap\, (k,l) =0\, \textrm{  and  } i < k\nonumber\\
(i,j) < (k,l)\, \textrm{  if  }\,  (i,j)\, \subset\, (k,l)
\end{align}
Such an ordering induces a canonical ordering on the sets of vertices $V(\gamma_{T_{1}}),\, V(\gamma_{T_{2}})$.
\begin{align}
v_{1,2,(12)} < v_{3,4,(34)}\, \textrm{   in   }\, \gamma_{T_{1}}\\
v_{2,3,(23)} < v_{4,1,(41)}\, \textrm{   in   }\, \gamma_{T_{2}}
\end{align}
We can now extend these definitions to arbitrary triangulations for all $n$. But rather than giving a general set of conditions, we simply illustrate the algorithm (to assign ordering on set of all vertices of a graph) through one more example. The reader is encouraged to verify that a systematic algorithm can be written down for any $n$.

Given a complete triangulation $T$ of the $n$-gon, we now argue that it's representation as parenthisizing $n$ letters generates a canonical complete ordering on the set of all the vertices.
\setlist{nolistsep}
\begin{enumerate}[noitemsep]
\item Given the parenthetisized expression for $T$, any bracket that contains $k$ elements $(i_{1},\, \dots, i_{k})$  must also contain an additional $k-3$ parentheses nested inside the outermost parenthesis. We can then associate a set $S_{(i_{1} \dots i_{k})}$ of vertices of $\gamma_{T}$ (with cardinality $k-1$) to this expression.  
\item Between any two vertices in $S_{(i_{1}, \dots, i_{k})}$ there is then a ordering determined by criteria laid out in eqn.(\ref{criteria}). 
\end{enumerate}
We then have a complete ordering on set of all the vertices in $V(\gamma_{T})$. Let us denote the ordered vertices as  $(v_{1}, \dots,\, v_{n-2})$. We can now order the set of all internal edges  as follows.
\begin{align}
e_{v_{i}, v_{j}}\, <\, e_{v_{k},v_{l}}\quad \textrm{if}\quad v_{i} < v_{k}
\end{align}
We note that this algorithm makes all the internal edges directed and puts ordering on the set $E_{\textrm{int}}$. 

Let us take  a specific example. Consider the following triangulation for $n=6$.
\begin{align}
T_{1} = \{\, (1,3), (3,5), (1,5)\, \}
\end{align}
$T_{1}$ can be mapped to the following bracketing of the sequence $\{1,\, \dots\, 6\}$.
\begin{align}
T_{1}\, =\,   \textcolor{blue}{(} (1 2)\, (3 4)\, (5 6) \textcolor{blue}{)}
\end{align}
The bracket fixes a unique order on the set of all internal vertices.
\begin{align}
\begin{tikzpicture}[scale=0.55,baseline={([yshift=-.5ex]current bounding box.center)}]
\draw (-2.414,-1) -- (-1.414,0) -- (-2.414,1);
\draw (-1.414,0) -- (0,0);
\draw (1,1) -- (0,0) -- (1,-1);
\draw (2.414,1) -- (1,1) -- (1,2.414);
\draw (2.414,-1) -- (1,-1) -- (1,-2.414);
\node at (-2.75,-1) {1};
\node at (-2.75,1) {2};
\node at (0.6,2.4) {3};
\node at (2.8,1) {4};
\node at (2.8,-1) {5};
\node at (0.6,-2.4) {6};
\node at (-0.7,0.45) {$e_a$};
\node at (0.95,0.25) {$e_{b}$};
\node at (0.1,-0.8) {$e_{c}$};
\draw[-stealth] (-0.55-1.414,0.55) -- (-0.56-1.414,0.56);
\draw[-stealth] (-0.55-1.414,-0.55) -- (-0.56-1.414,-0.56);
\draw[-stealth] (1.8,1) -- (1.81,1);
\draw[-stealth] (1,1.8) -- (1,1.81);
\draw[-stealth] (1.8,-1) -- (1.81,-1);
\draw[-stealth] (1,-1.8) -- (1,-1.81);
\draw[-stealth] (-0.7,0) -- (-0.69,0);
\draw[-stealth] (0.55,0.55) -- (0.56,0.56);
\draw[-stealth] (0.55,-0.55) -- (0.56,-0.56);
\end{tikzpicture}
\hspace{2cm} 
\begin{matrix}
v_{1,2,(12)} < v_{3,4,(34)} < v_{5,6,(56)} < v_{(12)(34)(56)} \\
\\
e_a < e_b < e_c
\end{matrix} \\
e_a \equiv e_{v_{1,2,(12)}\,v_{(12)(34)(56)}} ~, \quad e_b \equiv e_{v_{(12)(34)(56)}\,v_{3,4,(34)}} ~, \quad e_c \equiv  e_{v_{(12)(34)(56)}\,v_{5,6,(56)}}
\end{align}

This in turn fixes the orientation on the edges and a complete ordering on the set of internal edges.  

Hence we have placed additional data on any triangulation of the $n$-gon. That is, with each vertex $T$ of the associahedron we associate,  
\begin{align}
T_{\textrm{directed}}\, :=&\nonumber\\
&\hspace*{-0.8in}T \oplus \textrm{\big( Each dissection marked by an arrow}\nonumber\\
&\textrm{and a complete ordering on the set of chords that constitute $T$.\, \big)}
\end{align}
As there is a canonical map from $T\, \rightarrow\,T_{\textrm{directed}}$ that is defined by the triangulation itself, we will continue to denote the marked triangulation $T_{\textrm{directed}}$ as  $T$.  

\subsection{Complete Ordering on a Pseudo-Triangulation}\label{coopt}
Similarly, given any pseudo-triangulation $PT$ of an $n$-gon with a puncture, one can assign a complete ordering on the set of all the vertices and set of all the chords of the dual graph $\gamma_{PT}$. The simplest way to understand the ``ordering map" is via the ``forward-limit" construction as any $\gamma_{PT}$  with $n$ external vertices can be obtained from a tree-level graph with $n+2$ vertices by gluing the two additional edges. Rather than elaborating on the algorithm for general $n$, we illustrate it with an example. Interested reader can easily extend this discussion to a one loop planar graph with arbitrary number of external legs. 

We thus consider a pseudo-triangulation $PT_{0}$ of a punctured $n=4$-gon whose dual graph $\gamma_{PT_{0}} =: \gamma_{1}^{13\, \textrm{t}}$ is as shown in \eqref{coon41lg}.\footnote{The superscript is used to denote the fact that there is a tadpole on the propagator labelled by the chord $(1,3)$.} The corresponding loop integrand can also be interpreted as a forward limit of a $6$ point tree-level graph with the ordering $(1,2, 0_{L}, 0_{R}, 3, 4)$ as shown in \eqref{coon41lg}. All the external edges of this graph are outgoing with momenta on the two additional edges being $l, -l$ as shown.  We can now apply the algorithm devised for 3-valent planar graphs and immediately deduce the total ordering on set of vertices and map the set of edges to set of directed edges. 
 \begin{align}
\begin{tikzpicture}[scale=0.75,baseline={([yshift=-.5ex]current bounding box.center)}]
\draw (-1,1) -- (0,0) -- (-1,-1);
\draw (0,0) -- (2,0) -- (3,-1);
\draw (2,0) -- (3,1);
\draw (1,0) -- (1,0.8);
\draw (1,1.3) circle [radius=0.5];
\node at (-1.3,-1) {1};
\node at (-1.3,1) {2};
\node at (3.3,1) {3};
\node at (3.3,-1) {4};
\draw[-stealth] (-0.55,-0.55) -- (-0.56,-0.56);
\draw[-stealth] (-0.55,0.55) -- (-0.56,0.56);
\draw[-stealth] (2.55,-0.55) -- (2.56,-0.56);
\draw[-stealth] (2.55,0.55) -- (2.56,0.56);        
\draw[-stealth] (0.5,0) -- (0.51,0);
\draw[-stealth] (1.51,0) -- (1.5,0);
\draw[-stealth] (1,0.35) -- (1,0.34);
\draw[-stealth] (1.1,1.8) -- (1.11,1.8);
\node at (0.4,-0.4) {$e_{12}$};
\node at (1.6,-0.4) {$e_{34}$};
\node at (1.3,0.4) {$e$};
\node at (0.2,1.4) {$e_l$};
\node[gray] at (-0.6,0) {\small $v_{12}$};
\node[gray] at (2.6,0) {\small $v_{34}$};
\node[gray] at (1,1.1) {\small $v$};
\end{tikzpicture} \hspace{2cm}
\begin{matrix}
v_{12} < v < v_{34} < v_{1234} \\
\\
e_{12} < e_{l} < e < e_{34}
\end{matrix} \label{coon41lg}
\end{align}

We end this section with a couple of remarks on the relevance of the ordering and orientation we have introduced, on the set of all vertices.  
\setlist{nolistsep}
\begin{itemize}[noitemsep]
 \item The ordering on set of all vertices is sufficient to turn all the internal edges into directed edges. This is a necessary structure required to define the Corolla differential in section \ref{rngpdr}, which will ``spin up" bi-adjoint $\phi^{3}$ amplitude to obtain Yang-Mills amplitude. 
\item The action of Corolla differential on a 3-valent planar graph will result in a set of graphs with marked edges. The ordering on  $E_{\textrm{int}}$ is then necessary to prove that the direct sum of all such graphs generate the graph homology.\footnote{It has been proved in \cite{berghoff2008} that the homology class itself is invariant under a permutation of ordering. However the ordering is essential to prove that any representative of the homology is anhilated by so called graph differentials. These essential technicalities have been analysed in \cite{kreimer1208,berghoff2008} and we will not elaborate on them further.}
\item We used a structure inherent to the associahedron to define a ordering on set of all the vertices and set of all the edges. It is important to note for a given graph $\gamma$ one can also define such an ordering by simply giving a cyclic ordering on the set of all the corollas in $\gamma$, \cite{vogtmann}. A detailed comparison between the two prescriptions is outside the scope of this paper. 
\end{itemize}
\subsection{Review of a new graph polynomial and it's Differential Representation.}\label{rngpdr}
In this section, we review the definition of Corolla polynomial. Corolla polynomial is defined for any tri-valent graph with arbitrary number of cycles but we restrict ourself with either tree-level or one loop planar graphs with tri-valent vertices. Although not nearly as well known some of the other graph polynomials, (e.g. the Symanzik polynomials), Corolla polynomial has been studied in the literature by physicists and mathematicians. For a study of some of the fascinating properties of this polynomial, we refer the reader to \cite{kreimer-yeats}. 

Let $\gamma_{n}$ be a three-regular graph with $n$ external edges and no cycles.\footnote{Recall from the previous section that $\gamma_{n}$ has additional structures : ordering on the set of internal edges and clock-wise orientation at each vertex.} Then the Corolla polynomial associated to $\gamma$ is defined as, 
\begin{align}
C_{\gamma_{n}}\, =\, C^{(0)}_{\gamma_{n}}\, -\, \delta_{l,1}\, C^{(1)}_{\gamma_{n}}
\end{align}
where $0\, \leq\, l\, \leq\, 1$ is the number of cycles in $\gamma_{n}$. 
$C^{(0)}$ and $C^{(1)}$ are defined as follows. 
\begin{align}
C^{(0)}_{\gamma_{n}}\, &=\, \prod_{v\, \in\, V(\gamma_{n})}\, { C}_{v} ~, \\
\text{where} \qquad { C}_{v}\, &=\, \sum_{I=1}^{3}\, a_{h_{v}^{I}}
\end{align}

In order to define $C^{(1)}$ we introduce some notations first.
 \setlist{nolistsep}
\begin{enumerate}[noitemsep]
\item If $L_{\gamma_{n}}$ is the unique cycle in $\gamma_{n}$ then the set vertices adjacent to it will be denoted as $V(L_{\gamma_{n}})$. 
\item The half-edge incident on $v^{\prime}\, \in\, V(L_{\gamma_{n}})$ which is not inside $L_{\gamma_{n}}$ will be denoted as $l_{v^{\prime}}$. 
\item Given the cyclic ordering at $v^{\prime}$, the corresponding edges that precede and follow $l_{v^{\prime}}$ will be denoted as $l_{v^{\prime}}^{\pm}$. 
\item $\forall\, v^{\prime}\, \in\, V(L_{\gamma_{n}})$, we define 
\end{enumerate}
\begin{align}\nonumber
b_{v^{\prime}}\, :=\, a_{l_{v^{\prime}}^{+}}\, +\, a_{l_{v^{\prime}}^{-}}.
\end{align}
In summary, The Corolla polynomial is then a difference of two polynomials for all cubic graphs with number of closed cycles $\leq\, 1$. 
\begin{align}
C_{\gamma_{n}}\, =\, C^{(0)}_{\gamma_{n}}\, -\, C^{(1)}_{\gamma_{n}}
\end{align}
\begin{align}\label{c0andc1}
C^{(0)}_{\gamma_{n}}\, &=\, \prod_{v\, \in\, V(\gamma_{n})}\, C_{v}\nonumber\\
C^{(1)}_{\gamma_{n}}\, &=\, \prod_{v\, \in\, V(\gamma_{n}) - V(L_{\gamma_{n}})}\, C_{v}\, \prod_{v^{\prime}\  \in V(L_{\gamma_{n}})}\, b_{v^{\prime}}
\end{align}
The geometric interpretation of $-\, C^{(1)}_{\gamma_{n}}$ is that it maps the gluon loop integrand  to the ghost loop integrand. In other words it maps the three-gluon vertex adjacent to the cycle to a gluon-ghost-ghost vertex. 

We end this section by assigning  4-vectors to all the edges of $\gamma_{n}$ which will be used in defining a differential operator representation for the Corolla polynomial. 
\setlist{nolistsep}
\begin{enumerate}[noitemsep]
\item Every external edge $e\, \in\, E_{\textrm{ext}}$ will be labelled by a space-time vector $\xi_{i}^{\mu}$.
\item  If $\gamma_{n}$ has no cycles and if $T_{\gamma_{n}}$ is the corresponding triangulation, then for an internal edge $e^{\prime}$ which is dual to a chord $(i,j)\, \in\, T_{\gamma_{n}}$, we assign a label $\xi^{\mu}_{i,j}$. 
\item If $\gamma_{n}$ contains exactly one cycle and if $PT_{\gamma_{n}}$ is the corresponding (pseudo)-triangulation, then an internal edge $e^{\prime}$ which is dual to any of the chords $(I,J) := (i,j),\, (i,0), (i,i) \in\, PT_{\gamma_{n}}$, we assign a label $\xi^{\mu}_{I,J}$.\footnote{Recall that in this case the set of chords also include  $(i,i+1)$ which are not homotopic to any of the boundaries of the $n$-gon.}
\end{enumerate}
The scalar products of these vector labels generate the off-shell kinematic space ${\cal K}_{n}^{\textrm{os}}$ and projection to the kinematic space is obtained by simply using the following replacement rules.
\begin{align}
\xi_{i}^{\mu}\rightarrow\, p_{i}^{\mu} ~, \qquad 
\xi^{\mu}_{i,j}\rightarrow\, p_{i}^{\mu}\, +\, \dots\, + p_{j-1}^{\mu}
\end{align}


\section{From triangulations and flips to $n-3$ ranked Multi-vector field on ${\cal K}_{n}$ } \label{sec: from triangulatiosn to MVF}

As every vertex $v_{T}$ of  $A_{n-3}$ is labelled by a complete triangulation $T$ of the $n$-gon, one can associate a unique Corolla polynomial $C_{T}$ to each vertex of the associahedron. This assignment provides us with the following $(n-3)$-rank multi-vector field (MVF) which we denote as ${\cal P}_{n}$. We use the letter $c$ to denote a generic chord that dissects an $n$-gon.  
\begin{align}\label{mvfcor}
{\cal P}_{n-3}\, :=\, \prod_{i=1}^{n} x_{i}\, \sum_{T}\, (-1)^{T}\, C^{(0)}_{T}\, \prod_{c\, \in\, T}\, \prod_{j=1}^{n}\frac{1}{x_{j}}\, \frac{\partial}{\partial x_{c}}
\end{align}
Consider a specific triangulation 
\begin{align}
T_{0}\, =\, \{\, (1,3),\, (1,4),\, \dots,\, (1,n-1)\, \}.
\end{align}
The set of vertices of the graph $\gamma_{T_{0}}$ can be labelled as, 
\begin{align}
V(\gamma_{T_{0}})\, :=\, \{\, ( 1,2,(12) ),\, ( (12),3, ((12),3) ),\, \dots,\, ( (1\; n-2), n-1, (1\; n-1) )\, \}
\end{align}
One can use this presentation of the vertices of $\gamma_{T_{0}}$ to express Corolla polynomial as a product of monomials 
\begin{align}
C^{(0)}_{T}\, =\, \prod_{v\, \in\, V(\gamma_{T})}\, C_{v}
\end{align}
so that, 
\begin{align}\label{corollavprod}
C^{(0)}_{T_{0}}\, =\, C^{(0)}_{ ( 1,2,(12) )}\, C^{(0)}_{ ( (12),3, ((12),3) )}\, \dots\, C^{(0)}_{( (1\; n-2), n-1, (1\; n-1) )}
\end{align}

We will now show that tree-level YM color-ordered amplitude is a scalar  in ${\cal K}_{n}^{\textrm{os}}$ which is obtained by a contraction of a multi-vector field with the pull-back of the  planar scattering form $\Omega_{n}$ defined in ${\cal K}_{n}$, \cite{nima1711}. Restriction of the scalar to ${\cal K}_{n}\, \subset\, {\cal K}_{n}^{\textrm{os}}\, \dots$.

\subsection{From d-log form to  Tree-level Amplitude}
In \cite{kreimer1208} Corolla polynomial was represented as a map from a meromorphic function in the space of Schwinger parameter to a differential operator of degree $n-2$ on ${\cal K}_{n}^{\textrm{os}}$.  

Directly inspired by the construction of \cite{kreimer1208},  we propose a new representation of the Corolla polynomial which will appear inside ${\cal P}_{n-3}$ defined in eqn.(\ref{mvfcor}). We then show that contraction of ${\cal P}_{n-3}$ with the d-log form $\Omega_{n-3}$ on ${\cal K}_{n}^{\textrm{os}}$  produces the tree-level  color ordered YM amplitude.\footnote{Strictly speaking we have to pull back the scalar obtained via contraction from ${\cal K}_{n}^{\textrm{os}}$ to ${\cal K}_{n}$.} Contraction of a  d-log form by a multi-vector field determined by the half-edges of the set of graphs is reminiscent of the dervation of color-dressed amplitude of bi-adjoint $\phi^{3}$ theory from $\Omega_{n-3}$, (See section \ref{cdfdln}). One technical difference of course is that whereas, ${\cal P}_{n-3}^{\textrm{color}}$ is a multi-vector field in the big kinematic space ${\cal K}_{n}^{\star}$, we now define a multi-vector field in ${\cal K}_{n}^{\textrm{os}}$.

Tthe contraction of ${\cal P}_{n-3}$ with the $\Omega_{n-3}$ is, 
\begin{align}
\langle\, {\cal P}_{n-}, \Omega_{n-3}\, \rangle\, =\, \prod_{i=1}^{n}x_{i} \sum_{T\, \in\, V(A_{n-3})}\, C^{(0)}_{T}\, \prod_{i=1}^{n}\, \frac{1}{x_{i}}\, \prod_{C\, \in\, v}\, \frac{1}{x_{C}}
\end{align}
Each vertex $T = \triangle_{1}^{T}\, \cup\, \dots\, \triangle_{n-2}^{T}$ of the associahedron corresponds to a unique set of tri-valent vertices that we denote as $\{ v_{1}^{T},\, \dots\, v_{n-2}^{T} \}$. As reviewed in section  \ref{rngpdr}, one can then associate to $T$, a corolla polynomial $P_{T}$
\begin{align}
C^{(0)}_{T}\, :=\, \prod_{i=1}^{n-2}\, C_{v_{i}^{T}}
\end{align}
If the vertex $v_{i}^{T}$ has the three incident half-edges $h_{i,v}^{1} < h_{i,v}^{2} < h_{i,v}^{3}$ than the corolla polynomial associated to $v_{i}^{T}$ is, 
\begin{align}
\sum_{I=1}^{3} a_{h_{i,v}^{I}}
\end{align}
We can in fact interpret the multi-vector field defined in eqn.(\ref{mvfcf})  as a specific presentation of Corolla polynomial.
\setlist{nolistsep}
\begin{enumerate}[noitemsep]
\item The multi-vector field is parametrized by the color factors $\tilde{c}(\gamma_{\alpha})$ which is a polynomial in the color space and can be written as
\begin{align}
\tilde{c}(\gamma)\, =\, \prod_{v\, \in\, V(\gamma)}\, \textrm{Tr}(\, T_{a_{v}}\, [\, T_{b_{v}},\, T_{c_{v}}\, ]\, )
\end{align}
\item Each tensor at the vertex is associated to the half-edges and there is a contraction over repeated color indices corresponding to gluing of half-edges. 
\item The color-dressed bi-adjoint ampliude is a scalar in ${\cal K}_{n}^{\textrm{os}}$ as opposed to color ordered amplitude which is a form in ${\cal K}_{n}$.  
\item The polynomial at each vertex which is dependent on the (half)-edges incident at $v$ is a specific representation of Corolla polynomial. Given a fixed orientation at every vertex and a fixed ordering on the set of edges, Corolla polynomial is determined by assigning two variables $(b_{e}, a_{e})$ to each edge $e$. Given a cyclic ordering of $e_{1}, e_{2}, e_{3}$ at $v$ which is $e_{1} < e_{2} < e_{3}$ Corolla polynomial is
\begin{align}
\tilde{c}_{v} := f^{a_{e_{1}}a_{e_{2}}a_{e_{3}}}
\end{align}
\end{enumerate}
We now define another representation of Corolla Polynomial as a multi-vector field in ${\cal K}_{n}$. In this representation, Corolla polynomial $C_{T}$ associated to a given triangulation $T$ is not a polynomial in color space but a differential operator which acts on meromorphic functions on ${\cal K}_{n}$. 

The starting point is a representation for Corolla monomial associated to a a cell $\triangle\, \in\, T$. Let,  the three 4-vector labels associated to boundary of $\triangle$ be $\{\xi_{1}, \xi_{2}, \xi_{12}\, \}$.  $\triangle$ is dual to a 3 point vertex depicted in \eqref{threevv}. 
\begin{align}
&\begin{tikzpicture}[baseline={([yshift=-.5ex]current bounding box.center)}]\label{threevv}
\draw (0,0) -- (1,1);
\draw (0,0) -- (1,-1);
\draw (0,0) -- (-1.5,0);
\draw[-stealth] (-0.74,0) -- (-0.75,0);
\draw[-stealth] (0.5,0.5) -- (0.51,0.51);
\draw[-stealth] (0.5,-0.5) -- (0.51,-0.51);
\node at (-0.9,-0.3) {${(\mu_1}) ~ ~ ~ \xi_1$};
\node at (0.7,0.9) {$(\mu_2) ~ ~ ~ \xi_2$};
\node at (0.7,-0.9) {$(\mu_{12}) ~ ~ ~ \xi_{12}$};
\end{tikzpicture}\\
\label{ccell}
    C_{\triangle} &:= \eta^{\mu_1\mu_2} \left(\frac{\xi_1^2}{2}\,\p_{\xi_1} - \frac{\xi_2^2}{2}\,\p_{\xi_2}\right)^{\mu_{12}} + \eta^{\mu_2\mu_{12}} \left(\frac{\xi_2^2}{2}\,\p_{\xi_2} - \frac{\xi_{12}^2}{2}\,\p_{\xi_{12}}\right)^{\mu_1} + \eta^{\mu_{12}\mu_1} \left(\frac{\xi_{12}^2}{2}\,\p_{\xi_{12}} - \frac{\xi_1^2}{2}\,\p_{\xi_1}\right)^{\mu_2}
\end{align}
One can use eqns. (\ref{corollavprod}, \ref{ccell}) to write the Corolla polynomials any complete triangulation $T$. 
\subsubsection*{Corolla contraction with the one form}
Let us consider the simplest example of $n=4$ and the corresponding 1 dimensional associahedron $A_{1}$. Vertices of $A_{1}$ are the two triangulations $T_{1} = (1,3)$, $T_{2} = (2,4)$. The  Corolla differential associated to $T_{1}$ on ${\cal K}_{4}$ can be written down as a product of two differentials $C_{v_{1} = (1,2,(12))}, C_{v_{2} = ((12), 3,4)}$. 

Following notation will be useful for writing the Corolla differential in a compact form.  
\begin{align}
(\eta\, \wedge\, \eta)^{\alpha_{1}\alpha_{2}\alpha_{3}\alpha_{4}}\, :=\, 
\eta^{\alpha_{1}\alpha_{2}}\, \eta^{\alpha_{3}\alpha_{4}}\, -\, \eta^{\alpha_{1}\alpha_{4}}\, \eta^{\alpha_{2}\alpha_{3}}
\end{align}
We can then use eqn.(\ref{ccell}) and after some algebra, both the Corolla differentials at the two vertices of the graph $\gamma_{T_{1}}$ can be written as, 
\begin{align}
C_{v_{1}}\, &=\, 
-\, \frac{1}{2} \big( [\, \xi_{1}^{2}\, (\eta\, \wedge\, \eta)^{\mu_{1}\mu_{2}\alpha\mu}  \frac{\partial}{\partial \xi_{1}^{\alpha}} + \xi_{2}^{2}\, (\eta\, \wedge\, \eta)^{\mu_{2}\mu\alpha\mu_{1}}\, \frac{\partial}{\partial \xi_{2}^{\alpha}} + 
\xi_{12}^{2}\, (\eta\, \wedge\, \eta)^{\mu\mu_{2}\alpha\mu_{1}}\,  \frac{\partial}{\partial \xi_{12}^{\alpha}} \big) \label{cv1 for s channel}\\
C_{v_{2}}\, &=\, -\, \frac{1}{2} \big( [\, \xi_{3}^{2}\, (\eta\, \wedge\, \eta)^{\mu_{3}\mu_{4}\alpha\mu}\, \frac{\partial}{\partial \xi_{3}^{\alpha}} + \xi_{4}^{2}\, (\eta\, \wedge\, \eta)^{\mu_{4}\mu\alpha\mu_{3}}\, \frac{\partial}{\partial \xi_{4}^{\alpha}} + 
\xi_{12}^{2}\, ( \eta\, \wedge\, \eta)^{\mu\mu_{3}\alpha\mu_{4}}\, \frac{\partial}{\partial \xi_{12}^{\alpha}} \big)  \label{cv2 for s channel}
\end{align}
Similarly, if we denote the vertices of the graph $\gamma_{T_{2}}$ as
\begin{align}
v_{1}^{\prime} = (2,3, (23)),\, v_{2}^{\prime} = (23, 4, 1)
\end{align}
One can obtain the $C_{v_1'}$ and $C_{v_1'}$ by performing the following shift in \eqref{cv1 for s channel}: $i\to i+1~, \forall i$.

\begin{equation}
\begin{tikzpicture}[scale=1,baseline={([yshift=-.5ex]current bounding box.center)}]
    \draw (1,1) -- (0,0) -- (-1,1);
    \draw (0,0) -- (0,-1.5);
    \draw (1,-2.5) -- (0,-1.5) -- (-1,-2.5);
    \draw[-stealth] (0.5,0.5) -- (0.51,0.51);
    \draw[-stealth] (-0.5,0.5) -- (-0.51,0.51);
    \draw[-stealth] (0.5,-2) -- (0.51,-2.01);
    \draw[-stealth] (-0.5,-2) -- (-0.51,-2.01);
    \draw[-stealth] (0,-0.8) -- (0,-0.81);
    \node at (-1.25,0.9) {$\xi_1$};
    \node at (-0.9,0.3) {$(\mu_1)$};
    \node at (1.25,0.9) {$\xi_2$};
    \node at (0.9,0.3) {$(\mu_2)$};
    \node at (-1.25,-2.4) {$\xi_4$};
    \node at (-0.9,-1.8) {$(\mu_4)$};
    \node at (1.25,-2.4) {$\xi_4$};
    \node at (0.9,-1.8) {$(\mu_4)$};
    \node at (-0.1,-0.75) {$(\mu_{12}) ~ ~ ~ \xi_{12}$};
\end{tikzpicture}
\qquad + \qquad 
\begin{tikzpicture}[scale=1,baseline={([yshift=-.5ex]current bounding box.center)}]
    \draw (0,0) -- (1,1);
    \draw (0,0) -- (1,-1);
    \draw (0,0) -- (-1.5,0);
    \draw (-2.5,1) -- (-1.5,0) -- (-2.5,-1);
    \draw[-stealth] (-0.74,0) -- (-0.75,0);
    \draw[-stealth] (0.5,0.5) -- (0.51,0.51);
    \draw[-stealth] (0.5,-0.5) -- (0.51,-0.51);
    \draw[-stealth] (-2,0.5) -- (-2.01,0.51);
    \draw[-stealth] (-2,-0.5) -- (-2.01,-0.51);
    \node at (-0.75,0.3) {$(\mu_{23})$};
    \node at (-0.75,-0.3) {$\xi_{14}$};
    \node at (0.7,0.9) {$(\mu_2) ~ ~ ~ \xi_2$};
    \node at (0.7,-0.9) {$(\mu_3) ~ ~ ~ \xi_3$};
    \node at (-2.2,0.9) {$\xi_1 ~ ~ ~ (\mu_1)$};
    \node at (-2.2,-0.9) {$\xi_4 ~ ~ ~ (\mu_4)$};
\end{tikzpicture}
\end{equation}
\begin{align}
{\cal P}_{4}\, =\, \prod_{i=1}^{4} x_{i} C_{v_{1}}C_{v_{2}} \prod_{j=1}^{4}\frac{1}{x_{i}}\, \partial_{x_{13}}\, -\, \prod_{i=1}^{4} x_{i}  C_{v^{\prime}_{1}}C_{v^{\prime}_{2}} \prod_{j=1}^{4}\frac{1}{x_{i}}\, \partial_{x_{24}}
\end{align}
A moment of meditation reveals that inside either of the Corolla differential $\hat{C}_{T_{1}}$ or $C_{T_{2}}$ the differential $\xi_i^2\,\p_{\xi_i}$ with respect to the 4-vector label associated to any of the external edges appears only in one unique monomial $C_{v}$ where $v\, \in\, V(\gamma_{T_{i}}) \vert i = \{1,2\}$. Thus, we can make the following replacement across all the vertices:  $\xi_i^2\,\p_{\xi_i} \to -2\xi_i$.  We also note that this observation remains valid for any genus $0$ tri-valent graph $\gamma$. 

As a result, a rather direct computation gives, 
\begin{align}
    &\langle\, {\cal P}_{1}, \Omega_{1}\, \rangle\, =\nonumber\\
    &\hspace{0.5cm}= \left[\eta^{\mu_1\mu_2} \left({\xi_1} - {\xi_2}\right)^{\nu} + \eta^{\mu_2\nu} \left({\xi_2} - {\xi_{12}}\right)^{\mu_1} + \eta^{\nu\mu_1} \left({\xi_{12}} - {\xi_1}\right)^{\mu_2}\right] \nonumber\\
    &\hspace{4cm}\times \left[\eta_{\nu}{}^{\mu_4}\left({\xi_{12}} + {\xi_4}\right)^{\mu_3} + \eta^{\mu_4\mu_3}\left(-{\xi_4} + {\xi_3}\right)_{\nu} + \eta^{\mu_3}{}_{\nu}\left(-{\xi_3} -{\xi_{12}}\right)^{\mu_4}\right]  \frac{1}{\xi_{12}^2}  \nonumber \\
    &\hspace{0.7cm} - (\eta\, \wedge\, \eta)^{\mu_1\mu_4\mu_{2}\mu_{3}}\nonumber \\
    &\hspace{0.7cm}+\left[ \eta^{\mu_1}{}_{\nu}\left( {\xi_1} + {\xi_{14}} \right)^{\mu_4} + \eta_{\nu}{}^{\mu_4}\left(-{\xi_{14}} - {\xi_4} \right)^{\mu_1} + \eta^{\mu_4\mu_1}\left({\xi_4}-{\xi_1}\right)_{\nu}\right] \nonumber \\
    &\hspace{4cm}\times \left[\eta^{\nu\mu_2} \left({\xi_{14}} - {\xi_2}\right)^{\mu_3} + \eta^{\mu_2\mu_3} \left({\xi_2} - {\xi_3}\right)^{\nu} + \eta^{\mu_3\nu} \left({\xi_3} - {\xi_{14}}\right)^{\mu_2}\right]  \frac{1}{\xi_{14}^2} \nonumber \\
    &\hspace{0.7cm}  + (\, \eta \wedge \eta\, )^{\mu_1\mu_3\mu_{2}\mu_4}
\end{align}
The function $\langle\, {\cal P}_{1}, \Omega_{1}\, \rangle$ can now be restricted to ${\cal K}_{4}$ and equals the color ordered Four point tree-level gluon amplitude, which we denote as ${\cal M}_{4}^{\mu_{1}\dots \mu_{4}}$. The uncontracted Lorentz indices are simply substitutes for polarisation vectors of the external gluons. 

One can compute $n$ point tree-level gluon amplitude in a similar fashion. Rather than going over the entire computation for arbitrary $n$, we now argue as to why the result of the contraction between ${\cal P}_{n-3}, \Omega_{n-3}$ will always generate tree-level gluon amplitude.
\subsection*{Computing the scalar $\langle\, {\cal P}_{n-3}, \Omega_{n-3}\, \rangle$}
Let us start with the fixed triangulation of an $n$-gon 
\begin{align}
T\, &=\, \{\, (1,3),\, (1,4),\, \dots,\, (1, n-1)\, \}
\end{align}
which is dual to the graph $\gamma_{T}$ as shown in \eqref{fig:raylike triangulation}.
\begin{align} \label{fig:raylike triangulation}
\gamma_T = \ \ 
\begin{tikzpicture}[scale=0.75,baseline={([yshift=-.5ex]current bounding box.center)}]
\draw (-1,0) -- (1.5,0);
\draw (0,0) -- (0,1);
\draw (1,0) -- (1,1);
\node at (2,0) {$\hdots$};
\draw (2.5,0) -- (4,0);
\draw (3,0) -- (3,1);
\node at (-1.3,0) {$1$};
\node at (0,1.3) {$2$};
\node at (1,1.3) {$3$};
\node at (3,1.3) {$(n-1)$};
\node at (4.3,0) {$n$};
\end{tikzpicture} 
\hspace{2cm}
\begin{tikzpicture}[scale=0.75,baseline={([yshift=-.5ex]current bounding box.center)}]
\def \qq {1.414};
\draw (1.5+\qq,-0.5) -- (1+\qq,-1) -- (1,-1) -- (0,0) -- (0,\qq) -- (1,\qq+1) -- (1+\qq,1+\qq) -- (2+\qq,\qq);
\node at  (2+\qq,\qq*0.6) {$\vdots$};
\draw (1,-1) -- (0,\qq);
\draw (1,-1) -- (1,\qq+1);
\draw (1,-1) -- (1+\qq,1+\qq);
\draw (1,-1) -- (2+\qq,\qq);
\draw[teal] (0.5,-1) -- (0.25,0.1) -- (0.65,1.1) -- (1.6,1.75);
\draw[teal] (1.6,1.75) -- (2.6,1.35);
\draw[teal] (2.6,1.35) -- (3.2,0.8);
\draw[teal] (0.25,0.1) -- (-0.5,1);
\draw[teal] (0.65,1.1) -- (0.2,2.4);
\draw[teal] (1.6,1.75) -- (1.6,2.8);
\draw[teal] (2.6,1.35) -- (3.2,2.3);
\node[teal] at (0.1,-1) {1};
\node[teal] at (-0.8,1) {2};
\node[teal] at (0.1,2.6) {3};
\node[teal] at (1.6,3.1) {$4$};
\node[teal] at (3.2,2.65) {5};
\end{tikzpicture}
\end{align}
We now prove that the contraction of Corolla differential with the corresponding term in $\wedge_{(i,j)\, \in\, T}\, d\, \ln\, x_{ij}$ generates ${\cal M}_{\gamma_{T}}^{\mu_{1}\, \dots\, \mu_{n}}(p_{1},\, \dots,\, p_{n})$.  As discussed previously,  the vertices of $\gamma_{T}$ are ordered as follows.  we define $i$-th vertex of $\gamma_{T}$ as  $v_{i}$. Then, 
\begin{align}
v_{i}\, =\, (e_{1,i}, e_{i+1}, e_{1,i+1})\nonumber\\
v_{1}\, <\, v_{2}\, <\, \dots\, <\, v_{n-3}\, <\, v_{n-2}
\end{align}
Recall that at $v_{n-2}$ all the edges are incoming and on any other vertex $v^{T}_{i}$, $e_{1,i}, e_{i+1}$ are incoming and $e_{1,i+1}$ is outgoing. 
The cyclic ordering on the three edges at $v^{T}_{i}$ is then given by
\begin{align}
e_{1,i}\, <\, e_{i+1}\, <\, e_{1,i+1}\, \textrm{and cyclic ordering}
\end{align}
We denote lorentz index on any(internal or external) edge $e_{v}$ at $v$ as $\mu_{e_{v}}$. 
The Corolla differential can now be written as,
\begin{align}
\hat{C}^{(0)}_{T}\, &=\, \prod_{i=1}^{n-2}\, \hat{C}_{{v_{i}^{T}}}\nonumber\\
&=\, \prod_{i=1}^{n-2}\, \sum_{I=1}^{3}\, \frac{1}{2}\, \eta^{\mu_{e_{v_{i}}^{I-1}}\, \mu_{e_{v_{i}}^{I+1}}}\, \sum_{\pm}\, \pm\, \epsilon_{e_{v_{i}}(I\, \pm\, 1)}\, \left(\, \xi_{e_{v_{i}}(I\, \pm\, 1)}^{2}\, \frac{\partial}{\partial \xi_{e_{v_{i}}(I\, \pm\, 1)}^{\mu_{e_{v_{i}}(I)}}}\, \right)
\end{align}
where, we recall that $\epsilon_{e_{v}}\, =\, \pm\, 1 ~,$
depending on whether the  half edge $e_{v}$ is outgoing (incoming) at the vertex $v$.

The contraction of ${\cal P}_{n}^{T}$ with $\wedge_{(i,j)\, \in\, T}\, d\ln\, x_{ij}$ produces a sum of terms which can be classified according to  number of poles $\frac{1}{\xi_{1i}^{2}}$. A specific term which has no poles at $\{\, \xi_{(i,j)}^{2}\, \vert\, \vert (i,j)\vert = n_{\textrm{marked}}\, \}$ can be represented as a $n_{\textrm{marked}}$ graph, as shown in (\ref{markedvertexi}).  Marked edges arise as follows. For concreteness let us zoom in on the action of a pair of Corolla differentials on the d $\ln$ form.

In the rest of this section, we denote the Lorentz index on the chord $(1,i-1)$ as $\mu_{\overline{i}}$.
\begin{align}
\prod_{k=1}^{n}\, \xi_{k}^{2}\, \dots\, \hat{C}_{(\overline{i},i,\overline{i+1})}\, \hat{C}_{(\overline{i+1}i+1\overline{i+2})}\, \dots\, \frac{1}{\prod_{j=1}^{n} \xi_{i}^{2}}\, \frac{1}{\xi_{1,i+1}^{2}}\, \dots
\end{align}
Where $\dots$ to the left and right of Corolla differentials supress the remaining Corolla operators and the $\dots$ to the right of $\frac{1}{\xi_{1,i+1}^{2}}$ indicate the remaining propagators in the d-$\ln$ form.

The action of the highlighted Corollas can be immediately evaluated and produce precisely one term which has no pole at $x_{1,i+1} = \xi_{1,i+1}^{2} = 0$.  
\begin{flalign}
\prod_{k=1}^{n}\, \xi_{k}^{2}\, \dots\, \hat{C}_{(\overline{i},i,\overline{i+1})}\, \hat{C}_{(\overline{i+1}i+1\overline{i+2})}\, \dots\, \frac{1}{\prod_{j=1}^{n} \xi_{i}^{2}}\, \frac{1}{xi_{1,i+1}^{2}}\, \dots\, 
\rightarrow\,  \frac{1}{2}\, ( \eta\, \wedge\, \eta)^{\mu_{\overline{i}} \mu_{\overline{i+2}}\mu_{i}\mu_{i+1}}\, \dots
\end{flalign}
We will denote such a term pictorially by marking the $(1,i+1)$ chord as shown in \eqref{fig: marked edge 1i+1}.
\begin{align} \label{fig: marked edge 1i+1}
&\begin{tikzpicture}[scale=0.75,baseline={([yshift=-.5ex]current bounding box.center)}]
\def \qq {1.414};
\draw (1.5+\qq,-0.5) -- (1+\qq,-1) -- (1,-1) -- (0,0) -- (0,\qq) -- (1,\qq+1) -- (1+\qq,1+\qq) -- (2+\qq,\qq);
\node at  (2+\qq,\qq*0.6) {$\vdots$};
\draw (1,-1) -- (0,\qq);
\draw (1,-1) -- (1,\qq+1);
\draw[purple] (1,-1) -- (1+\qq,1+\qq);
\draw (1,-1) -- (2+\qq,\qq);
\draw[teal] (0.5,-1) -- (0.25,0.1) -- (0.65,1.1) -- (1.6,1.75);
\draw[line width=1.2pt,teal] (1.6,1.75) -- (2.6,1.35);
\draw[teal] (2.6,1.35) -- (3.2,0.8);
\draw[teal] (0.25,0.1) -- (-0.5,1);
\draw[teal] (0.65,1.1) -- (0.2,2.4);
\draw[teal] (1.6,1.75) -- (1.6,2.8);
\draw[teal] (2.6,1.35) -- (3.2,2.3);
\node[teal] at (0.1,-1) {1};
\node[teal] at (-0.8,1) {2};
\node[teal] at (1.6,3.1) {$i$};
\node[teal] at (3.4,2.65) {$i+1$};
\end{tikzpicture}\\
\begin{tikzpicture}[scale=0.75,baseline={([yshift=-.5ex]current bounding box.center)}]
\draw (-1,0) -- (0.5,0);
\draw (0,1) -- (0,0);
\node at (1,0) {$\hdots$};
\node at (-1.3,0) {1};
\node at (0,1.3) {2};
\draw (1.5,0) -- (3.5,0);
\draw (2,0) -- (2,1);
\draw (3,0) -- (3,1);
\node at (1.8,1.3) {$i$};
\node at (3.2,1.3) {$i+1$};
\node at (4,0) {$\hdots$};
\draw (4.5,0) -- (6,0);
\draw (5,0) -- (5,1);
\node at (5,1.3) {$n-1$};
\node at (6.3,0) {$n$};
\end{tikzpicture} &\quad \longrightarrow \quad 
\begin{tikzpicture}[scale=0.75,baseline={([yshift=-.5ex]current bounding box.center)}]
\draw (-1,0) -- (0.5,0);
\draw (0,1) -- (0,0);
\node at (1,0) {$\hdots$};
\node at (-1.3,0) {1};
\node at (0,1.3) {2};
\draw (1.5,0) -- (2.15,0);
\draw[line width=1.2pt,purple] (2.15,0) -- (2.85,0);
\draw (2.85,0) -- (3.5,0);
\draw (2.15,0) -- (2,1);
\draw (2.85,0) -- (3,1);
\node at (1.8,1.3) {$i$};
\node at (3.2,1.3) {$i+1$};
\node at (4,0) {$\hdots$};
\draw (4.5,0) -- (6,0);
\draw (5,0) -- (5,1);
\node at (5,1.3) {$n-1$};
\node at (6.3,0) {$n$};
\end{tikzpicture}
\end{align}
On the other hand, the same Corolla pair also produce a term which has a simple pole at $\xi_{1,i+1}^{2}\, =\, 0$. 
\begin{align}\label{markedvertexi}
\frac{1}{x_{i,i+1}}\, V^{\mu_{\overline{i}}\mu_{i}\mu_{\overline{i+1}}}\, :=\, \frac{1}{x_{1,i+1}}\, \left(\, \eta^{\mu_{\overline{i+1}}\mu_{i}}\, \xi_{\overline{i+1}}^{\mu_{\overline{i}}}\, -\, \eta^{\mu_{\overline{i}}\mu_{\overline{i+1}}}\, \xi_{\overline{i+1}}^{\mu_{i}}\, \right)  \left(\, \eta^{\mu_{\overline{i+1}}\mu_{\overline{i+2}}}\, \xi_{\overline{i+1}}^{\mu_{i+1}}\, -\, \eta^{\mu_{i+1}\mu_{\overline{i+1}}}\, \xi_{\overline{i+1}}^{\mu_{\overline{i+2}}}\, \right)
\end{align}
Pictorially we denote this term by marking vertices adjacent to the chord $(1,i+1)$.  To summarise, the effect of Corolla differential on the $d\,\ln$ form results in a sum of terms which have either marked vertices or marked edges. Several features of these markings for tree-level graphs  are self evident but a reader interested in explicit proofs of these statements can refer to \cite{kreimer1208} for detailed proofs.

In light of the above discussion, the end result of the contraction $\langle {\cal P}_{n-3}, \Omega_{n-3} \rangle$ can be structurally understood in terms of a combinatorial structure that we refer to as marked triangulation $T_{\textrm{marked}}$. 

\subsubsection*{Marked Triangulations}\label{mtri}
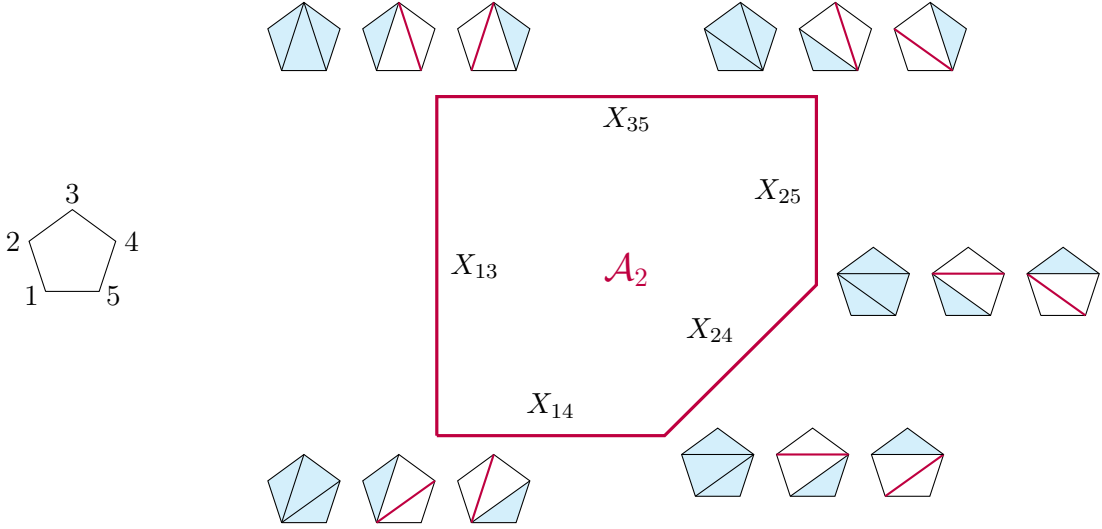
\begin{figure}[!tb]
	\centering
	\begin{tikzpicture}
	\draw[line width=1.2pt,purple] (0,0) -- (3,0) -- (5,2) -- (5,4.5) -- (0,4.5) -- (0,0);
	\node at (0.5,2.25) {$X_{13}$};
	\node at (2.5,4.2) {$X_{35}$};
	\node at (4.5,3.25) {$X_{25}$};
	\node at (3.6,1.4) {$X_{24}$};
	\node at (1.5,0.4) {$X_{14}$};
	\node[purple] at (2.5,2.2) {\Large $\mathcal{A}_2$};
	\begin{scope}[scale=0.6, xshift=-8cm, yshift=4cm]
	\coordinate (a1) at (-0.588,-0.809);
	\coordinate (a2) at (-0.951,0.3);
	\coordinate (a3) at (0,1);
	\coordinate (a4) at (0.951,0.3);
	\coordinate (a5) at (0.588,-0.809);
	\draw (a1) -- (a2) -- (a3) -- (a4) -- (a5) -- (a1);
	\node at (-0.9,-0.9) {1};
	\node at (-1.3,0.3) {2};
	\node at (0,1.35) {3};
	\node at (1.3,0.3) {4};
	\node at (0.9,-0.9) {5};
	\end{scope}
	\begin{scope}[scale=0.5, xshift=-3.5cm, yshift=-1.5cm]
	\coordinate (a1) at (-0.588,-0.809);
	\coordinate (a2) at (-0.951,0.3);
	\coordinate (a3) at (0,1);
	\coordinate (a4) at (0.951,0.3);
	\coordinate (a5) at (0.588,-0.809);
	\fill[cyan!15] (a1) -- (a2) -- (a3) -- (a4) -- (a5);
	\draw (a1) -- (a2) -- (a3) -- (a4) -- (a5) -- (a1);
	\draw (a1) -- (a3);
	\draw (a1) -- (a4);
	\end{scope}
	\begin{scope}[scale=0.5, xshift=-1cm, yshift=-1.5cm]
	\coordinate (a1) at (-0.588,-0.809);
	\coordinate (a2) at (-0.951,0.3);
	\coordinate (a3) at (0,1);
	\coordinate (a4) at (0.951,0.3);
	\coordinate (a5) at (0.588,-0.809);
	\fill[cyan!15] (a1) -- (a2) -- (a3);
	\draw (a1) -- (a2) -- (a3) -- (a4) -- (a5) -- (a1);
	\draw (a1) -- (a3);
	\draw[thick,purple] (a1) -- (a4);
	\end{scope}
	\begin{scope}[scale=0.5, xshift=1.5cm, yshift=-1.5cm]
	\coordinate (a1) at (-0.588,-0.809);
	\coordinate (a2) at (-0.951,0.3);
	\coordinate (a3) at (0,1);
	\coordinate (a4) at (0.951,0.3);
	\coordinate (a5) at (0.588,-0.809);
	\fill[cyan!15] (a1) -- (a4) -- (a5);
	\draw (a1) -- (a2) -- (a3) -- (a4) -- (a5) -- (a1);
	\draw (a1) -- (a4);
	\draw[thick,purple] (a1) -- (a3);
	\end{scope}
	\begin{scope}[scale=0.5, xshift=-3.5cm, yshift=10.5cm]
	\coordinate (a1) at (-0.588,-0.809);
	\coordinate (a2) at (-0.951,0.3);
	\coordinate (a3) at (0,1);
	\coordinate (a4) at (0.951,0.3);
	\coordinate (a5) at (0.588,-0.809);
	\fill[cyan!15] (a1) -- (a2) -- (a3) -- (a4) -- (a5);
	\draw (a1) -- (a2) -- (a3) -- (a4) -- (a5) -- (a1);
	\draw (a1) -- (a3);
	\draw (a3) -- (a5);
	\end{scope}
	\begin{scope}[scale=0.5, xshift=-1cm, yshift=10.5cm]
	\coordinate (a1) at (-0.588,-0.809);
	\coordinate (a2) at (-0.951,0.3);
	\coordinate (a3) at (0,1);
	\coordinate (a4) at (0.951,0.3);
	\coordinate (a5) at (0.588,-0.809);
	\fill[cyan!15] (a1) -- (a2) -- (a3);
	\draw (a1) -- (a2) -- (a3) -- (a4) -- (a5) -- (a1);
	\draw (a1) -- (a3);
	\draw[thick,purple] (a3) -- (a5);
	\end{scope}
	\begin{scope}[scale=0.5, xshift=1.5cm, yshift=10.5cm]
	\coordinate (a1) at (-0.588,-0.809);
	\coordinate (a2) at (-0.951,0.3);
	\coordinate (a3) at (0,1);
	\coordinate (a4) at (0.951,0.3);
	\coordinate (a5) at (0.588,-0.809);
	\fill[cyan!15] (a3) -- (a4) -- (a5);
	\draw (a1) -- (a2) -- (a3) -- (a4) -- (a5) -- (a1);
	\draw (a5) -- (a3);
	\draw[thick,purple] (a1) -- (a3);
	\end{scope}
	\begin{scope}[scale=0.5, xshift=8cm, yshift=10.5cm]
	\coordinate (a1) at (-0.588,-0.809);
	\coordinate (a2) at (-0.951,0.3);
	\coordinate (a3) at (0,1);
	\coordinate (a4) at (0.951,0.3);
	\coordinate (a5) at (0.588,-0.809);
	\fill[cyan!15] (a1) -- (a2) -- (a3) -- (a4) -- (a5);
	\draw (a1) -- (a2) -- (a3) -- (a4) -- (a5) -- (a1);
	\draw (a2) -- (a5);
	\draw (a3) -- (a5);
	\end{scope}
	\begin{scope}[scale=0.5, xshift=10.5cm, yshift=10.5cm]
	\coordinate (a1) at (-0.588,-0.809);
	\coordinate (a2) at (-0.951,0.3);
	\coordinate (a3) at (0,1);
	\coordinate (a4) at (0.951,0.3);
	\coordinate (a5) at (0.588,-0.809);
	\fill[cyan!15] (a1) -- (a2) -- (a5);
	\draw (a1) -- (a2) -- (a3) -- (a4) -- (a5) -- (a1);
	\draw (a2) -- (a5);
	\draw[thick,purple] (a3) -- (a5);
	\end{scope}
	\begin{scope}[scale=0.5, xshift=13cm, yshift=10.5cm]
	\coordinate (a1) at (-0.588,-0.809);
	\coordinate (a2) at (-0.951,0.3);
	\coordinate (a3) at (0,1);
	\coordinate (a4) at (0.951,0.3);
	\coordinate (a5) at (0.588,-0.809);
	\fill[cyan!15] (a3) -- (a4) -- (a5);
	\draw (a1) -- (a2) -- (a3) -- (a4) -- (a5) -- (a1);
	\draw (a3) -- (a5);
	\draw[thick,purple] (a2) -- (a5);
	\end{scope}
	\begin{scope}[scale=0.5, xshift=11.5cm, yshift=4cm]
	\coordinate (a1) at (-0.588,-0.809);
	\coordinate (a2) at (-0.951,0.3);
	\coordinate (a3) at (0,1);
	\coordinate (a4) at (0.951,0.3);
	\coordinate (a5) at (0.588,-0.809);
	\fill[cyan!15] (a1) -- (a2) -- (a3) -- (a4) -- (a5);
	\draw (a1) -- (a2) -- (a3) -- (a4) -- (a5) -- (a1);
	\draw (a2) -- (a5);
	\draw (a2) -- (a4);
	\end{scope}
	\begin{scope}[scale=0.5, xshift=14cm, yshift=4cm]
	\coordinate (a1) at (-0.588,-0.809);
	\coordinate (a2) at (-0.951,0.3);
	\coordinate (a3) at (0,1);
	\coordinate (a4) at (0.951,0.3);
	\coordinate (a5) at (0.588,-0.809);
	\fill[cyan!15] (a1) -- (a2) -- (a5);
	\draw (a1) -- (a2) -- (a3) -- (a4) -- (a5) -- (a1);
	\draw (a2) -- (a5);
	\draw[thick,purple] (a2) -- (a4);
	\end{scope}
	\begin{scope}[scale=0.5, xshift=16.5cm, yshift=4cm]
	\coordinate (a1) at (-0.588,-0.809);
	\coordinate (a2) at (-0.951,0.3);
	\coordinate (a3) at (0,1);
	\coordinate (a4) at (0.951,0.3);
	\coordinate (a5) at (0.588,-0.809);
	\fill[cyan!15] (a2) -- (a3) -- (a4);
	\draw (a1) -- (a2) -- (a3) -- (a4) -- (a5) -- (a1);
	\draw[thick,purple] (a2) -- (a5);
	\draw (a2) -- (a4);
	\end{scope}
	\begin{scope}[scale=0.5, xshift=7.4cm, yshift=-0.8cm]
	\coordinate (a1) at (-0.588,-0.809);
	\coordinate (a2) at (-0.951,0.3);
	\coordinate (a3) at (0,1);
	\coordinate (a4) at (0.951,0.3);
	\coordinate (a5) at (0.588,-0.809);
	\fill[cyan!15] (a1) -- (a2) -- (a3) -- (a4) -- (a5);
	\draw (a1) -- (a2) -- (a3) -- (a4) -- (a5) -- (a1);
	\draw (a1) -- (a4);
	\draw (a2) -- (a4);
	\end{scope}
	\begin{scope}[scale=0.5, xshift=9.9cm, yshift=-0.8cm]
	\coordinate (a1) at (-0.588,-0.809);
	\coordinate (a2) at (-0.951,0.3);
	\coordinate (a3) at (0,1);
	\coordinate (a4) at (0.951,0.3);
	\coordinate (a5) at (0.588,-0.809);
	\fill[cyan!15] (a1) -- (a4) -- (a5);
	\draw (a1) -- (a2) -- (a3) -- (a4) -- (a5) -- (a1);
	\draw (a1) -- (a4);
	\draw[thick,purple] (a2) -- (a4);
	\end{scope}
	\begin{scope}[scale=0.5, xshift=12.4cm, yshift=-0.8cm]
	\coordinate (a1) at (-0.588,-0.809);
	\coordinate (a2) at (-0.951,0.3);
	\coordinate (a3) at (0,1);
	\coordinate (a4) at (0.951,0.3);
	\coordinate (a5) at (0.588,-0.809);
	\fill[cyan!15] (a2)-- (a4) -- (a3);
	\draw (a1) -- (a2) -- (a3) -- (a4) -- (a5) -- (a1);
	\draw[thick,purple] (a1) -- (a4);
	\draw (a2) -- (a4);
	\end{scope}
	\end{tikzpicture}
	\caption{\emph{Layered} associahedron $\mathcal{A}_2$: Each vertex corresponds to an equivalence class of triangulations.}
	\label{fig:layered associahedron A2}
\end{figure}

Given any triangulation $T$ of an $n$-gon with $n-3$ chords, we will consider a set of triangulations in each of which a subset of these chords are colored red and a subset of cells of $T$ are marked with Blue.  All the chords and cells which are not in these subsets remain un-colored and the colorings are required to satisfy the following conditions. \footnote{Every graph obtained by action of Corolla on a three regular graph is marked in the sense described below \cite{kreimer1208}. As each tri-valent graph is dual to a triangulation, we refer to each triangulation dual to a marked graph as marked triangulation.}
\setlist{nolistsep}
\begin{enumerate}[noitemsep]
\item  No two red chords can be adjacent to each other. 
\item If a chord $(i,j)$ is red, then the two cells adjacent to it are uncolored.
\item If a cell is blue, then all the boundaries of this cell (which are chords or external edges of the $n$-gon) are uncolored.
\item As proved in \cite{kreimer1208}, the result of Corolla differential is a direct sum over marked triangulations such that total number of red and blue markings satisfy the following constraint:
\begin{align}\label{nbnrn-2}
n_{B} + 2 n_{R}\, =\, n-2~,
\end{align}
\end{enumerate}
where $n_{R}$ is the number of (non-adjacent) red chords and $n_{B}$ is the number of blue cells.\footnote{Hence in a marked triangulation based on $T$, either a cell is colored and if it un-colored then exactly one of it's edges is colored.} Look at \ref{fig:layered associahedron A2} for the case of $\mathcal{A}_2$.

Given a pair $(n_{B}, n_{R})$ that satisfy eqn.(\ref{nbnrn-2}), we  denote the set of marked triangulations based on $T$ as $T_{\textrm{marked}}^{n_{R}, n_{B}}$ and the set of corresponding graphs as $\gamma_{T}^{n_{R}, n_{B}}$. 

For the vertex of $A_{n-3}$ labelled by $T$, the scalar obtained via contraction of the multi-vector field with the $d\,\ln$ form $\wedge_{(i,j)\, \in\, T}\, d\ln\, x_{ij}$ can be written as a sum over $\oplus_{n_{B}=0}^{n-2} T_{\textrm{marked}}^{n_{R}, n_{B}}$ as follows.
\begin{align}
S_{n_{R}}\, =\, \{\, (\, ((i^{1}_{1}, j^{1}_{1}),\, \dots\, (i^{1}_{n_{R}}, j^{1}_{n_{R}})\, ), \dots,\, (\, (i^{K}_{1},  j^{K}_{1})\, \dots\, (i^{K}_{n_{R}}, j^{K}_{n_{R}})\, )\, \}
\end{align}
be the set of all possible subsets of $n_{R}\, \geq\, 1$  red-chord configurations. As an example, consider $n = 6$ and let us consider a triangulation 
\begin{align}
T\, =\, \{\, (1,3), (1, 4), (4, 6)\, \}
\end{align}
Then $0\, \leq\, n_{R}\, \leq\, 2$, and, 
\begin{equation}
\begin{aligned}
S_{n_{R} = 1}\, &=\, \{\, (13), (14), (46)\, \}\nonumber\\
S_{n_{R} = 2}\, &=\, \{\, \big( 13, 46 \big)\, \}
\end{aligned}
\end{equation}
\eqref{fig: n=6 marked triangulations} depicts all such possible markings on chords for $n=6$.
\begin{equation} \label{fig: n=6 marked triangulations}
\begin{aligned}
\text{6 ray-like triangulations: } &\qquad 
\begin{tikzpicture}[scale=0.5,baseline={([yshift=-.5ex]current bounding box.center)}]
\begin{scope}[xshift=0cm]
\coordinate (a1) at (-0.866,-0.5);
\coordinate (a2) at (-0.866,0.5);
\coordinate (a3) at (0,1);
\coordinate (a4) at (0.866,0.5);
\coordinate (a5) at (0.866,-0.5);
\coordinate (a6) at (0,-1);
\fill[cyan!15] (a1) -- (a2) -- (a3) -- (a4) -- (a5) -- (a6);
\draw (a1) -- (a2) -- (a3) -- (a4) -- (a5) -- (a6) -- (a1);
\draw (a1) -- (a3);
\draw (a1) -- (a4);
\draw (a1) -- (a5);
\end{scope}
\begin{scope}[xshift=3cm]
\coordinate (a1) at (-0.866,-0.5);
\coordinate (a2) at (-0.866,0.5);
\coordinate (a3) at (0,1);
\coordinate (a4) at (0.866,0.5);
\coordinate (a5) at (0.866,-0.5);
\coordinate (a6) at (0,-1);
\fill[cyan!15] (a1) -- (a4) -- (a5) -- (a6);
\draw (a1) -- (a2) -- (a3) -- (a4) -- (a5) -- (a6) -- (a1);
\draw[thick,purple] (a1) -- (a3);
\draw (a1) -- (a4);
\draw (a1) -- (a5);
\end{scope}
\begin{scope}[xshift=5.5cm]
\coordinate (a1) at (-0.866,-0.5);
\coordinate (a2) at (-0.866,0.5);
\coordinate (a3) at (0,1);
\coordinate (a4) at (0.866,0.5);
\coordinate (a5) at (0.866,-0.5);
\coordinate (a6) at (0,-1);
\fill[cyan!15] (a1) -- (a2) -- (a3);
\fill[cyan!15] (a1) -- (a5) -- (a6);
\draw (a1) -- (a2) -- (a3) -- (a4) -- (a5) -- (a6) -- (a1);
\draw (a1) -- (a3);
\draw[thick,purple] (a1) -- (a4);
\draw (a1) -- (a5);
\end{scope}
\begin{scope}[xshift=8cm]
\coordinate (a1) at (-0.866,-0.5);
\coordinate (a2) at (-0.866,0.5);
\coordinate (a3) at (0,1);
\coordinate (a4) at (0.866,0.5);
\coordinate (a5) at (0.866,-0.5);
\coordinate (a6) at (0,-1);
\fill[cyan!15] (a1) -- (a2) -- (a3) -- (a4);
\draw (a1) -- (a2) -- (a3) -- (a4) -- (a5) -- (a6) -- (a1);
\draw (a1) -- (a3);
\draw (a1) -- (a4);
\draw[thick,purple] (a1) -- (a5);
\end{scope}
\begin{scope}[xshift=11cm]
\coordinate (a1) at (-0.866,-0.5);
\coordinate (a2) at (-0.866,0.5);
\coordinate (a3) at (0,1);
\coordinate (a4) at (0.866,0.5);
\coordinate (a5) at (0.866,-0.5);
\coordinate (a6) at (0,-1);
\draw (a1) -- (a2) -- (a3) -- (a4) -- (a5) -- (a6) -- (a1);
\draw [thick,purple](a1) -- (a3);
\draw (a1) -- (a4);
\draw[thick,purple] (a1) -- (a5);
\end{scope}
\end{tikzpicture} \\
\text{6 `Z'-like triangulations: }&\qquad 
\begin{tikzpicture}[scale=0.5,baseline={([yshift=-.5ex]current bounding box.center)}]
\begin{scope}[xshift=0cm]
\coordinate (a1) at (-0.866,-0.5);
\coordinate (a2) at (-0.866,0.5);
\coordinate (a3) at (0,1);
\coordinate (a4) at (0.866,0.5);
\coordinate (a5) at (0.866,-0.5);
\coordinate (a6) at (0,-1);
\fill[cyan!15] (a1) -- (a2) -- (a3) -- (a4) -- (a5) -- (a6);
\draw (a1) -- (a2) -- (a3) -- (a4) -- (a5) -- (a6) -- (a1);
\draw (a2) -- (a4);
\draw (a1) -- (a4);
\draw (a1) -- (a5);
\end{scope}
\begin{scope}[xshift=3cm]
\coordinate (a1) at (-0.866,-0.5);
\coordinate (a2) at (-0.866,0.5);
\coordinate (a3) at (0,1);
\coordinate (a4) at (0.866,0.5);
\coordinate (a5) at (0.866,-0.5);
\coordinate (a6) at (0,-1);
\fill[cyan!15] (a1) -- (a4) -- (a5) -- (a6);
\draw (a1) -- (a2) -- (a3) -- (a4) -- (a5) -- (a6) -- (a1);
\draw[thick,purple] (a2) -- (a4);
\draw (a1) -- (a4);
\draw (a1) -- (a5);
\end{scope}
\begin{scope}[xshift=5.5cm]
\coordinate (a1) at (-0.866,-0.5);
\coordinate (a2) at (-0.866,0.5);
\coordinate (a3) at (0,1);
\coordinate (a4) at (0.866,0.5);
\coordinate (a5) at (0.866,-0.5);
\coordinate (a6) at (0,-1);
\fill[cyan!15] (a4) -- (a2) -- (a3);
\fill[cyan!15] (a1) -- (a5) -- (a6);
\draw (a1) -- (a2) -- (a3) -- (a4) -- (a5) -- (a6) -- (a1);
\draw (a2) -- (a4);
\draw[thick,purple] (a1) -- (a4);
\draw (a1) -- (a5);
\end{scope}
\begin{scope}[xshift=8cm]
\coordinate (a1) at (-0.866,-0.5);
\coordinate (a2) at (-0.866,0.5);
\coordinate (a3) at (0,1);
\coordinate (a4) at (0.866,0.5);
\coordinate (a5) at (0.866,-0.5);
\coordinate (a6) at (0,-1);
\fill[cyan!15] (a1) -- (a2) -- (a3) -- (a4);
\draw (a1) -- (a2) -- (a3) -- (a4) -- (a5) -- (a6) -- (a1);
\draw (a2) -- (a4);
\draw (a1) -- (a4);
\draw[thick,purple] (a1) -- (a5);
\end{scope} 
\begin{scope}[xshift=11cm]
\coordinate (a1) at (-0.866,-0.5);
\coordinate (a2) at (-0.866,0.5);
\coordinate (a3) at (0,1);
\coordinate (a4) at (0.866,0.5);
\coordinate (a5) at (0.866,-0.5);
\coordinate (a6) at (0,-1);
\draw (a1) -- (a2) -- (a3) -- (a4) -- (a5) -- (a6) -- (a1);
\draw[thick,purple] (a2) -- (a4);
\draw (a1) -- (a4);
\draw[thick,purple] (a1) -- (a5);
\end{scope}
\end{tikzpicture} \\
\text{2 `diamond'-like triangulations: } &\qquad 
\begin{tikzpicture}[scale=0.5,baseline={([yshift=-.5ex]current bounding box.center)}]
\begin{scope}[xshift=0cm]
\coordinate (a1) at (-0.866,-0.5);
\coordinate (a2) at (-0.866,0.5);
\coordinate (a3) at (0,1);
\coordinate (a4) at (0.866,0.5);
\coordinate (a5) at (0.866,-0.5);
\coordinate (a6) at (0,-1);
\fill[cyan!15] (a1) -- (a2) -- (a3) -- (a4) -- (a5) -- (a6);
\draw (a1) -- (a2) -- (a3) -- (a4) -- (a5) -- (a6) -- (a1);
\draw (a1) -- (a3);
\draw (a3) -- (a5);
\draw (a5) -- (a1);
\end{scope}
\begin{scope}[xshift=3cm]
\coordinate (a1) at (-0.866,-0.5);
\coordinate (a2) at (-0.866,0.5);
\coordinate (a3) at (0,1);
\coordinate (a4) at (0.866,0.5);
\coordinate (a5) at (0.866,-0.5);
\coordinate (a6) at (0,-1);
\fill[cyan!15] (a3) -- (a4) -- (a5);
\fill[cyan!15] (a1) -- (a5) -- (a6);
\draw (a1) -- (a2) -- (a3) -- (a4) -- (a5) -- (a6) -- (a1);
\draw[thick,purple] (a1) -- (a3);
\draw (a3) -- (a5);
\draw (a5) -- (a1);
\end{scope}
\begin{scope}[xshift=5.5cm]
\coordinate (a1) at (-0.866,-0.5);
\coordinate (a2) at (-0.866,0.5);
\coordinate (a3) at (0,1);
\coordinate (a4) at (0.866,0.5);
\coordinate (a5) at (0.866,-0.5);
\coordinate (a6) at (0,-1);
\fill[cyan!15] (a1) -- (a2) -- (a3);
\fill[cyan!15] (a1) -- (a5) -- (a6);
\draw (a1) -- (a2) -- (a3) -- (a4) -- (a5) -- (a6) -- (a1);
\draw (a1) -- (a3);
\draw[thick,purple] (a3) -- (a5);
\draw (a5) -- (a1);
\end{scope}
\begin{scope}[xshift=8cm]
\coordinate (a1) at (-0.866,-0.5);
\coordinate (a2) at (-0.866,0.5);
\coordinate (a3) at (0,1);
\coordinate (a4) at (0.866,0.5);
\coordinate (a5) at (0.866,-0.5);
\coordinate (a6) at (0,-1);
\fill[cyan!15] (a1) -- (a2) -- (a3);
\fill[cyan!15] (a3) -- (a4) -- (a5);
\draw (a1) -- (a2) -- (a3) -- (a4) -- (a5) -- (a6) -- (a1);
\draw (a1) -- (a3);
\draw (a3) -- (a5);
\draw[thick,purple] (a5) -- (a1);
\end{scope}
\end{tikzpicture}
\end{aligned}
\end{equation}

In each such subset in $S_{n_{R}}$, each red chord $c_{r}$  is adjacent to a pair of cells $\triangle_{c_{r}}^{-}, \triangle_{c_{r}}^{+}$ with clock wise orientation inside each cell. Thanks to the orientation on the edges of the graph dual to $T$, one has an ordering on the two cells that are adjacent to $c$ and without loss of generality we label the cells such that 
\begin{align}
\triangle_{c_{r}}^{-}\, <\, \triangle_{c_{r}}^{+}\, \forall\, c_{r}\, \in\, S_{n_{R}}
\end{align}
We also recall that with each cell we assign a triplet of Lorentz indices which are assigned to all edges of the polygon. For external edges, $e_{1}, \dots\, e_{n}$ these indices are $\mu_{1},\, \dots\, \mu_{n}$ and for all the chords in $T$ we assign indices $\mu_{c_{1}},\, \dots,\, \mu_{c_{n-3}}$.
We denote boundaries of each cell $\triangle$ as $(\partial\triangle)_{I} \vert 1\leq\, I\, \leq\, 3$.
Marking a chord $c$ implies that after the contraction $\frac{1}{x_{c}}$ will be replaced by 
\begin{align}
\frac{1}{2}\, (\eta\, \wedge\, \eta)^{\mu_{(\partial\triangle_{c}^{-})_{1}}\mu_{(\partial\triangle_{c}^{-})_{2}}\mu_{(\partial\triangle_{c}^{+})_{2}}\mu_{(\partial\triangle_{c}^{+})_{1}}}
\end{align}
Thus finally, contribution of set of all marked triangulations with $n_{R}$ red chords to the contraction is the function, 
\begin{align}
S^{\mu_{1}\, \dots\, \mu_{n}}_{T_{\textrm{marked}}^{n_{R}, n_{B}}}&=\nonumber\\ 
&\hspace*{-0.6in}\sum\limits_{\hspace*{-0.05in}(c_{1},\, \dots,\, c_{n_{R}})\, \in C_{n_{R}}}\hspace*{-0.05in}\prod\limits_{\, c\, \in\, (c_{1}, \dots,\, c_{n_{R}})}\frac{1}{2}\, (\eta\, \wedge\, \eta)^{\mu_{(\partial\triangle_{c}^{-})_{1}}\mu_{(\partial\triangle_{c}^{-})_{2}}\mu_{(\partial\triangle_{c}^{+})_{2}}\mu_{(\partial\triangle_{c}^{+})_{1}}}\hspace*{-0.5in} \prod\limits_{\triangle\, \in\, T\vert\, \partial\triangle\, \cap\, (c_{1}, \dots\, c_{n_{R}})\, =\, 0}\hspace*{-0.5in} V_{\triangle}^{\mu_{(\partial\triangle)_{1}}\mu_{(\partial\triangle)_{2}}\mu_{(\partial\triangle)_{3}}}
\end{align}
where 
$V_{\triangle}^{\mu_{(\partial\triangle)_{1}}\mu_{(\partial\triangle)_{2}}}$ is defined in eqn.(\ref{markedvertexi}). 

Hence the resulting scalar obtained by summing over all the vertices of $A_{n-3}$ is a scalar in ${\cal K}_{n}^{\textrm{os}}$,
\begin{align}\label{somtfr}
{\cal M}_{n}^{\mu_{1}\, \dots\, \mu_{n}}\, = \sum\limits_{T} \!\! \sum \limits_{\substack{n_R\,,\,n_B \\ n_R+2n_B=n-2}} \!\!\! S^{\mu_{1}\, \dots\, \mu_{n}}_{T_{\textrm{marked}}^{n_{R}, n_{B}}}.
\end{align}
We can pull back ${\cal M}_{n}^{\mu_{1}\, \dots\, \mu_{n}}$ to the on-shell Kinematic space that includes planar kinematic variables as well as scalars constructed out of  $n$ polarisations vectors and $\{p_{1},\, \dots,\, p_{n}\}$. The resulting function is the on-shell $n$ point gluon amplitude ${\cal M}_{n}((\epsilon_{1}, p_{1}),\, \dots,\, (\epsilon_{n}, p_{n}) )$.
\subsection{On Gauge Invariance of ${\cal M}_{n}$.}\label{gaugeinvarianceofmn}
How does action of Corolla differential guarantee that result is a gauge invariant object? A rigorous answer to this question was provided in \cite{kreimer1208}. The proof explored the relationship between BRST invariance of Feynman integrals in YM theory with homology of graph complexes, \cite{vogtmann}. This is a rich subject in it's own right and we refer the reader to \cite{kreimer1208, berghoff2008} for a detailed exploration.\footnote{Readers not interested in the homological aspects of Corolla action can safely skip this section as our subsequent computations do not depend on these aspects. However the importance of these results should also not be ignored.  That the homological proofs guarantee that action of Corolla differential on a $\phi^{3}$ amplitude leads to a function which is gauge invariant and unitary at all orders in perturbation theory.} 

More in detail, it was shown that at the graph theoretic level, action of Corolla differential generates element of a graph homology as defined in \cite{vogtmann}. Although this result holds for a graph with arbitrary number of cycles, we illustrate it for the simplest case of a planar tri-valent graph with no cycles. 

Let $\chi_{e}$ be an ``edge-marking operator" for any $e\, \in\, E(\gamma_{n})$ which is defined as follows.
\begin{align}
\chi_{e}\, \gamma_{n}&= 0\quad \textrm{if $e$ is adjacent to any other marked edge or a marked vertex.}\nonumber\\
&=\, \gamma_{n}^{e\, \rightarrow\, e_{\textrm{m}}}
\end{align}
where $e_{\textrm{m}}$ indicates a marked edge. 
Let 
\begin{align}\nonumber
\chi = \sum_{e \in E(\gamma)} \chi_{e}
\end{align}
It is then clear that the sum over all the graphs obtained from $\gamma_{n}$ and which contain at least one marked edge can written as,
\begin{align}
\oplus\, \gamma_{n}^{\textrm{marked}}\, =\  e^{\chi}\, \gamma_{n}
\end{align}
One can now define a graph differential $S$ as sum over two differentials $s_{1}, s_{2}$. These operators may look rather formal and we will attempt to give a physical picture for them below. But first we simply define them following \cite{kreimer1208}.

 In order to define such a differential, it is crucial that $E(\gamma)$ be ordered. These differentials are defined as, 
\begin{align}
s_{1}\, \gamma\, &:=\, \sum_{e\, \in\, E(\gamma)} (-1)^{\vert e^{\prime} < e \vert, e^{\prime} \textrm{marked}}\, \gamma_{\textrm{marked}}(e\, \rightarrow\, \cancel{e})\\
s_{2}\, \gamma &:=\, (-1)^{\vert E_{m} \cup E_{\cancel{e}}\vert}\, \sum_{e \in E_{\textrm{m}}}\, (-1)^{\{\vert e^{\prime} > e_{m} \vert e^{\prime} \in E_{m}\}}\, \sigma_{e} \gamma_{\textrm{marked}}\\
\text{with} \qquad \sigma_{e}\gamma_{\textrm{marked}} &= \gamma_{\textrm{marked}}(e\, \rightarrow\, \cancel{e})
\end{align}
It can now be checked that for $S = s_{1} + s_{2}$,
\begin{align}
S^{2}\, \gamma &= 0\nonumber\\
S e^{\chi}\, \gamma\, &=\, 0
\end{align}
And thus $S$ is a homological operator and $e^{\chi}(\gamma)\, \in\, H_{\star}(S)$. The action of $s_{1}, s_{2}$ involved a marking $\cancel{e}$ distinct from the action of $\chi$. Intuitively, $\cancel{e}$ can be understood as collapsing $e$ to obtain quartic vertex. And thus action of $s_{1}$ can be understood as collapsing an unmarked edge into four valent vertex and $s_{2}$ action can be understood as identifying a marked edge with a four valent vertex. The fact that $e^{\chi}\gamma$ generates graph homology is then necessary to ensure that for every graph with only three valent vertex, there is the appropriate number of graphs with quartic vertices consistent with gauge invariance.
We end this section with a remark.
\begin{itemize}
\item For a given $\gamma$, one can simply choose an ordering once and for all but in our case, the ordering has to be preserved under mutation. 
As we have seen previously, the ``associative structure" encapsulated by combinatorial associahedron precisely equips each tri-valent graph with such an ordering. 
\end{itemize}


\section{Transmutation as inverse of Corolla differential} \label{sec: transmutation}
In a beautiful paper, \cite{cheung:transmutation}, a map was derived which maps the color ordered (tree-level) Yang-Mills amplitude ${\cal A}_{\textrm{YM}}$ to the bi-adjoint $\phi^{3}$ amplitude with identical color ordering in the two color- channels. The map between the two theories was through a differential operator (known as transmutation)  in the space of all scalar invariants spanned by $p_{i} \cdot p_{j}, \epsilon_{i} \cdot p_{j}, \epsilon_{i} \cdot \epsilon_{j}$. Intuitively, transmutation operators undo the Corolla action! Given a specific color ordering for the Yang-Mills amplitude, the transmutation is a (product of) differential operators in the space spanned by $(\, p_{i} \cdot p_{j}, \epsilon_{i} \cdot p_{j}, \epsilon_{i} \cdot \epsilon_{j}\, )$ such that
\begin{align}
{\cal T}_{n}\, {\cal A}^{\textrm{co}}_{\textrm{YM}}(1, \dots, n)\, =\, {\cal A}_{\phi^{3}}^{\textrm{co}}(p_{1}, \dots, p_{n})
\end{align}
Definition of ${\cal T}_{n}$ can be found in \cite{cheung:transmutation} and will not be relevant for our purposes. The super-script on both the amplitudes stand for color-ordering. 

As ${\cal A}_{n}((\epsilon_{1}, p_{1}), \dots, (\epsilon_{n}, p_{n}) )$ is a scalar in the ``enlarged" Kinematic space parametrized by the planar kinematic variables and $ (\epsilon_{i} \cdot p_{j}, \epsilon_{i} \cdot \epsilon_{j})\vert_{1\leq\, i < j \leq\, n}$, we can now ask if there exists a transmutation-like operation that maps this function to $\Omega_{n-3}$ in ${\cal K}_{n}$. In the present scenario however, such a map is rather easy to construct.  

The sum over marked triangulations lead us to the following decomposition of the on-shell amplitude (see eqn.(\ref{somtfr}).),
\begin{align}
{\cal A}^{\textrm{YM}}_n =\, \sum_T\frac{n_T}{\prod_{(ij)\in T}X_{ij}} + \sum_{n_{R}\neq\, 0}\, \dots \label{general YM amplitude}
\end{align}
where the first term is the contribution of $n_{R} = 0$ obtained by contracting the three point vertices and multiplying the result with the valuation of $\phi^{3}$ amplitude over $\gamma_{T}$. The numerator factor $n_{T}$ is a function in the enlarged kinematic space. 
\begin{align}
n_{T} = n_{T}( \{X_{ij}, \epsilon_{i} \cdot p_{j}, \epsilon_{i} \cdot \epsilon_{j}\})
\end{align}
Transmutation which maps this amplitude back to the d-log form is in fact rather simple to write down. Let $Y_{ij} = \frac{1}{X_{ij}}\, \forall\, 1\, \leq\, i < j-1\, \leq\, n-2$. 
\begin{align}
\Upsilon\, :=\, \sum_{T} (-1)^{\sigma(T)}\, \wedge_{(ij) \in\, T} d \ln X_{ij}
\, \Upsilon_{T}  \end{align}
where 
\begin{align}
\Upsilon_{T}\, =\, \frac{1}{n_{T}} \prod_{(mn) \in T}\, \frac{\partial}{\partial Y_{mn}}
\end{align}
We note that any term in the Yang-Mills amplitude which arises from a marked triangulation corresponding to $T$ but with $n_{R}\, \neq\, 0$ is annihilated by the differential $ \prod_{(mn) \in T}\, \frac{\partial}{\partial Y_{mn}}$. In essence, $\Upsilon$ impliments the transmutation operation defined in \cite{cheung:transmutation}. However it does so at the level of individual vertices of the associahedron so as to generate the d-log form as opposed to the $\phi^{3}$ amplitude.  Of course, as is well known, the bi-adjoint amplitude is simply the pull back of d log form onto the ABHY associahedron in ${\cal K}_{n}$.
 

\section{From $\hat{D}_{n}$ polytope to one loop integrand of Yang-Mills amplitude} \label{sec: one loop YM using POmega}
In this section we extend the above construction to one loop integrands of gauge theories.  That is, we show that the contraction of the d $\ln$ form induced by the $\hat{D}_{n}$ polytope with the MVF ${\cal P}_{n}^{1-l}$ maps the bi-adjoint amplitude to the Yang-Mills amplitude. The  MVF in this case is defined using the Corolla polynomial associated to graphs with one cycle. This polynomial was reviewed in section (\ref{rngpdr}), and is a difference of two polynomials with the relative sign signifying the contribution of the ghost loop. (See eqn. (\ref{c0andc1}) for details.)
\begin{align}
{\cal P}_{n}^{1-l}\, =\, \sum_{PT\, \in\, V(\hat{D}_{n})}\, (-1)^{PT}\, \prod_{i=1}^{n} x_{i} [\, \hat{C}^{(0)}_{PT} - \hat{C}^{(1)}_{PT}\, ]\, \prod_{IJ\, \in\, PT}\, \prod_{i=1}^{n}\, \frac{1}{x_{i}}\, \partial_{x_{IJ}}
\end{align}
We will now argue that  $\langle\, {\cal P}_{n}^{1-l},\, \Omega_{n}^{1-l}\, \rangle$ is  the gauge invariant integrand for 1 loop gluon amplitude for Yang Mills theory. Just as action of the differential $\hat{C}^{(0)}$ generate 3 point and 4 point gluon vertices, the action of $\hat{C}^{(1)}$ ``transmutes" a gluon loop into a ghost loop which restores the gauge invariance at the integrand level. This new contribution at one loop is in fact due to  the 1 ghost-2 gluon vertex. We give a few examples of action of Corolla vector field on $\hat{D}_{n}$ canonical form and argue that sum over tadpole graphs in bi-adjoint scalar amplitude  cancel.  

The contraction $\langle P_{n}^{1-l}, \Omega_{n}^{1-l}\, \rangle$ can then be written as, 
\begin{align}
\Phi^{\mu_{1},\, \dots\, \mu_{n}}_{1-l}(p_{1}, \dots,\, p_{n}, l)\, =\, \sum_{PT\, \in\, V(\hat{D}_{n})}\, \hat{C}_{PT}^{\mu_{1}\, \dots\, \mu_{n}}\, \prod_{c\, \in\, PT}\, \frac{1}{x_{c}}
\end{align}
where
\begin{align}
\hat{C}_{PT}\, =\, \hat{C}^{(0)}_{PT}\, -\, \hat{C}^{(1)}_{PT}
\end{align}
At the outset, the action of Corolla on $d \ln$ form appears as complicated as simply evaluating tedious Feynman diagrams. However some intriguing simplifications occur due to (1) cancellation in the Corolla polynomial thanks to the cycle polynomial $C^{(1)}$ and (2) the sum over all the vertices of $\hat{D}_{n}$ polytope wherein each Feynman diagram is repeated twice.  The ``primitive" diagrams whose Corolla polynomials generate Corolla differentials for all one loop graph can be divided into polynomials associated to tree-level graphs, a tadpole vertex, a wheel graph with $n\, \geq\, 2$ external legs. The corresponding Corolla polynomials can be immediately evaluated. Recall that $a_{h^{I}_{e}}\, \vert\, I\, \in\, \{1,2\}$ is the variable which labels one of the two half edges $h^{1}_{e}, h^{2}_{e}$ which belong to the edge $e$ and which are such that
 \begin{align}
 h^{1}_{e}\, =\, (e, b(e)),\quad h^{2}_{e} = (e, f(e))
 \end{align}
 Then, 
\begin{align}\label{cpfdg1l}
C_{\textrm{tadpole}}( a_{e}, a_{e^{-}}, a_{e^{+}})\, &=\, a_{e}\nonumber\\
\hspace*{-0.7in}C_{\textrm{wheel, n=2}}&= 
(\, a_{e} + a_{h^{1}_{e^{-}}} + a_{h^{1}_{e^{+}}}\, ) (\, a_{e^{\prime}} + a_{h^{2}_{e^{-}}} + a_{h^{2}_{e^{+}}} ) - 
( a_{h^{1}_{e^{-}}} + a_{h^{1}_{e^{+}}})\, (a_{h^{2}_{e^{+}}} + a_{h^{2}_{e^{-}}} )\nonumber\\
&=\, a_{e} a_{e^{\prime}} + a_{e} ( a_{h^{2}_{e^{-}}} + a_{h^{2}_{e^{+}}} ) + a_{e^{\prime}} ( a_{h^{1}_{e^{-}}} + a_{h^{1}_{e^{+}}} )\nonumber\\
C_{\textrm{wheel, n=3}} =\, a_{e_{1}} a_{e_{2}} a_{e_{3}} &+ \sum_{i < j < k\vert \textrm{cyclic}} [ a_{e_{i}} a_{e_{j}} (a_{e_{k}^{-}} + a_{e_{k}^{+}})\, +\, a_{e_{i}} (a_{e_{j}^{-}} + a_{e_{j}^{+}} ) ( a_{e_{k}^{-}} + a_{e_{k}^{+}} ) ]
\end{align}
The labels on half edges are depicted in \eqref{fig: tadpole half edges}, \eqref{fig: bubble half edges} and \eqref{fig: triangle half edge}.
\begin{align} \label{fig: tadpole half edges}
\gamma_{\textrm{tadpole}} &= \quad \begin{tikzpicture}[rotate=90,scale=1.2,baseline={([yshift=-.5ex]current bounding box.center)}]
\draw (0,1.5) circle [radius=0.5];
\draw (0,0) -- (0,1);
\draw[-stealth] (-0.5,1.499) -- (-0.5,1.5);
\draw[-stealth] (0.5,1.5) -- (0.5,1.499);
\node at (-0.8,1.5) {$e^+$};
\node at (0.8,1.5) {$e^-$};
\draw[-stealth] (0,0.5) -- (0,0.51);
\node at (0.25,0.5) {$e$};
\end{tikzpicture}
\\
\label{fig: bubble half edges}
\gamma_{\textrm{wheel, n=2}}&= \quad \begin{tikzpicture}[scale=1.3,baseline={([yshift=-.5ex]current bounding box.center)}]
\draw (0,0) circle [radius=0.5];
\draw (0.5,0) -- (1.5,0);
\draw (-0.5,0) -- (-1.5,0);
\draw[-stealth] (0.04,-0.5) -- (0.05,-0.5);
\draw[-stealth] (0.04,0.5) -- (0.05,0.5);
\draw[-stealth] (0.99,0) -- (1,0);
\draw[-stealth] (-1.01,0) -- (-1,0);
\node at (1.75,0) {$e'$};
\node at (-0.6,0.6) {$h^{1}_{e^+}$};
\node at (-0.5,-0.6) {$h^{1}_{e^-}$};
\node at (0.6,-0.6) {$h^{2}_{e^-}$};
\node at (0.6,0.6) {$h^{2}_{e^+}$};
\node[gray] at (0,0.3) {$e^+$};
\node[gray] at (0,-0.3) {$e^-$};
\node at (-1.75,0) {$e$};
\end{tikzpicture}
\\
\label{fig: triangle half edge}
\gamma_{\textrm{wheel, n=3}} &= \quad \begin{tikzpicture}[scale=1.2,baseline={([yshift=-.5ex]current bounding box.center)}]
\draw (-1,0) -- (0,0) -- (1,1) -- (2,1);
\draw (0,0) -- (1,-1) -- (2,-1);
\draw (1,1) -- (1,-1);
\draw[-stealth] (-0.52,0) -- (-0.5,0);
\draw[-stealth] (1.5,1) -- (1.49,1);
\draw[-stealth] (1.5,-1) -- (1.49,-1);
\draw[-stealth] (1,-0.1) -- (1,0.1);
\draw[-stealth] (0.51,0.51) -- (0.5,0.5);
\draw[-stealth] (0.49,-0.49) -- (0.5,-0.5);
\node at (-1.3,0) {$e_1$};
\node at (2.3,-1) {$e_2$};
\node at (2.3,1) {$e_3$};
\node at (1.4,0.55) {$h_{e_3^+}$};
\node at (0.6,1.1) {$h_{e_3^-}$};
\node at (0,0.5) {$h_{e_1^+}$};
\node at (0,-0.5) {$h_{e_1^-}$};
\node at (1.4,-0.55) {$h_{e_2^-}$};
\node at (0.6,-1.1) {$h_{e_2^+}$};
\end{tikzpicture}
\end{align}
Before diving into explicit computation involving the contraction of Corolla MVF with $\Omega_{n}^{1-l}$, we derive two important structural results.
\setlist{nolistsep}
\begin{enumerate}[noitemsep]
\item Given two one loop graphs $\gamma_{1}, \gamma_{2}$ with tadpole vertices labelled by the chords $Y_{c},\, \tilde{Y}_{c}$ the action of the corresponding Corolla differentials are related by 
\begin{align}\label{cg1pcg2e0}
\hat{C}_{\gamma_{2}} \prod_{\textrm{prop}\in \gamma_{2}}\, \frac{1}{x_{\textrm{prop}}} 
+ \hat{C}_{\gamma_{1}} \prod_{\textrm{prop}\in \gamma_{1}}\, \frac{1}{x_{\textrm{prop}}} = 0 
\end{align}
\item  Action of Corolla differential on a labelled graph without a tadpole vertex is independent of the orientation of the loop momenta.  This implies that action of corolla differential on two vertices of $\hat{D}_{n}$ polytope which are ``mirror images" of each other and which do not contain chords of the type $(i,i)$ are equal. 
\end{enumerate}

The first result can be proved using explicit computation. For concreteness we consider $n = 4$ but the results can be trivially extended $\forall\, n$. 

To be specific, we consider the following four vertices of $\hat{D}_{4}$ polytope. Dual graphs  to these four vertices are forward limits of tree-level 6 point graphs with following ordering of external momenta. We denote these graphs as $\gamma_{i}^{13\, \textrm{t}}$, depicted in \eqref{fig: tadpoles on 4 point }. 
\begin{equation} \label{fig: tadpoles on 4 point }
\begin{aligned}
&\gamma_{1}^{13t} = \begin{tikzpicture}[scale=0.75,baseline={([yshift=-.5ex]current bounding box.center)}]
\draw (-1,1) -- (0,0) -- (-1,-1);
\draw (0,0) -- (2,0) -- (3,-1);
\draw (2,0) -- (3,1);
\draw (1,0) -- (1,0.8);
\draw (1,1.3) circle [radius=0.5];
\node at (-1.3,-1) {1};
\node at (-1.3,1) {2};
\node at (3.3,1) {3};
\node at (3.3,-1) {4};
\draw[-stealth] (-0.55,-0.55) -- (-0.56,-0.56);
\draw[-stealth] (-0.55,0.55) -- (-0.56,0.56);
\draw[-stealth] (2.55,-0.55) -- (2.56,-0.56);
\draw[-stealth] (2.55,0.55) -- (2.56,0.56);        
\draw[-stealth] (0.5,1.35) -- (0.5,1.36);        
\draw[-stealth] (1.5,1.35) -- (1.5,1.36);
\node at (0.0,1.3) {$-l$};
\node at (1.8,1.3) {$l$};
\end{tikzpicture} ~,\hspace{2cm}
\gamma_{2}^{13t} = \begin{tikzpicture}[scale=0.75,baseline={([yshift=-.5ex]current bounding box.center)}]
\draw (-1,1) -- (0,0) -- (-1,-1);
\draw (0,0) -- (2,0) -- (3,-1);
\draw (2,0) -- (3,1);
\draw (1,0) -- (1,0.8);
\draw (1,1.3) circle [radius=0.5];
\node at (-1.3,-1) {1};
\node at (-1.3,1) {2};
\node at (3.3,1) {3};
\node at (3.3,-1) {4};
\draw[-stealth] (-0.55,-0.55) -- (-0.56,-0.56);
\draw[-stealth] (-0.55,0.55) -- (-0.56,0.56);
\draw[-stealth] (2.55,-0.55) -- (2.56,-0.56);
\draw[-stealth] (2.55,0.55) -- (2.56,0.56);        
\draw[-stealth] (0.5,1.35) -- (0.5,1.36);        
\draw[-stealth] (1.5,1.35) -- (1.5,1.36);
\node at (0.2,1.3) {$l$};
\node at (1.9,1.3) {$-l$};
\end{tikzpicture} \\
&\gamma_{3}^{13t} = \begin{tikzpicture}[scale=0.75,baseline={([yshift=-.5ex]current bounding box.center)}]
\draw (-1,1) -- (0,0) -- (-1,-1);
\draw (0,0) -- (2,0) -- (3,-1);
\draw (2,0) -- (3,1);
\draw (1,0) -- (1,-0.8);
\draw (1,-1.3) circle [radius=0.5];
\node at (-1.3,-1) {1};
\node at (-1.3,1) {2};
\node at (3.3,1) {3};
\node at (3.3,-1) {4};
\draw[-stealth] (-0.55,-0.55) -- (-0.56,-0.56);
\draw[-stealth] (-0.55,0.55) -- (-0.56,0.56);
\draw[-stealth] (2.55,-0.55) -- (2.56,-0.56);
\draw[-stealth] (2.55,0.55) -- (2.56,0.56);        
\draw[-stealth] (0.5,-1.35) -- (0.5,-1.36);        
\draw[-stealth] (1.5,-1.35) -- (1.5,-1.36);
\node at (0,-1.3) {$-l$};
\node at (1.8,-1.3) {$l$};
\end{tikzpicture} ~,\hspace{2cm}
\gamma_{4}^{13t} = \begin{tikzpicture}[scale=0.75,baseline={([yshift=-.5ex]current bounding box.center)}]
\draw (-1,1) -- (0,0) -- (-1,-1);
\draw (0,0) -- (2,0) -- (3,-1);
\draw (2,0) -- (3,1);
\draw (1,0) -- (1,-0.8);
\draw (1,-1.3) circle [radius=0.5];
\node at (-1.3,-1) {1};
\node at (-1.3,1) {2};
\node at (3.3,1) {3};
\node at (3.3,-1) {4};
\draw[-stealth] (-0.55,-0.55) -- (-0.56,-0.56);
\draw[-stealth] (-0.55,0.55) -- (-0.56,0.56);
\draw[-stealth] (2.55,-0.55) -- (2.56,-0.56);
\draw[-stealth] (2.55,0.55) -- (2.56,0.56);        
\draw[-stealth] (0.5,-1.35) -- (0.5,-1.36);        
\draw[-stealth] (1.5,-1.35) -- (1.5,-1.36);
\node at (0.2,-1.3) {$l$};
\node at (1.9,-1.3) {$-l$};
\end{tikzpicture} 
\end{aligned}
\end{equation}
Then, we have 
 \begin{align}\label{cot1234}
 \gamma_{1}^{13\, \textrm{t}} \rightarrow\, (\, p_{1}, p_{2}, -l, l, p_{3}, p_{4}\, )\nonumber\\
 \gamma_{2}^{13\, \textrm{t}}\, \rightarrow\,  (\, p_{1}, p_{2}, l, -l, p_{3}, p_{4}\, )\nonumber\\
 \gamma_{3}^{13\, \textrm{t}} \, \rightarrow\, (\, p_{1}, p_{2}, p_{3}, p_{4}, l, -l\, )\nonumber\\
 \gamma_{4}^{13\, \textrm{t}}\, \rightarrow\, (\, p_{1}, p_{2}, p_{3}, p_{4}, -l, l\, )
 \end{align}
Using eqn.(\ref{cpfdg1l}), we can write down the action of Corolla on $\gamma_{1}$ and $ \gamma_{2}$ as follows. 
\begin{align}
\hat{C}_{\gamma_{i}} = \hat{C}_{v_{1}}\, \hat{C}_{v_{2}}\hat{C}_{w_{i}} \hat{C}_{v_{3}}
\end{align}
where 
\begin{align}
\hat{C}_{w_{1}}\, =\, \frac{1}{2}\, \eta^{\mu_{+}\mu_{-}} \left(\, - (l^{+})^{2}\, \frac{\partial}{\partial l_{+}^{\alpha}} + (l^{-})^{2}\, \frac{\partial}{\partial l_{-}^{\alpha}}\, \right)\\
\hat{C}_{w_{2}}\, =\, \frac{1}{2}\, \eta^{\mu_{+}\mu_{-}} \left(\, - (l^{-})^{2}\, \frac{\partial}{\partial l_{-}^{\alpha}} + (l^{+})^{2}\, \frac{\partial}{\partial l_{+}^{\alpha}}\, \right)
\end{align}
We thus see that
\begin{align}
\hat{C}_{w_{1}} + \hat{C}_{w_{2}} = 0
\end{align}
This result can be trivially extended to any pair of $n$-point graphs with two tadpole subgraphs which differ only in the orientation of the loop momenta proving assertion in eqn.(\ref{cg1pcg2e0}).  As a result, in the contraction of $\langle {\cal P}_{n}^{1-l}, \Omega_{n}^{1-l}\, \rangle$ we simply drop all those vertices of $\hat{D}_{n}$ polytope which correspond to graphs with tadpole vertices. 

An alert reader may worry that as the set of vertices of $\hat{D}_{n}$ always contain a pair of vertices with the same underlying graph but with loop momenta running clockwise and counter-clockwise respectively, perhaps the contraction of ${\cal P}_{n}^{1-l}$ with the d$\ln$ form trivially vanishes. We will now show that this is not the case and for all the vertices whose underlying graphs do not contain tadpole vertices, switching the direction of loop momenta keeps the Corolla action invariant. Once again, the simplest example is sufficient to display this general property. We thus consider $n=2$ and radiative correction to the propagator, as shown in \eqref{fig: tadpole with mom labels}.\footnote{The super-script indicates that these are bubble graphs.} 
Let us assign the following four vector labels to all the edges of the two graphs.
\begin{align}
\gamma^{\textrm{b}} = \ \ \ \begin{tikzpicture}[scale=0.8,baseline={([yshift=-.5ex]current bounding box.center)}]
\draw (0,0) circle [radius=0.5];
\draw (0.5,0) -- (1.5,0);
\draw (-0.5,0) -- (-1.5,0);
\draw[-stealth] (0.04,-0.5) -- (0.05,-0.5);
\draw[-stealth] (-0.04,0.5) -- (-0.05,0.5);
\draw[-stealth] (0.99,0) -- (1,0);
\draw[-stealth] (-1.01,0) -- (-1,0);
\node at (1.85,0) {$p_2$};
\node at (0,0.85) {$l_-$};
\node at (0,-0.85) {$l_+$};
\node at (-1.85,0) {$p_1$}; 
\end{tikzpicture} \ \ \ \rightarrow\, (p_{1}, l_{-}, l_{+}, p_{2}) \label{fig: tadpole with mom labels}
\end{align}
The corresponding two vertices of $\hat{D}_{2}$ can be obtained from $\gamma^{\textrm{b}}$ by  substituting 
\begin{align}\label{l-l+ooo}
v_{1}  &:= (\, p_{1} = p, p_{2} = -p, l_{-} = l, l_{+} = p_{1} + l\, )\quad \textrm{or}\nonumber\\
v_{2} &:= (\, p_{1} = p, p_{2} = - p, l_{-} = - (p_{1} + l), l_{+} = - l\, )
\end{align}
after the action of Corolla differentials on the two graphs.

The corolla polynomial for the underlying graph is given in eqn.(\ref{cpfdg1l}) using which the corresponding differentials can be immediately written down.  We will spare the reader from the relatively straightforward algebra which can be verified and simply write down the action of the Corolla differential on $\gamma^{\textrm{b}}$.
\begin{align}
\hat{C}^{\mu_{1}\mu_{2}}_{\gamma^{\textrm{b}}}\, =\, \frac{D^{2} + 8}{4}\, \sum_{\eta, \eta^{\prime} = \pm}\left( \eta\, \eta^{\prime}\,  l_{\eta}^{2}\, \frac{\partial}{\partial l_{\eta}^{\mu_{1}}} \ l_{\eta^{\prime}}^{2}\, \frac{\partial}{\partial l_{\eta^{\prime}}^{\mu_{2}}}\, \right) 
\end{align}
It can now be immediately verified that, 
\begin{align}
\prod_{i=1}^{2}\, p_{i}^{2}\ \hat{C}^{\mu_{1}\mu_{2}} \prod_{j=1}^{2}\, \frac{1}{p_{j}^{2}}\, \frac{1}{l_{-}^{2} l_{+}^{2}}\, \big|_{v_{1}} = \prod_{i=1}^{2} \,p_{i}^{2}\ \hat{C}^{\mu_{1}\mu_{2}} \prod_{j=1}^{2}\, \frac{1}{p_{j}^{2}}\, \frac{1}{l_{-}^{2} l_{+}^{2}}\, \bigg|_{v_{2}}
\end{align}
where $v_{1}, v_{2}$ are the two momenta assignments given in eqn.(\ref{l-l+ooo}). Hence we finally have the result of the contraction of Corolla induced MVF with the d log form defined by $\hat{D}_{2}$ polytope.
\begin{align}
\tcboxmath{\frac{1}{2}\, \langle\, {\cal P}_{2}^{1-l}, \Omega_{2}^{1-l}\, \rangle\, =\,  \prod_{i=1}^{2} p_{i}^{2} \hat{C}^{\mu_{1}\mu_{2}} \prod_{j=1}^{2}\, \frac{1}{p_{j}^{2}}\, \frac{1}{l_{-}^{2} l_{+}^{2}}\, \bigg|_{v_{1}}} 
\end{align}

\begin{figure}[!ht]
	\centering
	\begin{tikzpicture}[scale=0.5,baseline={([yshift=-.5ex]current bounding box.center)}]
	\draw[line width=1pt, purple] (-3.464,0) -- (0,-2) -- (3.464,0) -- (3.464,4) -- (0,6) -- (-3.464,4) -- (-3.464,0);
	\node[purple] at (0,2) {\large $\hat{D}_2$};
	\begin{scope}[scale=0.9,xshift=-8.3cm, yshift=-1cm]
	\fill[cyan!15] (0,1) .. controls (-1,0) .. (0,-1) .. controls (1,0) .. (0,1);
	\draw (0,1) .. controls (-1,0) .. (0,-1);
	\draw (0,1) .. controls (1,0) .. (0,-1);
	\fill[gray!70!black] (0,0) circle [radius=0.1];
	\draw (0,-1) .. controls (0.2,-0.3) .. (0.1,0);
	\draw (0,-1) .. controls (1.4,1) and (-1.4,1) .. (0,-1);
	\end{scope} 
	\begin{scope}[scale=0.9,xshift=-6.3cm, yshift=-1cm]
	\draw (0,1) .. controls (-1,0) .. (0,-1);
	\draw (0,1) .. controls (1,0) .. (0,-1);
	\fill[gray!70!black] (0,0) circle [radius=0.1];
	\draw[thick,purple] (0,-1) .. controls (0.2,-0.3) .. (0.1,0);
	\draw (0,-1) .. controls (1.4,1) and (-1.4,1) .. (0,-1);
	\end{scope} 
	\begin{scope}[scale=0.9,xshift=-4.5cm, yshift=-1cm]
	\fill[gray!70!black] (0,0) circle [radius=0.1];
	\draw (0,1) .. controls (-1,0) .. (0,-1);
	\draw (0,1) .. controls (1,0) .. (0,-1);
	\draw (0,-1) .. controls (0.2,-0.3) .. (0.1,0);
	\draw[thick,purple] (0,-1) .. controls (1.4,1) and (-1.4,1) .. (0,-1);
	\end{scope} 
	\begin{scope}[scale=0.9,xshift=-8.3cm, yshift=5.5cm]
	\fill[cyan!15] (0,1) .. controls (-1,0) .. (0,-1) .. controls (1,0) .. (0,1);
	\draw (0,1) .. controls (-1,0) .. (0,-1);
	\draw (0,1) .. controls (1,0) .. (0,-1);
	\fill[gray!70!black] (0,0) circle [radius=0.1];
	\draw (0,-1) .. controls (0.2,-0.3) .. (0.1,0);
	\draw (0,1) .. controls (-0.2,0.3) .. (-0.1,0);
	\end{scope}
	\begin{scope}[scale=0.9,xshift=-6.3cm, yshift=5.5cm]
	\draw (0,1) .. controls (-1,0) .. (0,-1);
	\draw (0,1) .. controls (1,0) .. (0,-1);
	\fill[gray!70!black] (0,0) circle [radius=0.1];
	\draw[thick,purple] (0,-1) .. controls (0.2,-0.3) .. (0.1,0);
	\draw (0,1) .. controls (-0.2,0.3) .. (-0.1,0);
	\end{scope}
	\begin{scope}[scale=0.9,xshift=-4.5cm, yshift=5.5cm]
	\draw (0,1) .. controls (-1,0) .. (0,-1);
	\draw (0,1) .. controls (1,0) .. (0,-1);
	\fill[gray!70!black] (0,0) circle [radius=0.1];
	\draw (0,-1) .. controls (0.2,-0.3) .. (0.1,0);
	\draw[thick,purple] (0,1) .. controls (-0.2,0.3) .. (-0.1,0);
	\end{scope}
	\begin{scope}[scale=0.9,xshift=-2cm, yshift=8cm]
	\fill[cyan!15] (0,1) .. controls (-1,0) .. (0,-1) .. controls (1,0) .. (0,1);
	\draw (0,1) .. controls (-1,0) .. (0,-1);
	\draw (0,1) .. controls (1,0) .. (0,-1);
	\fill[gray!70!black] (0,0) circle [radius=0.1];
	\draw (0,1) .. controls (-0.2,0.3) .. (-0.1,0);
	\draw (0,1) .. controls (-1.4,-1) and (1.4,-1) .. (0,1);
	\end{scope}
	\begin{scope}[scale=0.9,xshift=0cm, yshift=8cm]
	\draw (0,1) .. controls (-1,0) .. (0,-1);
	\draw (0,1) .. controls (1,0) .. (0,-1);
	\fill[gray!70!black] (0,0) circle [radius=0.1];
	\draw[thick,purple] (0,1) .. controls (-0.2,0.3) .. (-0.1,0);
	\draw (0,1) .. controls (-1.4,-1) and (1.4,-1) .. (0,1);
	\end{scope}
	\begin{scope}[scale=0.9,xshift=1.8cm, yshift=8cm]
	\draw (0,1) .. controls (-1,0) .. (0,-1);
	\draw (0,1) .. controls (1,0) .. (0,-1);
	\fill[gray!70!black] (0,0) circle [radius=0.1];
	\draw (0,1) .. controls (-0.2,0.3) .. (-0.1,0);
	\draw[thick,purple] (0,1) .. controls (-1.4,-1) and (1.4,-1) .. (0,1);
	\end{scope}
	\begin{scope}[scale=0.9,xshift=4.5cm, yshift=5.5cm]
	\fill[cyan!15] (0,1) .. controls (-1,0) .. (0,-1) .. controls (1,0) .. (0,1);
	\draw (0,1) .. controls (-1,0) .. (0,-1);
	\draw (0,1) .. controls (1,0) .. (0,-1);
	\fill[gray!70!black] (0,0) circle [radius=0.1];
	\draw (0,1) .. controls (-1.4,-1) and (1.4,-1) .. (0,1);
	\draw (0,1) .. controls (0.2,0.3) .. (0.1,0);
	\end{scope}
	\begin{scope}[scale=0.9,xshift=6.5cm, yshift=5.5cm]
	\draw (0,1) .. controls (-1,0) .. (0,-1);
	\draw (0,1) .. controls (1,0) .. (0,-1);
	\fill[gray!70!black] (0,0) circle [radius=0.1];
	\draw[thick,purple] (0,1) .. controls (-1.4,-1) and (1.4,-1) .. (0,1);
	\draw (0,1) .. controls (0.2,0.3) .. (0.1,0);
	\end{scope}
	\begin{scope}[scale=0.9,xshift=8.3cm, yshift=5.5cm]
	\draw (0,1) .. controls (-1,0) .. (0,-1);
	\draw (0,1) .. controls (1,0) .. (0,-1);
	\fill[gray!70!black] (0,0) circle [radius=0.1];
	\draw (0,1) .. controls (-1.4,-1) and (1.4,-1) .. (0,1);
	\draw[thick,purple] (0,1) .. controls (0.2,0.3) .. (0.1,0);
	\end{scope}
	\begin{scope}[scale=0.9,xshift=4.5cm, yshift=-1cm]
	\fill[cyan!15] (0,1) .. controls (-1,0) .. (0,-1) .. controls (1,0) .. (0,1);
	\draw (0,1) .. controls (-1,0) .. (0,-1);
	\draw (0,1) .. controls (1,0) .. (0,-1);
	\fill[gray!70!black] (0,0) circle [radius=0.1];
	\draw (0,1) .. controls (0.2,0.3) .. (0.1,0);
	\draw (0,-1) .. controls (-0.2,-0.3) .. (-0.1,0);
	\end{scope}
	\begin{scope}[scale=0.9,xshift=6.5cm, yshift=-1cm]
	\draw (0,1) .. controls (-1,0) .. (0,-1);
	\draw (0,1) .. controls (1,0) .. (0,-1);
	\fill[gray!70!black] (0,0) circle [radius=0.1];
	\draw[thick,purple] (0,1) .. controls (0.2,0.3) .. (0.1,0);
	\draw (0,-1) .. controls (-0.2,-0.3) .. (-0.1,0);
	\end{scope}
	\begin{scope}[scale=0.9,xshift=8.3cm, yshift=-1cm]
	\draw (0,1) .. controls (-1,0) .. (0,-1);
	\draw (0,1) .. controls (1,0) .. (0,-1);
	\fill[gray!70!black] (0,0) circle [radius=0.1];
	\draw (0,1) .. controls (0.2,0.3) .. (0.1,0);
	\draw[thick,purple] (0,-1) .. controls (-0.2,-0.3) .. (-0.1,0);
	\end{scope}
	\begin{scope}[scale=0.9,xshift=-2cm, yshift=-3.5cm]
	\fill[cyan!15] (0,1) .. controls (-1,0) .. (0,-1) .. controls (1,0) .. (0,1);
	\draw (0,1) .. controls (-1,0) .. (0,-1);
	\draw (0,1) .. controls (1,0) .. (0,-1);
	\fill[gray!70!black] (0,0) circle [radius=0.1];
	\draw (0,-1) .. controls (-0.2,-0.3) .. (-0.1,0);
	\draw (0,-1) .. controls (1.4,1) and (-1.4,1) .. (0,-1);
	\end{scope}
	\begin{scope}[scale=0.9,xshift=0cm, yshift=-3.5cm]
	\draw (0,1) .. controls (-1,0) .. (0,-1);
	\draw (0,1) .. controls (1,0) .. (0,-1);
	\fill[gray!70!black] (0,0) circle [radius=0.1];
	\draw[thick,purple] (0,-1) .. controls (-0.2,-0.3) .. (-0.1,0);
	\draw (0,-1) .. controls (1.4,1) and (-1.4,1) .. (0,-1);
	\end{scope}
	\begin{scope}[scale=0.9,xshift=1.8cm, yshift=-3.5cm]
	\draw (0,1) .. controls (-1,0) .. (0,-1);
	\draw (0,1) .. controls (1,0) .. (0,-1);
	\fill[gray!70!black] (0,0) circle [radius=0.1];
	\draw (0,-1) .. controls (-0.2,-0.3) .. (-0.1,0);
	\draw[thick,purple] (0,-1) .. controls (1.4,1) and (-1.4,1) .. (0,-1);
	\end{scope}
	\end{tikzpicture}    \caption{$\hat{D}_2$ polytope with each vertex corresponding to a set of marked pseudo-triangulations.}
	\label{fig: Dhat 2 with marked edges}
\end{figure}
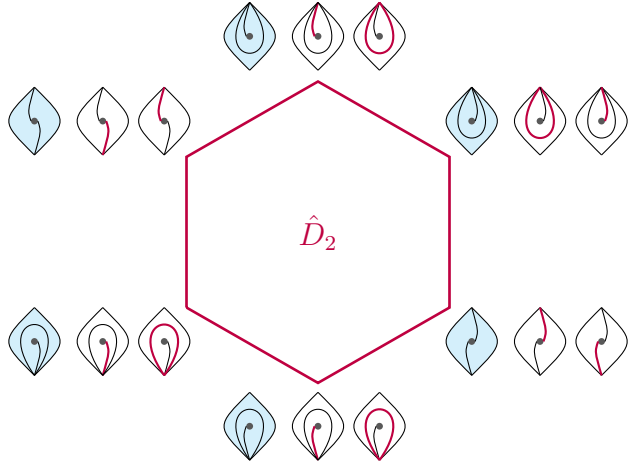

As the vertices in $V(\hat{D}_{n})$ that contain tadpole subgraphs do not contribute in the contracted scalar, we simply need to analyse the action of Corolla differential on rest of the vertices. In this case, the action can be cleanly captured by defining marked pseudo-triangulation which is exactly analogous to the concept of marked triangulation introduced in section (\ref{mtri}). As the idea is exactly analogous, we simply illustrate it diagrammatically. In the figure \ref{fig: Dhat 2 with marked edges}, we show how the action of the Corolla differential can be realised as transforming a pseudo-triangulation into a set of marked pseudo-triangulation where either vertices are marked or edges which are not adjacent to marked vertices are marked.\footnote{Of course the crucial difference in the explicit expression of marked triangles from marked pseudo-triangles is that the former corresponds to a three gluon vertex, whereas the latter corresponds to a combination of the three-gluon vertex and a two ghost one gluon vertex if any one of the boundaries of the pseudo-triangle is an arc that ends at the puncture.} 

\subsection{Explicit Example: planar one loop two points}
We will now illustrate the results obtained above with some explicit examples. We work out the simplest one-loop amplitude case, planar two-point, in detail. The $\hat{D}_2$ polytope, is a two-dimensional polygon as there are two internal propagators. There are six edges (faces) corresponding to six curves, and six vertices corresponding to six graphs. Figure \ref{fig:Dhat 2 polytope} depicts the g-vectors and triangulations for this case.

Accounting for the doubling of graphs, we restrict ourselves to the $\sum t_i \leq 0$ region. As a result, we have three graphs to account for:
\begin{align} 
    \begin{tikzpicture}[scale=1.3,baseline={([yshift=-.5ex]current bounding box.center)}]
        \draw (-1,0) -- (1,0);
        \draw (0,0) -- (0,1);
        \draw (0,1.5) circle [radius=0.5];
        \draw[-stealth] (-0.51,0) -- (-0.5,0);
        \draw[-stealth] (0.51,0) -- (0.5,0);
        \draw[-stealth] (0,0.5) -- (0,0.51);
        \draw[-stealth] (0.5,1.49) -- (0.5,1.51);
        \draw[-stealth] (-0.5,1.51) -- (-0.5,1.49);
        \node at (-0.9,-0.25) {$\xi_1\ (\mu_1)$};
        \node at (0.9,-0.25) {$\xi_2\ (\mu_2)$};
        \node at (0,0.6) {$(\nu)~ ~ ~ ~ p_\Sigma$};
        \node at (-1,1.5) {$l_1\ (\nu_1)$};
    \end{tikzpicture}
    \ + \
    \begin{tikzpicture}[scale=1.3,baseline={([yshift=-.5ex]current bounding box.center)}]
        \draw (0,0) circle [radius=0.5];
        \draw (0.5,0) -- (1.5,0);
        \draw (-0.5,0) -- (-1.5,0);
        \draw[-stealth] (0.04,-0.5) -- (0.05,-0.5);
        \draw[-stealth] (-0.04,0.5) -- (-0.05,0.5);
        \draw[-stealth] (0.99,0) -- (0.98,0);
        \draw[-stealth] (-0.99,0) -- (-0.98,0);
        \node at (1.5,0.25) {$\xi_2\ (\mu_2)$};
        \node at (0,-0.75) {$l_{2}\ (\nu_2)$};
        \node at (0,0.75) {$l_1\ (\nu_{1})$};
        \node at (-1.5,0.25) {$\xi_1\ (\mu_1)$};
    \end{tikzpicture}
    \ + \ 
    \begin{tikzpicture}[scale=1.3,baseline={([yshift=-.5ex]current bounding box.center)}]
        \draw (-1,0) -- (1,0);
        \draw (0,0) -- (0,-1);
        \draw (0,-1.5) circle [radius=0.5];
        \draw[-stealth] (-0.51,0) -- (-0.5,0);
        \draw[-stealth] (0.51,0) -- (0.5,0);
        \draw[-stealth] (0,-0.5) -- (0,-0.51);
        \draw[-stealth] (-0.5,-1.49) -- (-0.5,-1.51);
        \draw[-stealth] (0.5,-1.51) -- (0.5,-1.49);
        \node at (-0.9,0.25) {$\xi_1\ (\mu_1)$};
        \node at (0.9,0.25) {$\xi_2\ (\mu_2)$};
        \node at (0,-0.6) {$(\nu)~ ~ ~ ~ p_\Sigma$};
        \node at (1,-1.5) {$l_2\ (\nu_2)$};
    \end{tikzpicture} 
    \label{one loop tikz}
\end{align}

We can readily write the canonical form for this as follows:
\begin{align}
    \Omega = \dd X_{l_1} \wedge \dd X_{p_\Sigma} - \dd X_{l_1} \wedge \dd X_{l_2} + \dd X_{p_\Sigma} \wedge \dd X_{l_2} + \hdots \label{canonical form for one loop}
\end{align}
$\hdots$ refer to the rest of the three triangulations in $\sum t_i\geq 0$ region. 

Let us write the differential operator valued polyvector $\mathcal{P}_2^{1-l}$. The momentum labels are in accord with \eqref{one loop tikz}.
\begin{align}
    &\frac{1}{\xi_1^2\,\xi_2^2}\,\mathcal{P}_2^{1-l}  \nonumber\\
    &\hspace{-0.5cm}= \left[ \eta^{\nu\mu_2}\left(\frac{\xi_2^2}{2}\,\p_{\xi_2} + \frac{p_\Sigma^2}{2}\,\p_{p_{\Sigma}}\right)^{\mu_1} + \eta^{\mu_1\nu}\left(-\frac{p_\Sigma^2}{2}\,\p_{p_{\Sigma}} - \frac{\xi_1^2}{2}\,\p_{\xi_1} \right)^{\mu_2} + \eta^{\mu_2\mu_1}\left( \frac{\xi_1^2}{2}\,\p_{\xi_1} -  \frac{\xi_2^2}{2}\,\p_{\xi_2}\right)^{\nu}\right]\nonumber\\
    &\hspace{6cm} \times (4-2D)\left(\frac{l_1^2}{2}\,\p_{l_1}\right)_{\nu} \ \frac{1}{\xi_1^2\,\xi_2^2}\ \p_{X_{l_1}}\wedge \p_{X_{p_\Sigma}} \nonumber \\
    &\hspace{-0.5cm}-\bigg\{\left[\eta^{\nu_1\nu_2}\left( -\frac{l_2^2}{2}\,\p_{l_2}-\frac{l_1^2}{2}\,\p_{l_1} \right)^{\mu_1} + \eta^{\nu_2\mu_1}\left(-\frac{\xi_1^2}{2}\,\p_{\xi_1}+\frac{l_2^2}{2}\,\p_{l_2}\right)^{\nu_1} + \eta^{\mu_1\nu_1}\left( \frac{l_1^2}{2}\,\p_{l_1} + \frac{\xi_1^2}{2}\,\p_{\xi_1} \right)^{\nu_2}\right] \nonumber \\
    &\qquad\times \left[\eta^{\mu_2}{}_{\nu_2}\left( \frac{l_2^2}{2}\,\p_{l_2} + \frac{\xi_2^2}{2}\,\p_{\xi_2} \right)_{\nu_1} + \eta_{\nu_1\nu_2}\left( -\frac{l_1^2}{2}\,\p_{l_1} -\frac{l_2^2}{2}\,\p_{l_2} \right)^{\mu_2} + \eta_{\nu_1}{}^{\mu_2}\left(-\frac{\xi_2^2}{2}\,\p_{\xi_2} + \frac{l_1^2}{2}\,\p_{l_1}\right)_{\nu_2} \right]  \nonumber \\
    &  \qquad- \left(\frac{l_2^2}{2}\,\p_{l_2}\right)^{\mu_1}\left(\frac{l_1^2}{2}\,\p_{l_1}\right)^{\mu_2} - \left(\frac{l_2^2}{2}\,\p_{l_2}\right)^{\mu_2}\left(\frac{l_1^2}{2}\,\p_{l_1}\right)^{\mu_1} \bigg\} \frac{1}{\xi_1^2\,\xi_2^2}\ \p_{X_{l_1}}\wedge \p_{X_{l_2}}\nonumber \\
    &\hspace{-0.5cm}+\left[\eta^{\mu_2\nu}\left( -\frac{p_\Sigma^2}{2}\,\p_{p_\Sigma} - \frac{\xi_2^2}{2}\,\p_{\xi_2} \right)^{\mu_1} + \eta^{\nu\mu_1}\left(\frac{\xi_1^2}{2}\,\p_{\xi_1} + \frac{p_\Sigma^2}{2}\,\p_{p_\Sigma}\right)^{\mu_2} + \eta^{\mu_1\mu_2}\left(\frac{\xi_2^2}{2}\,\p_{\xi_2} - \frac{\xi_1^2}{2}\,\p_{\xi_1}\right)^{\nu}\right] \nonumber \\
    &\hspace{6cm}\times (4-2D)\left( \frac{l_2^2}{2}\,\p_{l_2} \right)_\nu\ \frac{1}{\xi_1^2\,\xi_2^2}\ \p_{X_{p_\Sigma}}\wedge \p_{X_{l_2}} \label{polyvector for one loop}
\end{align}
Note that the ghosts have been accounted for. We have omitted the three `mirror image' contributions from the $\sum t_i>0$ region.

Now, we can obtain the one-loop integrand by contracting the polyvector \eqref{polyvector for one loop} with the canonical form \eqref{canonical form for one loop} as follows:
\begin{align}
    & \frac{1}{2}\langle \mathcal{P}_2^{1-l}\,,\,\Omega_2^{1-l} \rangle \nonumber \\
    &= (4-2D)\big(l_2-l_1\big)_\nu\left[\eta^{\nu\mu_2}\big(-\xi_2-p_{\Sigma}\big)^{\mu_1} + \eta^{\mu_1\nu}\big(p_{\Sigma}+\xi_1\big)^{\mu_2} + \eta^{\mu_2\mu_1}\big(-\xi_1+\xi_2\big)^{\nu}\right] \frac{1}{l_1^2\,p_{\Sigma}^2} \nonumber \\
    &\quad +\bigg( \left[\eta^{\nu_1\nu_2}\big( l_2+l_1 \big)^{\mu_1} + \eta^{\nu_2\mu_1}\big(\xi_1-l_2\big)^{\nu_1} + \eta^{\mu_1\nu_1}\big(-l_1-\xi_1\big)^{\nu_2}\right] \nonumber \\
    &\hspace{3cm} \times \left[\eta^{\mu_2}{}_{\nu_2}\big( -l_2-\xi_2 \big)_{\nu_1} + \eta_{\nu_1\nu_2}\big(l_1+l_2\big)^{\mu_2} + \eta_{\nu_1}{}^{\mu_2}\big(\xi_2-l_1\big)_{\nu_2}\right]\bigg) \frac{1}{l_1^2\,l_2^2} \nonumber \\
    &\quad + (1-D)\frac{1}{l_1^2}\,\eta^{\mu_1\mu_2} +  (1-D)\frac{1}{l_2^2}\,\eta^{\mu_1\mu_2} \nonumber \\
    &\quad -\left(l_2{}^{\mu_1}\,l_1{}^{\mu_2} + l_2{}^{\mu_2}\,l_1{}^{\mu_1}\right)\,\frac{1}{l_1^2\,l_2^2} \label{one loop integrand final}
\end{align}
At this point, we can impose momentum conservation, and this gives us the integrand of the one-loop two-point amplitude. Since we have not imposed momentum conservation and on-shell conditions yet in \eqref{one loop integrand final}, these graphs can be embedded in any bigger graphs.

In \eqref{one loop integrand final}, the first term contains the tadpole contributions. Note that there are no contact vertices appearing due to tadpole graphs. The second term is the trivalent gluon bubble graph, corresponding to none of the lines marked. The next two terms in the penultimate line are the diagrams with a contact vertex, corresponding to $l_1$ and $l_2$ marked respectively. The last term is the ghost contribution. 

\begin{figure}[!h]
    \centering
    \includegraphics[scale=0.3]{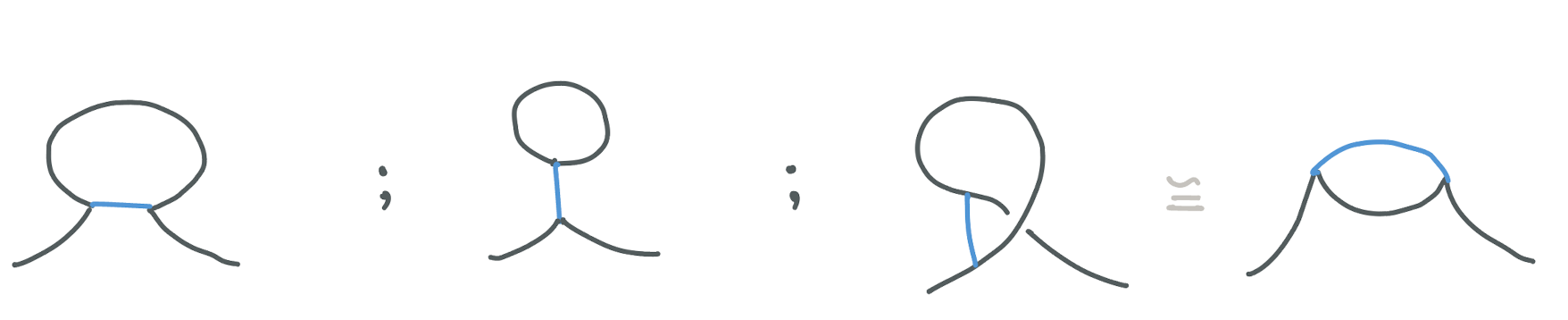}
    \caption{Underlying structure of the contact vertex for a 'wart' diagram under usual Feynman rules}
    \label{fig:wart diagram}
\end{figure}

Note that the doubling of terms, $l_1$ and $l_2$ being marked, are consistent with the usual Feynman diagrams. Figure \ref{fig:wart diagram} depicts the underlying structure of the one-loop, two-point contact interaction graph. The middle diagram, with a line in the tadpole diagram marked, is zero. The other two terms are precisely $l_i$ being marked.
\section{Curve integral formula for gluons} \label{sec: curve integral}
The Corolla differential defined in the kinematic space is directly inspired by the definition given by KSVS of the graph differential $\hat{C}_{\gamma}^{\textrm{sch}}$ in the parametric space spanned by the Schwinger parameters. We review this seminal construction in section \ref{cgdksvs}.  $\hat{C}_{\gamma}^{\textrm{sch}}$ maps the top form in the parametric space which integrates to contribution of $\gamma$ to $\phi^{3}$ amplitude to a top form which integrates to contribution of $\gamma$ to Yang-Mills (and in general gauge theory) amplitude. 

On the other hand, in the recent works \cite{nima2309,nima2311}, the authors discovered and proved the so-called \emph{curve integral formula} for $\phi^{3}$ amplitudes, (See section \ref{ciffs} for a brief review). Clearly, it is natural to consider a ``fusion" of these two ideas and obtain a curve integral formula for gluon amplitudes. Results in earlier section should convince the reader that this should indeed be possible and in this section we give explicit curve integral formula for tree-level and (planar) 1-loop $n$ point gluon amplitude. As both the curve integral formula and the Corolla graph differential is well defined at arbitrary order in perturbation theory, generalisation of our results to higher loops should be immediate and will be pursued in the future. 


\subsection{Review of Corolla graph differential}\label{cgdksvs}
In \cite{kreimer1208}, Corolla polynomials on a fixed trivalent graph $\gamma$ were represented as differential operators-valued meromorphic maps in the parametric space  co-ordinatized by the Schwinger parameters.\footnote{Ordinarily in the amplitude computations, one assigns a Schwinger parameter $t_{e}$ to any internal edge $e$ of a $k$-valent graph both of whose end points are $k$-valent vertices.  However, following \cite{kreimer1208}, the parametric space we consider is ${\bf R}^{\vert E_{\textrm{int}} \vert + \vert E_{\textrm{out}} \vert}$ with Schwinger parameters also labelling external edges.}  A remarkable all loop parametric representation of Yang-Mills amplitude was then a result of the action of such a Corolla differential on $\phi^{3}$ Schwinger integrand, \cite{Golz:2017aoa}.  We now review this result. 

Let ${\cal U}_{\gamma},\, {\cal F}_{\gamma}$ be the First and Second Symanzik graph polynomials obtained from $\gamma$. We denote the contribution of a $\gamma$ to the (parametric representation) of $\phi^{3}$ amplitude as $\textrm{Val}_{\phi^{3}}(\gamma)$. 
\begin{align}\label{soag}
\textrm{Val}_{\phi^{3}}(\gamma)\, :=\, \int\, \prod_{i=1}^{E} dt_{i}\, \left(\frac{\pi^{L}}{{\cal U}_{\gamma}}\right)^{\frac{D}{2}}\, e^{-\, \frac{{\cal F}_{\gamma}}{{\cal U}_{\gamma}}}
\end{align}
where ${\cal U}_{\gamma},\, {\cal F}_{\gamma}$ are the first and second Symanzik polynomials which are generated by the set ${\cal T}_{1}$ of all the 1-spanning trees and  set ${\cal T}_{2}$ of all the 2 spanning trees of $\gamma$ respectively.\footnote{If $\gamma$ has no cycles, then ${\cal U}_{\gamma} := 1$ and ${\cal F}_{\gamma} :=\, \sum_{I\, \in\, \textrm{prop}} t_{I} P_{I}^{2}$,}
\begin{align}\label{ufgamma}
{\cal U}_{\gamma}\, &=\, \sum_{T_{1}\, \in\, {\cal T}_{1}}\, \prod_{e\, \notin\, T_{1}}\, t_{e}\nonumber\\
{\cal F}_{\gamma}\, &=\, \sum_{T_{2} = T_{2}^{\prime} \Pi T_{2}^{\prime\prime}\, \in\, {\cal T}_{2}}\, \prod_{e\notin\, T_{2}} t_{e} \bigg( \sum_{e^{\prime} \in T_{2}^{\prime}} q_{e^{\prime}}\bigg)^{2}
\end{align}
where $\sum_{e^{\prime} \in T_{2}} q_{e^{\prime}}^{\mu}$ is the physical momenta flowing into $\gamma$ from $T_{2}^{\prime}$. For more details on the parametric representation using Symanzik polynomials, we refer the reader to \cite{Bogner_Weinzierl, weinzierl, smirnovbook1, smirnovbook2}. 

The Corolla differential introduced by KSVS in \cite{kreimer1208} requires a minor re-writing of $\textrm{Val}_{\phi^{3}}(\gamma)$. We label all the edges of the graph with independent four-vector labels $\xi_{e}\, \forall e \in E_{\textrm{in}\cup\textrm{out}}$.  $\textrm{Val}_{\phi^{3}}$ can then be re-written as, 
\begin{align}\label{soagwem}
\textrm{Val}_{\phi^{3}}(\gamma)&=\nonumber\\
&\hspace*{-0.5in}\int_{0}^{\infty}\,  \prod_{i=1}^{E} dt_{i}\, \prod_{j=1}^{\vert E_{\textrm{out}}\vert} \xi_{j}^{2}\, \int_{0}^{\infty}\, d s_{j} \left(\frac{\pi^{L}}{{\cal U}_{\gamma}}\right)^{\frac{D}{2}}\, e^{-\, \frac{{\cal F}_{\gamma}(\{ \xi_{e^{\prime}}\})_{e^{\prime} \in E_{\textrm{in}}}}{{\cal U}_{\gamma}} - \sum_{m=1}^{\vert E_{\textrm{out}}\vert} s_{m} \xi_{m}^{2}}\vert_{\xi_{e} = p_{e}\, \forall e \in E_{\textrm{out}}\ ; \  \xi_{e^{\prime}} = q_{e^{\prime}} \forall e^{\prime} \in E_{\textrm{in}}}
\end{align}
where $q_{e^{\prime}}$ was introduced in eqn.(\ref{ufgamma}). 

Action of Corolla graph differential introduced in  \cite{kreimer1208} on $e^{-\frac{{\cal F}}{{\cal U}}}$ leads to the Yang-Mills integrand in the parametric space $\textrm{Val}_{\textrm{YM}}$.  The discovery of the Corolla polynomial as a generator of  graph homology (which  leads to gauge invariant perturbative expansion of Yang-Mills amplitude) is a rather beautiful chapter in  parametric representation of gauge theory Feynman graphs. For a nice exposition of these ideas we refer the reader to \cite{prinz} and for application of the idea to QED see \cite{Golz:2017aoa}).

We will denote the Corolla graph-differential in the parametric space  $\hat{C}_{\gamma}^{\textrm{sch}}$ to distinguish it from $\hat{C}_{\gamma}$ which is defined in the kinematic space. Parametric space is the positive quadrant in  ${\bf R}^{\vert E_{\textrm{in}}\vert + \vert E_{\textrm{out}}\vert }$ where $E_{\textrm{in(out)}}$ is the set of internal (external) edges of $\gamma$ all of which begin and end in a trivalent vertex. This space is co-ordinatized by the Schwinger parameters $\{t_{1}, \dots,\, t_{\vert E_{\textrm{in}}\vert}, s_{1}, \dots, s_{\vert E_{\vert\textrm{out}\vert}}\}$. 

As we have already defined a Corolla differential $\hat{C}_{\gamma}$ in the kinematic space, $\hat{C}_{\gamma}^{\textrm{sch}}$ defined by KSVS can be understood via the following replacement in $\hat{C}_{\gamma}$.  Given the Corolla differential $\hat{C}_{\gamma}$ on the kinematic space.
\begin{align}\label{cgdiss}
\hat{C}_{\gamma}(\vec{t}, \vec{s})\, :=\, \hat{C}_{\gamma}\vert_{ \xi_{e^{\prime}}^{2}\, \rightarrow\, \frac{1}{t_{e^{\prime}}}\ , \  p_{e}^{2}\, \rightarrow\, \frac{1}{s_{e}}}
\end{align}
The parametric representation of the Yang-Mills amplitude is then given by the following master formula. Given a $\gamma$, let 
Let ${\cal E}\, \subset\, E_{\textrm{int}}$ be a subset of internal edges no two of which share a common vertex. Let ${\cal E}^{c}$ be the compliment of ${\cal E}$ in $E_{\textrm{int}}$. For an $n$-point tri-valent graph with no cycles, cardinality of any such ${\cal E}$ lies between
\begin{align}
0\, \leq\, \vert {\cal E}\vert\, \leq\, \lceil{\frac{n}{2}\rceil}
\end{align}
Let $f : {\bf R}^{\vert E_{\textrm{tot}}\vert}\, \rightarrow\, {\bf R}$ be a function of the parametric space over internal as well as external edges which has (1) simple poles at most on $t_{e} \vert e\, \in\, {\cal E}_{k}$ for some subset ${\cal E}_{k}$ with cardinality $k$ and, (2) is  regular on $E_{\textrm{tot}} - {\cal E}_{k}$.\footnote{That is, $f_{{\cal E}_{k}}$ is regular in the limit $t_{e^{\prime}}\, \rightarrow\, 0\, \forall e^{\prime} \in {\cal E}_{k}^{c}$.} 

We can now consider
\begin{align}\label{rr1}
\textrm{Res}(f)&=\, \sum_{k \geq\, 0}\, \sum_{{\cal E}_{k} \subset E_{\textrm{int}}}\, \prod_{e \in {\cal E}_{k}}\, \oint\, d t_{e} f \nonumber\\
&=\, \sum_{k \geq\, 0} \sum_{{\cal E}_{k} \subset E_{\textrm{int}}} f_{{\cal E}_{k}}\\
\label{rr2} \text{where } \quad &f_{{\cal E}_{k}} : {\bf R}^{\vert {\cal E}_{k}^{c}\vert}\, \rightarrow\, {\bf R}
\end{align}
Hence one can now evaluate the natural pairing
\begin{align}\label{rr3}
\int_{{\bf R}^{\vert {\cal E}_{k}^{c}\vert}} \wedge_{e^{\prime}} d t_{e^{\prime}}  \textrm{Res}(f_{{\cal E}_{k}})  
\end{align}
Given any function $f : {\bf R}^{\vert E_{\textrm{tot}} \vert}\, \rightarrow\, {\cal R}$ which has at most simple poles on subsets of edges no two of which share a vertex, one can define
\begin{align}\label{rrmain}
\textrm{Reg-Res}\, f := \sum_{k \geq\, 0}\, \int_{{\bf R}^{\vert {\cal E}_{k}^{c}\vert}} \bigwedge_{e^{\prime}} \dd t_{e^{\prime}}  \textrm{Res}(f_{{\cal E}_{k}}) 
\end{align}
Thus, we have
\begin{align}\label{mastergamma}
&{\cal A}^{\textrm{YM}}_{n}(p_{1}, \dots, p_{n})\nonumber\\
& \qquad =\sum_{\gamma}\, \textrm{Color}_{\gamma}\, \frac{1}{{\cal U}_{\gamma}^{\frac{D}{2}}}\, \textrm{Res-Reg}(\, \hat{C}_{\gamma}^{\textrm{sch}} e^{-\frac{{\cal F}(\xi_{e_{1}}, \dots, \xi_{e_{\vert E\vert_{\textrm{int}}}})}{{\cal U}} - \sum_{i=1}^{n_{\textrm{int}}} s_{i} \xi_{i}^{2}})\, \bigg|_{\xi_{i} = p_{i},\, \xi_{e} = K_{e}\, \forall\, e \in E_{\textrm{int}}}
\end{align}
where $\textrm{Color}_{\gamma}$ is the color factor.  This striking result holds at all loops and in fact has been used to obtained parametric representation of a wide class of gauge theories including gauge theories coupled to matter from parametric integrals of $\phi^{3}$ amplitudes. We refer the reader to \cite{prinz, Golz:2017aoa} for application of this formula to obtain a wider class of gauge theory amplitudes from $\phi^{3}$ amplitude. 

We now note that although the Residue in the (Reg-Res) formula computes the residue of the meromorphic function $f_{\gamma}(t_{e_{1}}\, \dots,\, t_{e_{\vert E_{\textrm{int}} \vert}})$ on simple poles over a subset of \emph{mutually disjoint} (internal) edges, we could equally well consider set of subets of internal edges (as opposed to demanding that the edges be mutually disjoint.)  This is because the Corolla action results in a integrand in the parametric space which never has poles on two adjacent edges. This is simply because given a sequence of set of edges $e, e^{\prime}, e^{\prime\prime}$ such that either\\
(1) all three of them meet in a vertex or, 
\begin{align}
e \cap e^{\prime} \neq\, 0\, \textrm{and}\, e^{\prime} \cap e^{\prime\prime} \neq\, 0 \nonumber
\end{align}
Then action of Corolla differential  on a function that depends on $t_{e}, t_{e^{\prime}}, t_{e^{\prime\prime}}$, will always result in a meromorphic function which in the first case have no double poles involving two of $(t_{e}, t_{e^{\prime}}, t_{e^{\prime\prime}})$ and in the second case can only have double pole in $(t_{e}, t_{e^{\prime\prime}})$ but not in $(t_{e}, t_{e^{\prime}})$ or $(t_{e^{\prime}}, t_{e^{\prime\prime}})$. This proves the above assertion. 

The formula derived by KSVS ``spins up" $\textrm{Val}_{\phi^{3}}(\gamma)$ to obtain $\textrm{Val}_{YM}(\gamma)$ and of course one can sum over all the 3-valent graphs to obtain the full amplitude at a given loop order. However, in  \cite{nima2309,nima2311}, the authors proved that the generic (single trace as well as multi-trace color factors) $n$ point $L$ loop colored $\phi^{3}$ amplitude  in the parametric space  can also be expressed in terms of the so-called first and second \emph{surface} Symanzik polynomials. The topology of the underlying surface is fixed by the number of boundary points, genus and number of marked points, the latter corresponding to the degree of non-planarity in the dual graph. 

In light of the fact that individual Feynman diagrams can be evaluated using graph Symanzik polynomials, this is a striking result. In a nut-shell, any $n$-point, $L$ loop, multi-trace amplitude is an integral of a top form in the parametric space  where the form is completely fixed by the ${\bf g}$-vector fan, (see section \ref{gvfacr} for a review of $g$ vector fans for an $n$ point graph with $L\, \leq\ 1$ and no marked points) and the coordinates of the Kinematic space. Hence, the ${\bf g}$-vector fan, which gave a unique realization of the positive geometries also defines a top form in the $R^{\vert E \vert}$ that leads to parametric representation of $\phi^{3}$ amplitude.  This formula was called the curve integral formula in \cite{nima2309}.  As KSVS's original discovery of the Corolla differential was in obtaining the Schwinger parametrization of a diagram for gluon amplitude from a $\phi^{3}$ diagram,  it is a rather natural question to ask if we can ``fuse" these two seminal ideas together and obtain curve integral formula for gluon amplitude. This is to be expected from results in the previous section in the kinematic space, but the details of the resulting ``curve integral formula for gluon amplitude" perhaps also have an intrinsic value. In the rest of the paper, we derive such a formula  for tree-level and one loop amplitudes and  provide some explicit examples.  As we observe, the Corolla differential which appears inside the curve integrand depends non-trivially on the ${\bf g}$ vector fan. 

We will analyze the tree level amplitude in section \ref{sec: tree level corolla for ym} and work out some examples in the following section. In section \ref{sec: curve integral at loops}, we present the curve integral formula at planar one loop.  We present some rudimentary examples in the following section.
\subsection{Review of the curve integral formula for scalars}\label{ciffs}
In this section, we briefly review the seminal results derived in \cite{nima2309, nima2311} for  $\phi^{3}$ amplitudes.  Although the results derived in these papers are completely general in the sense that they generate a full $n$ point amplitude at all orders in perturbation theory including non-planar diagrams, we will restrict ourselves to the parametric representation for  tree-level and planar one loop $\phi^{3}$ amplitudes which will generate corresponding Yang-Mills amplitude using a formula inspired from eqn.(\ref{mastergamma}).

\subsubsection{The central idea}
The headlight functions are (piece-wise) linear forms on the vector space spanned by ${\bf g}$ vectors and are labeled by the curves. They are fixed through the following defining equation.
\begin{align}
\alpha_{c'}(g_c) = \delta_{c c'}~. 
\end{align}
In \cite{nima2309, nima2311}, the authors present a beautiful algorithm to compute the headlight functions for arbitrary processes. 

The definition of the headlight function dictates that $\alpha_c$ `turns on' only in those cones which share positive span of  $\vec{g}_c$ as a common boundary. Tree-level color ordered massless $\phi^{3}$ amplitude can now be written in a compact form as, 
\begin{align}
{\cal A}_{n}^{\phi^{3}} = \int_{\mathbb{R}^{\vert E_{\text{int}}\vert}} \dd^{\vert E_{\text{int}}\vert}\vec{t}\  e^{-\sum_c\alpha_cX_c} \label{general tree curve integral}
\end{align}
where for $c = [i, j]$, $X_{c} = (p_{i} + \dots + p_{j-1})^{2}$ and $\vert E_{\textrm{int}}\vert  = n-3$. 

Equipped with such a remarkably simple formula, it is easy to see how this reproduces the tree-level amplitude. This is essentially due to the fact that the fan spanned by set of all ${\bf g}$ vectors is complete and essential.\footnote{A fan is complete if the $g$ vectors span ${\bf R}^{n-3}$ and is essential if the intersection of all the cones is the origin, \cite{pppp1906}.} As a result, the above integral over $n-3$ dimensional parametric space splits into sum over top dimensional  cones each of which is bounded by  (positive span) of $\{{\bf g}_{c}\}$ over the curves which generate a complete triangulation of the $n$-gon. 
\begin{align}
{\cal A}_{n}^{\tr \phi^{3}}\, = \sum_{T}\, \int \dd^{n-3}\vec{t}\, e^{-\sum_{c\, \in\, T}\, \alpha_{c} X_{c}}
 \label{curve integral in a cone}
 \end{align}
In a top-dimensional cone (indexed by a complete triangulation), the volume form can written in terms of the headlight functions as follows:
\begin{align}
\dd^{n-3}\vec{t}\,\big|_{T} &= \bigwedge_{c\in T}\dd \alpha_c  \label{t measure in terms of alpha}\\
\dd^{\vert E_{\text{int}}\vert}\vec{t} &= \sum_T\bigwedge_{c\in T}\dd \alpha_c
\end{align}
As an example, for $n=5$ and with a reference triangulation $T_{0} = \{13,14\}$ the corresponding $g$-vector fan spans ${\bf R}^{2}$.
\begin{align}\label{n=5gfr2}
g_{13} = (1,0), \ \ \ g_{14} = (0,1),\ \ \ g_{25} = (-1,0),\ \ \ g_{35} = (0,-1),\ \ \ g_{24} = (-1,1)
\end{align}
and the headlight functions are shown in \eqref{fig:headlight functions for tree level 5 points}. 
Using \eqref{t measure in terms of alpha} in \eqref{curve integral in a cone}, we can integrate the curve integral formula in each cone:
\begin{align}
\int_{T} \dd^{n-3}\vec{t}\  e^{-\sum_c\alpha_c X_c} =     \int\, \dd^{n-3}\vec{t} \ e^{-\sum_{c^{\prime} \in T}\alpha_{c^{\prime}}X_{c^{\prime}}}\, = \prod_{c\in T} \frac{1}{X_c}
\end{align}
Thus, the integral over the entire fan gives us a sum over all cones, i.e., sum over all trivalent graphs, and we obtain the tree-level trivalent scalar amplitude. 

The four-point tree-level example is quite instructive, as the $g$-vectors $g_{13}, g_{24}$ span ${\bf R}$ and the headlight functions are precisely the heavyside functions:
\begin{align}
g_{13} = 1, g_{24} = -1
\end{align}
and 
\begin{align}
\alpha_{13} = t\,\theta(t) = \mxx{0,t} ~ ~, &\hspace{2cm} \alpha_{24} = -t\,\theta(-t) = \mxx{0,-t}\\
\int_{-\infty}^{\infty} \dd^{n-3}\vec{t} \ e^{-\sum_c \alpha_{c}\,X_{c} } &= \int_{-\infty}^{\infty} \dd t\ e^{- \alpha_{13}\,X_{13}- \alpha_{24}\,X_{24}} \nonumber\\
&=\int_0^{\infty} \dd t\ e^{-t\,\theta(t)\,X_{13}} + \int_{-\infty}^0 \dd t\ e^{t\,\theta(-t)\,X_{24}} = \frac{1}{X_{13}} + \frac{1}{X_{24}} ~. 
\end{align}
We now re-write the curve integral formula in the parametric space ${\bf R}^{\vert E_{\textrm{int}}\vert + \vert E_{\textrm{out}}\vert}$ as follows. Let $s_{i}\vert i\, \in\, \{1,\, \dots,\, \vert E_{\textrm{ext}}\vert\}$ be the set of additional parameters.  Let
\begin{align}\label{extheadlight}
\alpha_{c} = \textrm{max}(0, s_{c})\, \forall\, c\, \in\, E_{\textrm{ext}}
\end{align}
Let  $x_{c}$ be independent real variables $\forall\, c\, \in\, E$.
The curve integral formula in eqn.(\ref{general tree curve integral}) can be re-written  as follows. 
\begin{align}\label{cifsm}
{\cal A}_{n}^{\tr \phi^{3}} (p_{1}, \dots, p_{n})\, =\, \prod_{c^{\prime} \in E_{\textrm{ext}}}\, x_{c^{\prime}} \int_{{\bf R}^{n-3 + n}}\, d^{n-3}\vec{t}\, d^{n}s\, e^{-\sum_{c} \alpha_{c} (\vec{t}, \vec{s}) x_{c}}\ \bigg|_{x_{c} = X_{c}}
\end{align}
\begin{align}
X_{c} = 0\, \forall\, c\, \in\, E_{\textrm{out}} 
\end{align}
\subsubsection{Planar one-loop integrated curve integrated formula}
We now review the curve integral formula for planar $\tr \phi^{3}$ amplitude at one loop, \cite{nima2309, nima2311}.  The parametric representation of $\textrm{Val}_{\phi^{3}}(\gamma)$ is in  terms of Symanzik polynomials which is obtained by writing the loop integrand in Schwinger representation and integrating over loop momenta. One of the beautiful corollaries of the curve integral formula is the discovery that the ``graph theoretic definition of Symanzik graph-polynomials in terms of spanning trees in fact extends to curve integral formulae to all orders in perturbation theory. Specifically, at one loop,  integrating over the loop momenta leads to a curve integral formula for one loop amplitude in terms of so-called surface Symanzik polynomials. we briefly review this result which will hopefully make the reader more acquainted with the formula. The structural ingredients involved in this derivation also help us in deriving curve integral formula for planar, one loop gluon amplitudes.

Let us recall that in the case of $n$ external particles, there are $n(n+1)$ number of curves which generate pseudo-triangulation of an $n$ gon with a hole in the center. These are, $C_{ij}$, $S_i$ and $S_i'$. (See figures \ref{fig:spiral curves} and \ref{fig:nonspiral curves}.)   The headlight functions corresponding to these curves were computed in \cite{nima2309, nima2311} and can be written as, 
\begin{align}
C_{ik} &= i\,L\,t_{i}\,R\,t_{i+1} \,R\hdots R\,t_{k-1}L\,k ~,\nonumber\\
&\hspace{4cm}\alpha_{C_{ik}} = f_{i,(k-2)} + f_{(i+1),(k-1)} - f_{(i+1),(k-2)} - f_{i,(k-1)} ~.\\
S_i' &= iR\,\big(t_{i-1}L\,t_{i-2} L\,t_{i-3}L\hdots L \,t_{i-n}\,L\big)^{\infty} ~, \nonumber\\
&\hspace{4cm}\alpha_{S_{i+1}'} = -t_i+f_{i,(i+n-2)} - f_{(i+1),(i+n-1)} ~. \label{headlight functions for spiral curves} \\
f_{i,k} &= \textrm{max}(\big(t_{i}+t_{i+1}+\hdots+t_{k}\big)\ , \ \big(t_{i+1}+t_{i+2}+\hdots+t_{k}\big) \ , \ \hdots \ ,\ \big(t_{k-1}+t_k\big)\ , \ t_k\ , \ 0)
\end{align}
We can hence  write the one-loop planar $n$-point scalar amplitude as follows:
\begin{align}
{\cal A}_{n\, L=1}^{\tr \phi^3} = \int_{\sum t_i<0} \dd^n\vec{t} \int \dd^D\ell\ \exp\left(-\hspace{-0.4cm}\sum_{\substack{i\in [1,n] \\ j\in [i+2,i+n]}}\hspace{-0.4cm}\alpha_{C_{ij}}X_{ij} - \sum_{i=1}^n\alpha_{S_i'}X_{S_i'} \right) \label{curve integral for one loop scalars}
\end{align}
We have labelled the loop momentum as $\ell$. Note that we have restricted the integral to the half of the parametric space bounded by $\sum t_i =0$, to account for the doubling of the graphs. \footnote{The co-dimension one hyper-plane is also called $\triangle$-plane in \cite{nima2309, nima2311}.} The headlight functions for the spirals $S_i$ vanish in the particular Schwinger region $\sum t_i<0$ as can be verified from explicit computation. For the benefit of the reader, we recall that (figures \ref{fig:nonspiral curves},\ref{fig:spiral curves}).
\begin{align}
X_{ij} &= (p_i+\hdots +p_{j-1})^2 \equiv (z_j-z_i)^2 ~,\\
X_{S_{i}'} &= (-l_{i-1})^2 \equiv (\ell + z_i)^2 ~,
\end{align}
where we have defined the $z_i$, the dual coordinates to momenta,
\begin{align}
p_i = z_{i+1} - z_{i} ~, \qquad z_{n+1} = z_1.
\end{align}
The total momentum conservation of $p_i$ becomes the translation symmetry of $z_i$. We can fix one of the $z_i$ for concreteness as follows:
\begin{align}
z_1 &= 0 ~, \quad z_2 = p_1 ~, \quad z_3 = p_1 + p_2 ~, \quad \hdots \quad ~, \quad z_k = \sum_{i=1}^{k-1}p_i \\ 
l_n &= \ell ~, \qquad l_1 = \ell + p_1 ~, \quad\hdots \quad ~, \quad l_k = \ell + \sum_{i=1}^k p_i = \ell + z_{k+1}
\end{align}
Looking back at \eqref{curve integral for one loop scalars}, we can carry out the loop integral to obtain:
\begin{align}
{\cal A}_{n, L=1}^{\tr \phi^{3}}&= \int_{\sum t_i<0}\dd^n\vec{t} \exp\left(-\sum_{ij}\alpha_{C_{ij}}\,z_{ij}^2\right)\int \dd^D\ell\ \exp\left( - \sum_{i=1}^n\alpha_{S_i'}(\ell+z_i)^2 \right) \nonumber\\
&= \int_{\sum t_i\leq0}\dd^n\vec{t} \ \left(\frac{-\pi}{\sum_{i}t_i}\right)^{D/2}\exp\left(-\sum_{ij}\alpha_{C_{ij}}z_{ij}^2 - \frac{\left(\sum_{i}\alpha_{S_i'}z_i\right)^2}{\sum_i t_i} - \sum_{i}\alpha_{S_i'}z_i^2\right) \label{integrated one loop planar scalar} ~,
\end{align}
This is the explicit form for \eqref{surface symanzik as one loop}.  Note that we made use of the following fact while simplifying the integral:
\begin{align}
\sum_i\alpha_{S_i'} = -\sum_i t_i ~.
\end{align}
${\cal A}_{n}^{\phi^{3}\, 1-l}$ in eqn.(\ref{integrated one loop planar scalar}) is in fact a parametric representation of the planar one loop amplitude terms of so-called surface Symanzik polynomials.  It can be shown that the above formula can be written as, 
\begin{align}
{\cal A}_{n}^{\phi^{3}\, (L=1)}(\vec{x})\, =\, \int_{{\bf R}^{n}}\, \dd^{n}\vec{t}\, \left(\frac{\pi}{{\cal U}^{D/2}}\right)\, e^{-\, \frac{{\cal F}}{{\cal U}}\, + \sum_{c} \alpha_{c} x_{c}} \label{surface symanzik as one loop}
\end{align}
where given a surface $S$ with $n$ marked points on the boundary and an annulus in the center, the surface Symanzik polynomials are defined in terms of the headlight functions as follows. Let ${\cal S}_{1}$ be the set of curves that cut ${\cal S}$ into a disk and let ${\cal S}_{2}$ be the set of all pair of curves that cut $S$ into two disjoint disks. 
\begin{align}
{\cal U} &= \sum_{c\, \in\, S_{1}}\,  \alpha_{c}  \\
{\cal F} &=   \sum_{(c,c^{\prime}) \in {\cal S}_{2}} \alpha_{c} \alpha_{c^{\prime}}\, \left(\sum_{i} p^{\mu}\right)^{2} 
\end{align}
As $\textrm{Val}_{\phi^{3}}$, one can trivially re-write these curve integral formulae in the same way eqn.(\ref{soag}) was re-written as \ref{soagwem}. Of course this makes no difference for $\phi^{3}$ amplitude, but is central to the Corolla action. 

\begin{align}\label{mnphi31l}
{\cal A}_{n}^{\phi^{3} (L=1)}(\vec{x})\, =\, \prod_{i=1}^{n} p_{i}^{2}\, \int d^{n}\vec{t}\, \int_{-\infty}^{0}\, d^{n}\vec{s}\, \left(\frac{\pi}{{\cal U}^{\frac{D}{2}}}\right)\, e^{-\, \frac{{\cal F}}{{\cal U}} + \sum_{w} \alpha_{w} x_{w}}
\end{align}
where the sum in the exponent is over \emph{all the curves} including the external edges for which the headlight functions are defined in eqn.(\ref{extheadlight}). 

Once again, the physical amplitude is obtained by substituting 
\begin{align}
x_{c} = X_{c}\, \forall\, c
\end{align}
Although we have restricted our review to the curve integral formulae for obtaining tree-level and (planar) $\phi^{3}$ amplitudes, it is important to emphasize that the striking power of these formulae lie in their ``universality". Namely the complete $\phi^{3}$ amplitude at all orders in the perturbation theory including the non-planar contributions can be obtained using curve integral formulae.

The Corolla differential operators shall have dependence on the legs with `loop' momenta as well. We need to morph \eqref{integrated one loop planar scalar} into an appropriate form such that all the internal and external legs's momenta are explicitly present. Let us label the momenta corresponding to the curve $C_{ij}$ as $\xi_{ij}$, and to the curve $S_{i+1}'$ as $-\xi_{i0}$. After imposing the momentum conservation, we shall identify $\xi_{ij}$ with $z_{ij}$. $\xi_{i0}$ shall appear in certain linear combinations only so that after imposing momentum conservation, we have only $z_{ij}$ dependence. We make the following replacement:
\begin{align}
z_{ij} \to \xi_{ij} ~, \hspace{2cm} z_i \to (\xi_{i-1,0} - \xi_{n,0})
\end{align}
One can check that \eqref{integrated one loop planar scalar} upon this substitution becomes:
\begin{align}
{\cal A}^{\phi^{3}}_n = \int_{\sum t_i\leq0}\dd^n\vec{t} \ \left(\frac{-\pi}{\sum_{i}t_i}\right)^{D/2}\exp\left(-\sum_{ij}\alpha_{C_{ij}}\xi_{ij}^2  -\frac{ \big( \sum_i\alpha_{S_{i+1}'}\,\xi_{i,0} \big)^2 }{\sum_i t_i} -\sum_i\alpha_{S_{i+1}'}\,\xi_{i,0}^2  \right) \bigg|_{\,\xi_C = \text{phys}}
\end{align}
Including the external particles as well, we rewrite the curve integral formula appropriate for the action of Corolla:
\begin{align}
{\cal A}^{\phi^{3}}_n &= \int \dd^ns\int_{\sum t_i\leq0}\dd^n\vec{t} \ \left(\frac{-\pi}{\sum_{i}t_i}\right)^{D/2} \nonumber \\
&\hspace{0.5cm}\times\exp\left(-\sum_{k \in \text{Ext}}s_k\,\xi_k^2-\sum_{ij}\alpha_{C_{ij}}\xi_{ij}^2  -\frac{ \big( \sum_i\alpha_{S_{i+1}'}\,\xi_{i,0} \big)^2 }{\sum_i t_i} -\sum_i\alpha_{S_{i+1}'}\,\xi_{i,0}^2  \right) \bigg|_{\,\xi_C = \text{phys}}~. \label{one loop amplitude with unphys}
\end{align}


\subsection{Curve integral for gluons at tree-level} \label{sec: tree level corolla for ym}
We now derive the second primary result of this paper: The curve integral formulae for planar (color-ordered) gluon amplitude. We start with the tree-level amplitudes. In the next section, we will derive the curve integral formula for the planar one-loop gluon amplitude. Throughout this section, we will denote the Corolla differentials associated to a triangulation as $\hat{C}_{T}$ instead of $\hat{C}^{(0)}_{T}$. As we only focus on deriving tree-level gluon amplitudes, we drop the super-script $(0)$ from the Corolla differentials. 

The starting point is a (trivial re-writing of) the tree-level curve integral formula for $\phi^{3}$ amplitude given in eqn.(\ref{cifsm}). As we will show below, the corresponding curve integral formula in Yang-Mills theory involves action of Corolla  differential  which is a weighted sum over Corolla graph differentials. 
\begin{align}\label{caftlymci}
\mathcal{A}^{\textrm{YM}}_n\big[1,\hdots,n\big] &=  \prod_{i\,\in \text{ ext}}\xi_i^2\, \big(\text{Reg-Res}\big)\, \frac{\sum_{T}\left(\prod_{c\in T} \alpha_c\right)\hat{C}_T}{\sum_T \left(\prod_{c\in T} \alpha_c\right)} \ \exp \left(- \sum_{w}\alpha_{w}\,\xi_w^2 \right) \bigg|_{\xi_{w}=p_{\text{phys}}} ~.
\end{align}
The sum over chords $c\, \in T$ is sum over all the dissection chords of the triangulation $T$ and sum over $w$ in the exponent include sum over all possible dissections of the $n$-gon as well as sum over external sides $[i,i+1]\, \vert\, 1\leq\, i\, \leq\, n-1$ of the $n$-gon. The differential $\hat{C}_{T}$ was introduced in eqn.(\ref{cgdiss}). In fact, $\hat{C}_{T} = \hat{C}_{\gamma(T)}$ where $\gamma$ is the tree-level planar graph dual to $T$. 

We shall expand on this formula (including the (Res-Reg) operation) below. The (Reg-Res) procedure required to extract the specific curve-integrand is a generalisation of the procedure for a fixed graph. This procedure is explained in section (\ref{trre}) below.


\subsubsection*{The Corolla operator for a particular cone }
Eqn.(\ref{caftlymci}) simply reduces to sum over all the triangulations and hence equals $\sum_{\gamma} \textrm{Val}_{\textrm{YM}}(\gamma)$.  This is simply due to the following identity.
\begin{align}
\int d^{n-3}\vec{t}\, \frac{F_{T}(\vec{t})\, \prod_{c \in T}\, \alpha_{c}}{\sum_{T}\prod_{c^{\prime} \in T} \alpha_{c^{\prime}}}\, e^{-\, \sum_{c^{\prime\prime}} \alpha_{c^{\prime\prime}} x_{c^{\prime\prime}}} = \sum_{T}\, \int\, d^{n-3}\vec{t}\, F_{T}(\vec{t})\, e^{-\, \sum_{c \in T} \alpha_{c} x_{c}} \label{integral restricting in a particular cone}
\end{align}
As a result, eqn.(\ref{caftlymci}) can be re-written as, 
\begin{align}
\mathcal{A}^{\textrm{YM}}_{n}\big[1,\hdots,n\big] &=\, \prod_{i\,\in \text{ ext}}\xi_i^2\int\, d^{n-3}\vec{t}\,  d^{n}\vec{s}\, \sum_{T} \big(\text{Reg-Res}\big)_T\ \hat{C}_{T}\ e^{-\sum_{C\in \text{int, ext}}\alpha_{C}\!(\vec{t})\,\xi^2_{C}}\ \bigg|_{\xi_{i,C} = p_{\text{phys}}}\nonumber\\
&=\, \sum_{\gamma} \textrm{Val}_{\textrm{YM}}(\gamma) ~. \label{premature color ordered ampl}
\end{align}
For the benefit of the reader, we recall that $x_{c} = \xi_{i}^{2}$ for $c = [i,i+1]$. 

The last equality follows from the identification of sum over $T$ with sum over all the trivalent graphs.  Of course for a single graph, the Corolla graph differential defined in KSVS is known, (eqn.(\ref{cgdiss}) ).  In light of the fact that the exponent $ - t_{c} \xi_{c}^{2}$ in the standard Schwinger parametrisation is replaced with $-\alpha_{c} x_{c}$ in the curve integral formula. It is natural to consider the Corolla differential operator where $\frac{1}{t_{e}}$ is replaced by $\frac{1}{\alpha_{c}}$. 

We write the differential operator $\hat{C}_v$ for a particular vertex $v$, with three external particles below for completeness:
\begin{align}
\begin{tikzpicture}[scale=0.8,baseline={([yshift=-.5ex]current bounding box.center)}]
\draw (0,0) -- (1,1);
\draw (0,0) -- (1,-1);
\draw (0,0) -- (-1.5,0);
\draw[-stealth] (-0.75,0) -- (-0.74,0);
\draw[-stealth] (0.5,0.5) -- (0.49,0.49);
\draw[-stealth] (0.5,-0.5) -- (0.49,-0.49);
\node at (-1.5,0.4) {$\xi_1 \ \ (\mu_1)$};
\node at (1.1,1) {$\xi_2 \ \ (\mu_2)$};
\node at (1.1,-1) {$\xi_3 \ \ (\mu_3)$};
\end{tikzpicture} \hspace{2cm} 
\hat{C}_v = \eta^{\mu_1\mu_2} \left(-\frac{1}{2\alpha_1}\,\p_{\xi_1}+\frac{1}{2\alpha_2}\,\p_{\xi_2}\right)^{\mu_3} + \text{cyclic}\label{cgd3pt}
\end{align}
Note that the conventions for the orientation of vertices, etc. remain the same as in the earlier sections. We repeat that in all the graphs, the orientation of all the vertices shall be consistently the same, and the direction of momentum flow in the shared legs of any two graphs shall be the same.
 
\subsubsection{The (Reg-Res) extraction}\label{trre}
The result of Corolla graph differential  in the Parametric space is a meromorphic functions which evaluates to $\textrm{Val}_{\gamma}(\textrm{YM})$ after applying the (Res-Reg) procedure, (see eqns (\ref{rr1}, \ref{rr2}, \ref{rr3}). This procedure hinges upon isolating poles of a graph dependent function $f_{\gamma}(t_{1}, \dots, t_{\vert E_{\textrm{tot}}\vert})$ over a subset of Schwinger parameters associated to set of internal edges which are mutually disconnected. However the curve integral formula treats space of all graphs democratically and hence we need to generalise eqn. (\ref{rrmain}) accordingly. 

Our generalisation is based on the simple observation that the (Res-Reg) ``extraction" is  primarily based on the set of red chords (which are dual to marked edges of a graph obtained after action of Corolla differential) in a marked triangulation. In light of the discussion in the section \ref{mtri}, it immediate follows that each of the colored chords is accompanied with a $1/\alpha_C$ in the curve integrand which does not get cancelled by $\alpha_{C}$ in the numerator once Corolla differential applies. This observation in conjunction with the fact that any generalisation of eqn(\ref{rrmain}), when restricted to a particular triangulation $T$ should simply reduce to the Res-Reg formula for functions defined on $\gamma_{T}$ (once we change the variables from $\alpha_{C\in T} \to t_i$) is sufficient to generalise the formula to the full curve integral.  That is, our generalisation should be such that, 
\begin{align}
\big(\text{Reg-Res}\big)_T =\, (\textrm{Reg-Res})_{\gamma(T)}
~, \label{reg res for each cone}
\end{align}
We now write down the (Res-Reg) procedure for the curve integral.  As the volume form in the parametric space can be written as,  (see eqn. (\ref{t measure in terms of alpha}) )
\begin{align}
\dd^E\vec{t}|_T = \bigwedge_{I\in T} \dd \alpha_I
\end{align}
we can write the (Reg-Res) as follows: Let ${\cal A}_{[1]}$ be the set of all co-dimension one facets of the associahedron and let $\chi$ be the set of all proper subsets of ${\cal A}_{[1]}$ including an empty set. We note that a given $\chi$ can also involve facets which are not adjacent to each other and hence these set of all such subsets is larger than (union over all the tri-valent planar graphs of) set of proper subsets of internal edges of a given graph. However as the Corolla differential will only poles over faces which are adjacent to each other, any subset which involve co-dimension one facets which do not share a common vertex of the associahedron will not contribute to the sum. 
\begin{align}\label{regresmaster}
\big(\text{Reg-Res}\big)\, \rightarrow\, \sum_{\chi \subseteq \mathcal{A}^{[1]}}\int\, d^{n-3}\vec{t}\ d^{n}\vec{s}\, \prod_{c\, \in\, \chi}\delta(\alpha_{c}(\vec{t}))\, \underset{\{\alpha_{c\in (\mathcal{A}^{[1]}/\chi)\, \cup E_{\textrm{ext}}}\}}{\text{Regular}} \ \underset{\{\alpha_{c\in \chi}\}\to 0}{\text{Residue}}
\end{align}
The extraction formula may look rather convoluted but it can be verified that it simply reduces to eqn.(\ref{rrmain}) inside each cone in the parametric space. Several comments are in order.
\begin{itemize}
\item As the action of Corolla never induces a pole in any of the Schwinger parameters $\{s_{1},\, \dots,\, s_{n}\}$, the (Res-Reg) extraction can be applied over the headlight functions associated to dissection chords of the $n$-gon and finally integrate the integrand with respect to $d^{n}\vec{s}$ followed by multiplication with $\prod_{i=1}^{n}\, \xi_{i}^{2}$. This elementary observation helps us in writing detailed computations and will be used throughout subsequent sections where we work out several examples. 
\item Examples given in sections (\ref{efoptl}, \ref{efiptl}) help with the concrete application of eqn.(\ref{regresmaster}). 
\end{itemize}
To summarize, combining eqns (\ref{cgd3pt}, \ref{regresmaster}) we obtain the curve integral formula for color-ordered tree-level gluon amplitudes is given in eqn.(\ref{premature color ordered ampl}). 
\subsubsection{Example: Four point tree level}\label{efoptl}
Let us work out the $n=4$ case in detail. There is only one internal edge for tree level $n=4$, so the g-vector fan lives in 1-dimensional space. There are two \emph{curves} corresponding to two: $s$ ($X_{13}=\xi_{13}^2$) and $t$ ($X_{24}=\xi_{24}^2$) channel graphs.

The headlight functions are as follows:
\begin{align}
\alpha_{13} = \mxx{0,t} = t\,\theta(t) ~, \qquad \alpha_{24} = -t+\mxx{0,t} = -t\,\theta(-t) ~.
\end{align}
Since the top dimensional cones are made up of single $g$-vectors, we have:
\begin{align}
\frac{\sum_{T}\left(\prod_{I\in T} \alpha_I\right)\hat{C}_T}{\sum_T \left(\prod_{I\in T} \alpha_I\right)} = \frac{\alpha_{13}\,\hat{C}_{(13)} + \alpha_{24}\,\hat{C}_{(24)}}{\alpha_{13}+\alpha_{24}} = \begin{dcases} \hat{C}_{(13)} & t>0 \\ \hat{C}_{(24)} & t<0\end{dcases}
\end{align}
There are two complete triangulations, leading to the two graphs shown below:
\begin{align}
\gamma_{(13)} \ \ : \quad& \begin{tikzpicture}[scale=0.5,baseline={([yshift=-.5ex]current bounding box.center)}]
\draw (-1,1) -- (0,0) -- (-1,-1);
\draw (0,0) -- (1.5,0) -- (2.5,-1);
\draw (1.5,0) -- (2.5,1);
\node at (-1.4,-1) {1};
\node at (-1.4,1) {2};
\node at (2.9,1) {3};
\node at (2.9,-1) {4};
\draw[-stealth] (-0.55,-0.55) -- (-0.56,-0.56);
\draw[-stealth] (-0.55,0.55) -- (-0.56,0.56);
\draw[-stealth] (2.05,-0.55) -- (2.06,-0.56);
\draw[-stealth] (2.05,0.55) -- (2.06,0.56);        
\draw[-stealth] (0.75,0) -- (0.76,0);
\node at (0.75,-0.5) {$\xi_{13}$};
\end{tikzpicture} ~,\hspace{2cm}
\gamma_{(24)} \ \ : \quad \begin{tikzpicture}[scale=0.5,baseline={([yshift=-.5ex]current bounding box.center)}]
\draw (-1,1) -- (0,0) -- (-1,-1);
\draw (0,0) -- (1.5,0) -- (2.5,-1);
\draw (1.5,0) -- (2.5,1);
\node at (-1.4,-1) {2};
\node at (-1.4,1) {3};
\node at (2.9,1) {4};
\node at (2.9,-1) {1};
\draw[-stealth] (-0.55,-0.55) -- (-0.56,-0.56);
\draw[-stealth] (-0.55,0.55) -- (-0.56,0.56);
\draw[-stealth] (2.05,-0.55) -- (2.06,-0.56);
\draw[-stealth] (2.05,0.55) -- (2.06,0.56);        
\draw[-stealth] (0.75,0) -- (0.76,0);
\node at (0.75,-0.5) {$\xi_{24}$};
\end{tikzpicture} \label{fig: four point tree level}
\end{align}
Here are the Corolla differential operators for the graphs shown in \eqref{fig: four point tree level}:
\begin{align}
\hat{C}_{(13)} &= \left[\eta^{\mu_1\mu_2} \left(\frac{1}{2\alpha_1}\,\p_{\xi_1} - \frac{1}{2\alpha_2}\,\p_{\xi_2}\right)^{\nu} + \eta^{\mu_2\nu} \left(\frac{1}{2\alpha_2}\,\p_{\xi_2} - \frac{1}{2\alpha_{13}}\,\p_{\xi_{13}}\right)^{\mu_1} + \eta^{\nu\mu_1} \left(\frac{1}{2\alpha_{13}}\,\p_{\xi_{13}} - \frac{1}{2\alpha_1}\,\p_{\xi_1}\right)^{\mu_2}\right] \nonumber\\
& \times \left[\eta_{\nu}{}^{\mu_4}\left(\frac{1}{2\alpha_{13}}\,\p_{\xi_{13}} + \frac{1}{2\alpha_4}\,\p_{\xi_4}\right)^{\mu_3} + \eta^{\mu_4\mu_3}\left(-\frac{1}{2\alpha_4}\,\p_{\xi_4} + \frac{1}{2\alpha_3}\,\p_{\xi_3}\right)_{\nu} + \eta^{\mu_3}{}_{\nu}\left(-\frac{1}{2\alpha_3}\,\p_{\xi_3} - \frac{1}{2\alpha_{13}}\,\p_{\xi_{13}}\right)^{\mu_4}\right]  \\
\hat{C}_{(24)}&=\left[ \eta^{\mu_1}{}_{\nu}\left( \frac{1}{2\alpha_1}\,\p_{\xi_1} + \frac{1}{2\alpha_{24}}\,\p_{\xi_{24}} \right)^{\mu_4} + \eta_{\nu}{}^{\mu_4}\left(-\frac{1}{2\alpha_{24}}\,\p_{\xi_{24}} - \frac{1}{2\alpha_4}\,\p_{\xi_4} \right)^{\mu_1} + \eta^{\mu_4\mu_1}\left(\frac{1}{2\alpha_4}\,\p_{\xi_4}-\frac{1}{2\alpha_1}\,\p_{\xi_1}\right)_{\nu}\right] \nonumber \\
&\times \left[\eta^{\nu\mu_2} \left(\frac{1}{2\alpha_{24}}\,\p_{\xi_{24}} - \frac{1}{2\alpha_2}\,\p_{\xi_2}\right)^{\mu_3} + \eta^{\mu_2\mu_3} \left(\frac{1}{2\alpha_2}\,\p_{\xi_2} - \frac{1}{2\alpha_3}\,\p_{\xi_3}\right)^{\nu} + \eta^{\mu_3\nu} \left(\frac{1}{2\alpha_3}\,\p_{\xi_3} - \frac{1}{2\alpha_{24}}\,\p_{\xi_{24}}\right)^{\mu_2}\right]
\end{align}
The action of these operators on the Schwinger exponential is as follows: 
\begin{align}
\alpha_{13}\,\hat{C}_{(13)}e^{-\sum_i\alpha_i\xi_i^2 -\alpha_{13}\xi_{13}^2-\alpha_{24}\xi_{24}^2} &= \left(
\begin{tikzpicture}[scale=0.4,baseline={([yshift=-.5ex]current bounding box.center)}]
\draw (-1,1) -- (0,0) -- (1.5,0) -- (2.5,1);
\draw (-1,-1) -- (0,0);
\draw (1.5,0) -- (2.5,-1);
\node at (-1.4,-1) {1};
\node at (-1.4,1) {2};
\end{tikzpicture} \hspace{0.3cm} 
+ \frac{1}{\alpha_{13}}\  \begin{tikzpicture}[scale=0.4,baseline={([yshift=-.5ex]current bounding box.center)}]
\draw (-1,1) -- (0,0);
\draw[thick,red] (0,0)-- (0.8,0);
\draw (-1,-1) -- (0,0);
\draw (0.8,0) -- (1.8,1);
\draw (0.8,0) -- (1.8,-1);
\node at (-1.4,-1) {1};
\node at (-1.4,1) {2};
\end{tikzpicture} \right)\alpha_{13}\, e^{-\sum_i\alpha_i\xi_i^2 -\alpha_{13}\xi_{13}^2} \\
\alpha_{24}\,\hat{C}_{(24)}e^{-\sum_i\alpha_i\xi_i^2 -\alpha_{13}\xi_{13}^2-\alpha_{24}\xi_{24}^2} &= \left(
\begin{tikzpicture}[scale=0.4,baseline={([yshift=-.5ex]current bounding box.center)}]
\draw (-1,1) -- (0,0) -- (1,1);
\draw (0,0) -- (0,-1.5);
\draw (-1,-2.5) -- (0,-1.5) -- (1,-2.5);
\node at (-1.4,0.8) {2};
\node at (1.4,0.8) {3};
\end{tikzpicture} \hspace{0.1cm} 
+ \frac{1}{\alpha_{24}}\  \begin{tikzpicture}[scale=0.4,baseline={([yshift=-.5ex]current bounding box.center)}]
\draw (-1,1) -- (0,0) -- (1,1);
\draw[thick,red] (0,0) -- (0,-0.8);
\draw (-1,-1.8) -- (0,-0.8) -- (1,-1.8);
\node at (-1.4,0.8) {2};
\node at (1.4,0.8) {3};
\end{tikzpicture} \right) \alpha_{24}\, e^{-\sum_i\alpha_i\xi_i^2 -\alpha_{24}\xi_{24}^2}
\end{align}
Note that these graphs denote only the numerator factors, and the external legs are labeled $[1,2,3,4]$ clockwise.

Finally, the (Reg-Res) term in this integrand is, 
 \begin{align}
\big(\text{Reg-Res}\big) = \int \dd^4 s_i \left[\left(\int_{-\infty}^{\infty} \dd t\  \underset{\alpha_{13}\,,\, \alpha_{24}}{\text{Regular}}  \right) + \underset{\alpha_{24}}{\text{Regular}}\ \underset{\alpha_{13}\to 0}{\text{Residue}}+ \underset{\alpha_{13}}{\text{Regular}}\ \underset{\alpha_{24}\to 0}{\text{Residue}} \right] 
\end{align}
Here, we made use of the fact that there shall be no poles in $\alpha_i$ for external legs. This simplified the contributions from external legs to (Reg-Res).
\begin{align}
{\cal A}^{\textrm{YM}}_4&\big[1,2,3,4\big] = \prod_{i\,\in \text{ ext}}\xi_i^2\,\big(\text{Reg-Res}\big)\ \frac{\sum_{T}\left(\prod_{I\in T} \alpha_I\right)\hat{C}_T}{\sum_T \left(\prod_{I\in T} \alpha_I\right)}\ e^{-\sum_i\alpha_i\xi_i^2 -\alpha_{13}\xi_{13}^2-\alpha_{24}\xi_{24}^2} \nonumber \\
&= \int \dd \alpha_{13} 
\begin{tikzpicture}[scale=0.4,baseline={([yshift=-.5ex]current bounding box.center)}]
\draw (-1,1) -- (0,0) -- (1.5,0) -- (2.5,1);
\draw (-1,-1) -- (0,0);
\draw (1.5,0) -- (2.5,-1);
\node at (-1.4,-1) {1};
\node at (-1.4,1) {2};
\end{tikzpicture} \  e^{-\alpha_{13}\xi_{13}^2} 
+ \int \dd \alpha_{24} \begin{tikzpicture}[scale=0.4,baseline={([yshift=-.5ex]current bounding box.center)}]
\draw (-1,1) -- (0,0) -- (1,1);
\draw (0,0) -- (0,-1.5);
\draw (-1,-2.5) -- (0,-1.5) -- (1,-2.5);
\node at (-1.4,0.8) {2};
\node at (1.4,0.8) {3};
\end{tikzpicture}\, e^{-\alpha_{24}\xi_{24}^2} \nonumber\\ 
&\hspace{2cm} + \quad \underset{\alpha_{13}\to 0}{\text{Residue}} \left( \frac{e^{-\alpha_{13}\xi_{13}^2}}{\alpha_{13}}  \begin{tikzpicture}[scale=0.4,baseline={([yshift=-.5ex]current bounding box.center)}]
\draw (-1,1) -- (0,0);
\draw[thick,red] (0,0)-- (0.8,0);
\draw (-1,-1) -- (0,0);
\draw (0.8,0) -- (1.8,1);
\draw (0.8,0) -- (1.8,-1);
\node at (-1.4,-1) {1};
\node at (-1.4,1) {2};
\end{tikzpicture} \right) + \quad \underset{\alpha_{24}\to 0}{\text{Residue}} \left( \frac{e^{-\alpha_{24}\xi_{24}^2}}{\alpha_{24}}  \begin{tikzpicture}[scale=0.4,baseline={([yshift=-.5ex]current bounding box.center)}]
\draw (-1,1) -- (0,0) -- (1,1);
\draw[thick,red] (0,0) -- (0,-0.8);
\draw (-1,-1.8) -- (0,-0.8) -- (1,-1.8);
\node at (-1.4,0.8) {2};
\node at (1.4,0.8) {3};
\end{tikzpicture} \right) \\
&= \frac{1}{X_{13}} \begin{tikzpicture}[scale=0.4,baseline={([yshift=-.5ex]current bounding box.center)}]
\draw (-1,1) -- (0,0) -- (1.5,0) -- (2.5,1);
\draw (-1,-1) -- (0,0);
\draw (1.5,0) -- (2.5,-1);
\node at (-1.4,-1) {1};
\node at (-1.4,1) {2};
\end{tikzpicture} \quad + \frac{1}{X_{24}} \begin{tikzpicture}[scale=0.4,baseline={([yshift=-.5ex]current bounding box.center)}]
\draw (-1,1) -- (0,0) -- (1,1);
\draw (0,0) -- (0,-1.5);
\draw (-1,-2.5) -- (0,-1.5) -- (1,-2.5);
\node at (-1.4,0.8) {2};
\node at (1.4,0.8) {3};
\end{tikzpicture} + \left(\begin{tikzpicture}[scale=0.4,baseline={([yshift=-.5ex]current bounding box.center)}]
\draw (-1,1) -- (0,0);
\draw[thick,red] (0,0)-- (0.8,0);
\draw (-1,-1) -- (0,0);
\draw (0.8,0) -- (1.8,1);
\draw (0.8,0) -- (1.8,-1);
\node at (-1.4,-1) {1};
\node at (-1.4,1) {2};
\end{tikzpicture} + \begin{tikzpicture}[scale=0.4,baseline={([yshift=-.5ex]current bounding box.center)}]
\draw (-1,1) -- (0,0) -- (1,1);
\draw[thick,red] (0,0) -- (0,-0.8);
\draw (-1,-1.8) -- (0,-0.8) -- (1,-1.8);
\node at (-1.4,0.8) {2};
\node at (1.4,0.8) {3};
\end{tikzpicture} \right) \label{four point amplitude}
\end{align}
We have integrated the $s_i$, the Schwinger like parameters for external particles, which yielded us $1/\xi_i^2$, and in the presence of $\xi_i^2$ gave us unity. 

Note that as the usual Feynman rules dictate, the vertex factor for a contact interaction comes from three channels. As we see in \eqref{contact interaction}, only two channels contribute to the planar amplitude.

\subsubsection{n=5}\label{efiptl}
The g-vector space is two-dimensional for $n=5$. There are five curves, hence five $g$-vectors, dual to the five faces (1-d edges) of the two-dimensional (pentagon) associahedron. We denote the five curves as $C_{k(k+2)}$ carrying the momenta $(p_k+p_{k+1})$. There are five cones and hence five trivalent graphs, corresponding to the five vertices of the associahedron. All the triangulations are \emph{ray-like}, so let us label them as $T(k)_{k\in[1,\hdots,5]}$ in accord with figure \eqref{fig:five point tree level}. 
\begin{align} \label{fig:five point tree level}
\begin{tikzpicture}[scale=1,baseline={([yshift=-.5ex]current bounding box.center)}]
\node at (-5,0) {$ T(k) \quad : $};
\draw (-1,-1) -- (0,0) -- (-1,1);
\draw (0,0) -- (4,0);
\draw (2,0) -- (2,1);
\draw (4,0) -- (5,1);
\draw (4,0) -- (5,-1);
\draw[-stealth] (1,0) -- (1.01,0);
\draw[-stealth] (3,0) -- (3+0.01,0);
\draw[-stealth] (-0.51,-0.51) -- (-0.5,-0.5);
\draw[-stealth] (-0.51,0.51) -- (-0.5,0.5);
\draw[-stealth] (4.51,0.51) -- (4.5,0.5);
\draw[-stealth] (4.51,-0.51) -- (4.5,-0.5);
\draw[-stealth] (2,0.53) -- (2,0.52);
\node at (-1.5,-1.1) {$\xi_{k-2}$};
\node at (-1.5,-0.6) {$(\mu_{k-2})$};
\node at (-1.5,1.1) {$\xi_{k-1}$};
\node at (-1.5,0.6) {$(\mu_{k-1})$};
\node at (2.2,1.2) {$\xi_{k} ~ ~ (\mu_k)$};
\node at (5.5,-1.1) {$\xi_{k+2}$};
\node at (5.5,-0.6) {$(\mu_{k+2})$};
\node at (5.5,1.1) {$\xi_{k+1}$};
\node at (5.5,0.6) {$(\mu_{k+1})$};
\node at (1,0.3) {$\xi_{(k-2)k}$};
\node at (1,-0.3) {$(\nu_1)$};
\node at (3.2,0.3) {$\xi_{(k+1)(k-2)}$};
\node at (3.2,-0.3) {$(\nu_2)$};
\end{tikzpicture}
\end{align}

For the \eqref{fig:five point tree level}, we write down the Corolla differential operator:
\begin{align}
&\hat{C}_{T(k)} \nonumber \\
&=\bigg\{\eta^{\nu_1\mu_{k-2}}\left( -\frac{1}{2\alpha_{k-2}}\,\p_{\xi_{k-2}} -\frac{1}{2\alpha_{(k-2)k}}\,\p_{\xi_{(k-2)k}} \right)^{\mu_{k-1}} 
+ \eta^{\mu_{k-1}\nu_1}\left( \frac{1}{2\alpha_{(k-2)k}}\,\p_{\xi_{(k-2)k}} + \frac{1}{2\alpha_{k-1}}\,\p_{\xi_{k-1}} \right)^{\mu_{k-2}} \nonumber \\
&\hspace{8.7cm} + \eta^{\mu_{k-2}\mu_{k-1}}\left( -\frac{1}{2\alpha_{k-1}}\,\p_{\xi_{k-1}} + \frac{1}{2\alpha_{k-2}}\,\p_{\xi_{k-2}}  \right)^{\nu_1}\bigg\} \nonumber\\
&\hspace{-0.5cm}\times \bigg\{ \eta_{\nu_1}{}^{\mu_k}\left( -\frac{1}{2\alpha_k}\,\p_{\xi_k} + \frac{1}{2\alpha_{(k-2)k}}\,\p_{\xi_{(k-2)k}} \right)_{\nu_2} + \eta^{\mu_k}{}_{\nu_2} \left( \frac{1}{2\alpha_{(k+1)(k-2)}}\,\p_{\xi_{(k+1)(k-2)}} + \frac{1}{2\alpha_k}\,\p_{\xi_k} \right)_{\nu_1}  \nonumber \\
& \hspace{6.5cm} + \eta_{\nu_2\nu_1}\left( -\frac{1}{2\alpha_{(k-2)k}}\,\p_{\xi_{(k-2)k}} -\frac{1}{2\alpha_{(k+1)(k-2)}}\,\p_{\xi_{(k+1)(k-2)}} \right)^{\mu_k}  \bigg\} \nonumber \\
&\hspace{-0.5cm}\times \bigg\{ \eta^{\mu_{k+1}\mu_{k+2}}\left( -\frac{1}{2\alpha_{k+2}}\,\p_{\xi_{k+2}} + \frac{1}{2\alpha_{k+1}}\,\p_{\xi_{k+1}} \right)^{\nu_2} + \eta^{\mu_{k+2}\nu_2}\left( -\frac{1}{2\alpha_{(k+1)(k-2)}}\,\p_{\xi_{(k+1)(k-2)}} + \frac{1}{2\alpha_{k+2}}\,\p_{\xi_{k+2}} \right)^{\mu_{k+1}}  \nonumber \\
&\hspace{6.5cm} + \eta^{\nu_2\mu_{k+1}}\left( -\frac{1}{2\alpha_{k+1}}\,\p_{\xi_{k+1}} + \frac{1}{2\alpha_{(k+1)(k-2)}}\,\p_{\xi_{(k+1)(k-2)}} \right)^{\mu_{k+2}}  \bigg\}
\end{align}

We can write the 5-point tree level YM amplitude as follows:
\begin{align}
\mathcal{A}_5\big[1,\hdots,5\big] &= \prod_{i=1}^5 \xi_i^2\, \big(\text{Reg-Res}\big)\, \frac{\sum_{T}\left(\prod_{I\in T} \alpha_I\right)\hat{C}_T}{\sum_T \left(\prod_{I\in T} \alpha_I\right)} \ \exp \left( - \sum_{C} \alpha_C\,\xi_C^2 \right) \nonumber \\
&= \prod_{i=1}^5 \xi_i^2\,\big(\text{Reg-Res}\big)\,\frac{\sum_{k=1}^5\alpha_{(k-2)k} \,\alpha_{(k+1)(k-2)} \,\hat{C}_{T(k)}}{\sum_{k=1}^5\alpha_{(k-2)k} \,\alpha_{(k+1)(k-2)}} \exp-\sum_{i=1}^5\left(\alpha_i\,\xi_i^2 + \alpha_{i(i+2)} \,\xi_{i(i+2)}^2\right)
\end{align}

Let us evaluate the action of Corolla differential operator $\hat{C}_T$ on the exponential:
\begin{align}
&\hat{C}_{T(k)}\exp-\sum_{i=1}^5\left(\alpha_i\,\xi_i^2 + \alpha_{i(i+2)} \,\xi_{i(i+2)}^2\right) \nonumber\\
&= \bigg\{\eta^{\nu_1\mu_{k-2}}\left( {\xi_{k-2}} +{\xi_{(k-2)k}} \right)^{\mu_{k-1}} 
+ \eta^{\mu_{k-1}\nu_1}\left( -{\xi_{(k-2)k}} - {\xi_{k-1}} \right)^{\mu_{k-2}}  + \eta^{\mu_{k-2}\mu_{k-1}}\left( {\xi_{k-1}} - {\xi_{k-2}}  \right)^{\nu_1}\bigg\}e^{-\sum\hdots} \nonumber\\
&\quad\times \bigg\{ \eta^{\mu_k}{}_{\nu_2} \left( -{\xi_{(k+1)(k-2)}} - {\xi_k} \right)_{\nu_1} + \eta_{\nu_2\nu_1}\left( {\xi_{(k-2)k}} +{\xi_{(k+1)(k-2)}} \right)^{\mu_k} + \eta_{\nu_1}{}^{\mu_k}\left( {\xi_k} - {\xi_{(k-2)k}} \right)_{\nu_2} \bigg\} \nonumber \\
&\quad\times \bigg\{ \eta^{\mu_{k+2}\nu_2}\left( {\xi_{(k+1)(k-2)}} - {\xi_{k+2}} \right)^{\mu_{k+1}} + \eta^{\nu_2\mu_{k+1}}\left( {\xi_{k+1}} - {\xi_{(k+1)(k-2)}} \right)^{\mu_{k+2}} + \eta^{\mu_{k+1}\mu_{k+2}}\left( {\xi_{k+2}} - {\xi_{k+1}} \right)^{\nu_2}  \bigg\} \nonumber\\
& + \frac{1}{\alpha_{(k+1)(k-2)}}\left(\eta^{\mu_{k}\mu_{k+2}}\,\eta_{\nu_1}{}^{\mu_{k+1}} - \eta^{\mu_{k}\mu_{k+1}}\,\eta_{\nu_1}{}^{\mu_{k+2}}\right)e^{-\sum\hdots} \nonumber \\    
&\qquad \times\bigg\{\eta^{\nu_1\mu_{k-2}}\left( {\xi_{k-2}} +{\xi_{(k-2)k}} \right)^{\mu_{k-1}} 
+ \eta^{\mu_{k-1}\nu_1}\left( -{\xi_{(k-2)k}} - {\xi_{k-1}} \right)^{\mu_{k-2}}  + \eta^{\mu_{k-2}\mu_{k-1}}\left( {\xi_{k-1}} - {\xi_{k-2}}  \right)^{\nu_1}\bigg\} \nonumber\\
&+\frac{1}{\alpha_{(k-2)k}}\left(\eta^{\mu_k\mu_{k-2}}\,\eta^{\mu_{k-1}}{}_{\nu_2} - \eta^{\mu_k\mu_{k-1}}\,\eta^{\mu_{k-2}}{}_{\nu_2}\right) \,e^{-\sum\hdots} \nonumber \\
&\quad \times \bigg\{ \eta^{\mu_{k+2}\nu_2}\left( {\xi_{(k+1)(k-2)}} - {\xi_{k+2}} \right)^{\mu_{k+1}} + \eta^{\nu_2\mu_{k+1}}\left( {\xi_{k+1}} - {\xi_{(k+1)(k-2)}} \right)^{\mu_{k+2}} + \eta^{\mu_{k+1}\mu_{k+2}}\left( {\xi_{k+2}} - {\xi_{k+1}} \right)^{\nu_2}  \bigg\}
\end{align}
The first term is the trivalent contribution. The next two terms correspond to contact interactions, with $\xi_{(k+1)(k-2)}$ and $\xi_{(k-2)k}$ being marked respectively. To summarise, we have obtained the appropriate numerator factors as follows:
\begin{align}
&\sum_{T}\left(\prod_{I\in T} \alpha_I\right)\hat{C}_T \ \exp \left(-\sum_{i=1}^n\alpha_i \,\xi_i^2 - \sum_{(ik)} \alpha_{ik}\,\xi_{ik}^2 \right) \nonumber \\
&\hspace{1cm}= 
\left(\sum_{k=1}^5\alpha_{(k-2)k} \,\alpha_{(k+1)(k-2)} \ \hat{C}_{T(k)}\right) \exp-\sum_{i=1}^5\left(\alpha_i\,\xi_i^2 + \alpha_{i(i+2)} \,\xi_{i(i+2)}^2\right)\nonumber \\
&\hspace{1cm}=
\alpha_{14}\,\alpha_{24}\left( 
\begin{tikzpicture}[scale=0.5,baseline={([yshift=-.5ex]current bounding box.center)}]  
\draw (-1.7,-0.7) -- (-1,0) -- (-1.7,0.7);
\draw (-1,0) -- (1,0);
\draw (0,0) -- (0,1);
\draw (1.7,-0.7) -- (1,0) -- (1.7,0.7);
\node at (-0.3,1) {1};
\node at (1.15,0.8) {2};
\end{tikzpicture}   \quad
+ \frac{1}{\alpha_{14}} \  
\begin{tikzpicture}[scale=0.5,baseline={([yshift=-.5ex]current bounding box.center)}]  
\draw (-1.3,-0.7) -- (-0.6,0) -- (-1.3,0.7);
\draw[thick,red] (-0.6,0) -- (0,0);
\draw (0,0) -- (1,0);
\draw (0,0) -- (0,1);
\draw (1.7,-0.7) -- (1,0) -- (1.7,0.7);
\node at (-0.3,1) {1};
\end{tikzpicture} \quad
+ \frac{1}{\alpha_{24}} \ 
\begin{tikzpicture}[scale=0.5,baseline={([yshift=-.5ex]current bounding box.center)}]  
\draw (-1.7,-0.7) -- (-1,0) -- (-1.7,0.7);
\draw (-1,0) -- (0,0);
\draw[thick,red] (0,0) -- (0.6,0);
\draw (0,0) -- (0,1);
\draw (1.3,-0.7) -- (0.6,0) -- (0.6+0.7,0.7);
\node at (-0.3,1) {1};
\end{tikzpicture}\right)\,e^{-\sum\hdots} + \text{cyclic} \label{action of corolla good equation} \\
&\hspace{1cm}= \left(\alpha_{14} \, \alpha_{24} \  \begin{tikzpicture}[scale=0.5,baseline={([yshift=-.5ex]current bounding box.center)}]  
\draw (-1.7,-0.7) -- (-1,0) -- (-1.7,0.7);
\draw (-1,0) -- (1,0);
\draw (0,0) -- (0,1);
\draw (1.7,-0.7) -- (1,0) -- (1.7,0.7);
\node at (-0.3,1) {1};
\end{tikzpicture} \quad + \alpha_{25} \, \alpha_{35} \  
\begin{tikzpicture}[scale=0.5,baseline={([yshift=-.5ex]current bounding box.center)}]  
\draw (-1.7,-0.7) -- (-1,0) -- (-1.7,0.7);
\draw (-1,0) -- (1,0);
\draw (0,0) -- (0,1);
\draw (1.7,-0.7) -- (1,0) -- (1.7,0.7);
\node at (-0.3,1) {2};
\end{tikzpicture} \quad+ \hdots\right) \,e^{-\sum \hdots} \nonumber \\
&\hspace{1cm}\qquad + \alpha_{45}\left( \begin{tikzpicture}[scale=0.5,baseline={([yshift=-.5ex]current bounding box.center)}]  
\draw (-1.7,-0.7) -- (-1,0) -- (-1.7,0.7);
\draw (-1,0) -- (0,0);
\draw[thick,red] (0,0) -- (0.6,0);
\draw (0,0) -- (0,1);
\draw (1.3,-0.7) -- (0.6,0) -- (0.6+0.7,0.7);
\node at (-0.3,1) {1};
\end{tikzpicture} \quad + \quad
\begin{tikzpicture}[scale=0.5,baseline={([yshift=-.5ex]current bounding box.center)}]  
\draw (-1.7,-0.7) -- (-1,0) -- (-1.7,0.7);
\draw (-1,0) -- (0,0);
\draw[thick,red] (0,0) -- (0.4,0.4);
\draw (0,0) -- (0.7,-0.7);
\draw (0.4,0.4+1) -- (0.4,0.4) -- (0.4+1,0.4);
\node at (0.1,1.25) {1};
\end{tikzpicture}
\right) \,e^{-\sum \hdots} + \hdots \label{contact interaction}
\end{align}
Note that these graphs denote only the numerator factors, and the external legs are labeled $[1,\hdots,n]$ clockwise. Using \eqref{integral restricting in a particular cone}, we can write the following,
\begin{align}
&\frac{\sum_{T}\left(\prod_{I\in T} \alpha_I\right)\hat{C}_T}{\sum_T \left(\prod_{I\in T} \alpha_I\right)} \ e^{-\sum_i\alpha_i \,\xi_i^2 - \sum_C \alpha_C\,\xi_C^2 } \bigg|_{\text{Cone} \{C_1,C_2\}} \nonumber\\
&\hspace{3cm} = \left( 
\begin{tikzpicture}[scale=0.45,baseline={([yshift=-.5ex]current bounding box.center)}]  
\draw (-1.7,-0.7) -- (-1,0) -- (-1.7,0.7);
\draw (-1,0) -- (1,0);
\draw (0,0) -- (0,1);
\draw (1.7,-0.7) -- (1,0) -- (1.7,0.7);
\node at (-0.6,-0.5) {\small $C_1$};
\node at (0.6,-0.5) {\small $C_2$};
\end{tikzpicture}   \ \
+ \frac{1}{\alpha_{C_1}} \  
\begin{tikzpicture}[scale=0.45,baseline={([yshift=-.5ex]current bounding box.center)}]  
\draw (-1.3,-0.7) -- (-0.6,0) -- (-1.3,0.7);
\draw[thick,red] (-0.6,0) -- (0,0);
\draw (0,0) -- (1,0);
\draw (0,0) -- (0,1);
\draw (1.7,-0.7) -- (1,0) -- (1.7,0.7);
\node at (0.6,-0.5) {\small $C_2$};
\end{tikzpicture} \ \
+ \frac{1}{\alpha_{C_2}} \ 
\begin{tikzpicture}[scale=0.45,baseline={([yshift=-.5ex]current bounding box.center)}]  
\draw (-1.7,-0.7) -- (-1,0) -- (-1.7,0.7);
\draw (-1,0) -- (0,0);
\draw[thick,red] (0,0) -- (0.6,0);
\draw (0,0) -- (0,1);
\draw (1.3,-0.7) -- (0.6,0) -- (0.6+0.7,0.7);
\node at (-0.3,1) {1};
\node at (-0.6,-0.5) {\small $C_1$};
\end{tikzpicture}\right) e^{-\sum\hdots} 
\end{align}
As we see in \eqref{contact interaction}, each $\alpha_I$ in the coefficient of a graph should result in a $X_I$ propagator. The (Reg-Res) procedure does exactly this. Denoting the set of all curves, i.e., the co-dimension one faces of the associahedron by $\mathcal{A}^{[1]}$, and using \eqref{regresmaster}, we have the following:
\begin{align}
\big(\text{Reg-Res}\big) &= \int \dd^5s\, \sum_{\chi \subset \mathcal{A}^{[1]}}\int_{\mathbb{R}^{2}|_{\{\alpha_{C\in \chi}\}=0}}\dd^{2-|\chi|}\vec{t}\ \underset{\{\alpha_{C\in \mathcal{A}^{[1]}/\chi}\}}{\text{Regular}} \ \underset{\{\alpha_{C\in \chi}\}\to 0}{\text{Residue}} \\
&= \int \dd^5s\, \int_{\mathbb{R}^2}\dd^2 \vec{t}\sum_{\chi \subset \mathcal{A}^{[1]}} \prod_{I\in \chi}\delta\big(\alpha_I(\vec{t})\big) \ \underset{\{\alpha_{C\in \mathcal{A}^{[1]}/\chi}\}}{\text{Regular}} \ \underset{\{\alpha_{C\in \chi}\}\to 0}{\text{Residue}}~,
\end{align}
Curves (propagators) belonging to $\chi$ are precisely the marked edges. Note that we have again used the fact that there shall be no poles for $\alpha_i$ corresponding to external legs, hence simplifying their contributions.
\begin{align}
\big(\text{Reg-Res}\big) = \int_{\mathbb{R}^2}\dd^2\vec{t}\ \ \underset{\{\alpha_C\}\ \forall C}{\text{Regular}} \ \  + \ \ \sum_{C'}\int_{\mathbb{R}^2} \dd^2\vec{t}\ \delta\big(\alpha_{C'}(\vec{t})\big) \ \underset{\{\alpha_C\}\ C\neq C'}{\text{Regular}}\ \underset{\alpha_{C'}\to 0}{\text{Residue}} \ \ + \hdots ~, \label{reg res for 5 points}
\end{align}
$\hdots$ denote the terms when $\chi$ contains more than one curve. Since no more than one propagator can be marked for $n=5$, these shall vanish. 

\begin{align} \label{fig:headlight functions for tree level 5 points}
\begin{tikzpicture}[scale=1.3,baseline={([yshift=-.5ex]current bounding box.center)}]  
\draw[-stealth] (0,0) -- (0,2);
\draw[-stealth] (0,0) -- (2,0);
\draw[-stealth] (0,0) -- (0,-2);
\draw[-stealth] (0,0) -- (-2,0);
\draw[-stealth] (0,0) -- (-2,2);
\node at (0.3,2.2) {$g_{14}$};
\node at (2.3,0) {$g_{13}$};
\node at (-2.3,-0) {$g_{25}$};
\node at (-0.3,-2) {$g_{35}$};
\node at (-2.3,2) {$g_{24}$};
\node at (1.2,1.4) {$\alpha_{13}=t_1$};
\node at (1.2,1) {$\alpha_{14}=t_2$};
\node at (1.2,-1.4) {$\alpha_{13}=t_1$};
\node at (1.2,-1) {$\alpha_{35}=-t_2$};
\node at (-1,2) {$\alpha_{14}=t_1+t_2$};
\node at (-0.8,1.6) {$\alpha_{24}=-t_1$};
\node at (-2,1) {$\alpha_{24} =t_2$};
\node at (-1.8,0.6) {$\alpha_{25}=-t_1-t_2$};
\node at (-1.2,-1) {$\alpha_{25}=-t_1$};
\node at (-1.2,-1.4) {$\alpha_{35}=-t_2$};
\end{tikzpicture}
\end{align}
We evaluate \eqref{reg res for 5 points} by restricting it to various cones. Inspecting \eqref{fig:headlight functions for tree level 5 points}, we can see that the volume form in the Schwinger space restricted in any cone can be written as follows
\begin{align}
\dd^2\vec{t}\,\big|_{\text{cone }(13)(14)} = \dd\alpha_{13}\wedge \dd\alpha_{14} ~, \qquad \dd^2\vec{t}\,\big|_{\text{cone }T} = \bigwedge_{I\in T}\dd \alpha_I 
\end{align}
Using these pieces of information, we can infer the following:
\begin{align}
&\int_{\text{Cone }\{C_1,C_2\}}\dd^2\vec{t}\ \ \delta\big(\alpha_{C_1}(\vec{t})\big)\, \ \underset{\{\alpha_C\}\ C\neq C_1}{\text{Regular}}\ \underset{\alpha_{C_1}\to 0}{\text{Residue}} \left( 
\begin{tikzpicture}[scale=0.45,baseline={([yshift=-.5ex]current bounding box.center)}]  
\draw (-1.7,-0.7) -- (-1,0) -- (-1.7,0.7);
\draw (-1,0) -- (1,0);
\draw (0,0) -- (0,1);
\draw (1.7,-0.7) -- (1,0) -- (1.7,0.7);
\node at (-0.6,-0.5) {\small $C_1$};
\node at (0.6,-0.5) {\small $C_2$};
\end{tikzpicture}   \ \
+ \frac{1}{\alpha_{C_1}} \  
\begin{tikzpicture}[scale=0.45,baseline={([yshift=-.5ex]current bounding box.center)}]  
\draw (-1.3,-0.7) -- (-0.6,0) -- (-1.3,0.7);
\draw[thick,red] (-0.6,0) -- (0,0);
\draw (0,0) -- (1,0);
\draw (0,0) -- (0,1);
\draw (1.7,-0.7) -- (1,0) -- (1.7,0.7);
\node at (0.6,-0.5) {\small $C_2$};
\end{tikzpicture} \ \
+ \frac{1}{\alpha_{C_2}} \ 
\begin{tikzpicture}[scale=0.45,baseline={([yshift=-.5ex]current bounding box.center)}]  
\draw (-1.7,-0.7) -- (-1,0) -- (-1.7,0.7);
\draw (-1,0) -- (0,0);
\draw[thick,red] (0,0) -- (0.6,0);
\draw (0,0) -- (0,1);
\draw (1.3,-0.7) -- (0.6,0) -- (0.6+0.7,0.7);
\node at (-0.3,1) {1};
\node at (-0.6,-0.5) {\small $C_1$};
\end{tikzpicture}\right) e^{-\sum\hdots} \nonumber \\
&= \int_{\text{Cone }\{C_1,C_2\}} \dd\alpha_{C_1}\,\dd\alpha_{C_2} \ \delta\big(\alpha_{C_1}\big)  
\left(\begin{tikzpicture}[scale=0.45,baseline={([yshift=-.5ex]current bounding box.center)}]  
\draw (-1.3,-0.7) -- (-0.6,0) -- (-1.3,0.7);
\draw[thick,red] (-0.6,0) -- (0,0);
\draw (0,0) -- (1,0);
\draw (0,0) -- (0,1);
\draw (1.7,-0.7) -- (1,0) -- (1.7,0.7);
\node at (0.6,-0.5) {\small $C_2$};
\end{tikzpicture} \right)  e^{-\sum \alpha_i \xi_i^2 - \alpha_{C_1}\xi_{C_1}^2 - \alpha_{C_2}\xi_{C_2}^2} \label{some random reg res penultimate eqn} \\
&= \frac{1}{\xi_{C_2}^2}\left(\begin{tikzpicture}[scale=0.45,baseline={([yshift=-.5ex]current bounding box.center)}]  
\draw (-1.3,-0.7) -- (-0.6,0) -- (-1.3,0.7);
\draw[thick,red] (-0.6,0) -- (0,0);
\draw (0,0) -- (1,0);
\draw (0,0) -- (0,1);
\draw (1.7,-0.7) -- (1,0) -- (1.7,0.7);
\node at (0.6,-0.5) {\small $C_2$};
\end{tikzpicture} \right)  e^{-\sum \alpha_i \xi_i^2} \label{final result of regres}
\end{align}
Note that in the exponential in \eqref{some random reg res penultimate eqn}, there are no other $\alpha_C\xi_C^2$, because in this particular cone, $\alpha_C$'s vanish. Also note that the numerator factors, generated by Corolla, denoted by the graph, are independent of the Schwinger parameters, and hence the headlight functions. 

Following the roadmap in \eqref{final result of regres}, we can piecewise integrate to obtain the correct propagators. Finally, one can readily integrate the Schwinger parameters for the external particles, and in the end, we obtain the five-point tree-level amplitude.

\subsection{Curve integral for gluons at loop level} \label{sec: curve integral at loops}
The graphs $T$'s are dual to the vertices of the $\hat{D}_n$ polytope. We can have $g$-vectors and hence headlight functions for $\hat{D}_n$ polytope analogous to the associahedron case for tree-level scattering. We can write the Corolla differential for each vertex of the polytope, and join them together just like the tree-level case. 

The real power of the Corolla construction reviewed in section \ref{cgdksvs}, lies in the fact that it can be used to convert the loop-integrated scalar graphs into loop-integrated gluon graphs. The latter being the usual trivalent graph, corresponding graphs including contact vertices, and the ghost contributions. Before the loop integral, we can just put a Corolla polynomial in the integrand of the scalar amplitude in Schwinger parameterization, changing all the scalar vertices into gluon vertices and appropriate contact vertices. This works at arbitrary loops. The neat thing is that since the Corolla differential operator has abstract momentum labels, we can pass the loop integral $\dd^Dl$ through the $\hat{C}_T$, and perform the loop integration prior to the action of $\hat{C}_T$. 

Concretely, a planar scalar one-loop graph will give us a single trace gluon contribution, as follows:
\begin{align}
\mathcal{A}_n^{(L=1)} &= \prod_{i\in E_{\text{ext}}}\xi_i^2\int \dd^D\ell\,\big(\text{Reg-Res}\big)\frac{\sum_{T} \left(\prod_{I \in T}\alpha_I(\vec{t})\right) \hat{C}_{T}}{\sum_{T} \left(\prod_{I \in T}\alpha_I(\vec{t})\right)} \left(e^{-\sum_{m\in \text{ext}}\alpha_mp_m^2}\  e^{-\sum_{C}\alpha_{C}\!(\vec{t})\,X_{C}}\right) ~. 
\end{align}
A remarkable property of Corolla graph differential discovered by KSVS was that the action commutes with loop integration. 
\begin{align}
\mathcal{A}_n\big[1,\hdots,n\big] &= \prod_{i\in E_{\text{ext}}}\xi_i^2 \big(\text{Reg-Res}\big)\frac{\sum_{T} \left(\prod_{I \in T}\alpha_I(\vec{t})\right) \hat{C}_{T}}{\sum_{T} \left(\prod_{I \in T}\alpha_I(\vec{t})\right)} \int \dd^D\ell \left(e^{-\sum_{ \text{ext}}\alpha_mp_m^2}\  e^{-\sum_{C}\alpha_{C}\!(\vec{t})\,X_{C}}\right) ~. \label{raw one loop amplitude}
\end{align}
Here, we make use of \eqref{one loop amplitude with unphys}, where we have carried out the loop integral, and written the result appropriately. So, the planar one-loop amplitude is given by the following:
\begin{align}
A_n^{\text{YM },(L=1)} &= \prod_{i\in E_{\text{ext}}}\xi_i^2 \ \big(\text{Reg-Res}\big)\, \left(\frac{-\pi}{\sum_{i}t_i}\right)^{D/2}\ \frac{\sum_{T} \left(\prod_{I \in T}\alpha_I(\vec{t})\right) \hat{C}_{T}}{\sum_{T} \left(\prod_{I \in T}\alpha_I(\vec{t})\right)} \nonumber \\
&\hspace{-0.5cm}\times\exp\left(-\sum_{k \in \text{ext}}s_k\,\xi_k^2-\sum_{ij}\alpha_{C_{ij}}\xi_{ij}^2  -\frac{ \big( \sum_i\alpha_{S_{i+1}'}\,\xi_{i,0} \big)^2 }{\sum_i t_i} -\sum_i\alpha_{S_{i+1}'}\,\xi_{i,0}^2  \right) \bigg|_{\,\xi_C = \text{phys}}~. \label{master one loop amplitude}
\end{align}

The Corolla differential operator comprises $1/2\alpha_C \,\p_{P_C}$. So, apart from the tree level propagators, dual to the $C_{ij}$'s, the ones involving loop momenta shall be as follows:
\begin{align}
\frac{1}{2\alpha_C} \,\p_{P_C} \xrightarrow{C \in \{S_{i+1}'\}} -\frac{1}{2\alpha_{S_{i+1}'}}\,\p_{\xi_{i,0}}
\end{align}
It shall be useful later on to see how this acts on the integrated exponential:
\begin{align}
\frac{-1}{2\alpha_{S_{k+1}'}}\,\p_{\xi_{k0}}{}^\nu\ &\exp\left(-\sum_{ij}\alpha_{C_{ij}}\xi_{ij}^2  -\frac{ \left( \sum_i\alpha_{S_{i+1}'}\,\xi_{i0} \right)^2 }{\sum_i t_i} -\sum_i\alpha_{S_{i+1}'}\,\xi_{i0}^2  \right) \nonumber \\
&\hspace{-1cm}= \left(\frac{\sum \alpha_{S_{i+1}'}\,\xi_{i0}}{\sum t_i} + \xi_{k0}\right)^\nu \exp\left(-\sum_{ij}\alpha_{C_{ij}}\xi_{ij}^2  -\frac{ \left( \sum_i\alpha_{S_{i+1}'}\,\xi_{i0} \right)^2 }{\sum_i t_i} -\sum_i\alpha_{S_{i+1}'}\,\xi_{i0}^2  \right) \label{derivatives}
\end{align}

For the differentials corresponding to curves $C_{ij}$, we have:
\begin{align}
\frac{1}{2\alpha_{C_{kl}}}\,\p_{\xi_{kl}}{}^\nu\ &\exp\left(-\sum_{ij}\alpha_{C_{ij}}\xi_{ij}^2  -\frac{ \left( \sum_i\alpha_{S_{i+1}'}\,\xi_{i0} \right)^2 }{\sum_i t_i} -\sum_i\alpha_{S_{i+1}'}\,\xi_{i0}^2  \right) = -\xi_{kl}{}^\nu\ \exp\bigg(\hdots\bigg) 
\end{align}

\subsubsection*{Spurious dependence on $\xi_{k0}$}
Even though there is an apparent dependence on $\ell$ in form of $\xi_{k0}$ even after loop integration, after the action of Corolla, we can set $\ell \to 0$. Here we argue how the dependence on $\ell$ is spurious. In a completely trivalent gluons graph $T$, all the derivatives in $\hat{C}_T$ are acting on the exponential. So using \eqref{derivatives}, the numerator factor can be obtained by the following replacement in $\hat{C}_T$:
\begin{align}
\frac{-1}{2\alpha_{S_{k+1}'}}\,\p_{\xi_{k0}}{}^\nu \to \left(\frac{\sum \alpha_{S_{i+1}'}\,\xi_{i0}}{\sum t_i} + \xi_{k0}\right)^\nu &= \frac{1}{\sum t_i}\sum_i\alpha_{S_{i+1}'}\big(\xi_{i0}-\xi_{k0}\big)^{\nu} \nonumber \\
&\qquad \qquad \qquad \xrightarrow{\text{mom. conserv}} \frac{1}{\sum t_i}\,\sum_{i}\alpha_{S_i'}\,z_{ki}{}^{\nu} \label{spurious dep}
\end{align}
Note that after imposing momentum conservation, the result depends only on the external momenta $z_i$. We obtain a contact vertex when the derivative acts on the neighboring $c_v$ rather than the exponential, resulting in a product of appropriate $\eta$'s. This is in accord with the basic statement that a contact vertex has no momentum dependence. So, even though there is an unsettling presence of $\xi_{k0}$ variables reminding us of the loop momenta in the intermediate steps, the final amplitude after imposing momentum conservation depends only on the external momenta, as expected.

\subsubsection{Example: Two point planar one-loop}
Here, we shall work out the simplest case, planar one-loop $n=2$. We proceed to write down the Corolla differential operators $C_{\mathcal{T}}$, and insert them in front of the loop-integrated Schwinger exponential.  

The headlight functions can be read off from \eqref{headlight functions for spiral curves}:
\begin{align}
\alpha_{S_1'} &= -t_2 - \mxx{0,t_1} + \mxx{0,t_2} \\
\alpha_{S_2'} &= -t_1 - \mxx{0,t_2} + \mxx{0,t_1}
\end{align}
So, in the relevant cone with the bubble, we have $\alpha_{S_{i}'}=-t_{i+1}$.

Recall that the momentum associated with the curve $S'_i$ is $-\xi_{i-1,0}$. So, the prefactor associated with $\p_{\xi_{i0}}$ shall be $-1/2\alpha_{S'_{i+1}}$. So, we make the following replacements to the differential operators from $P$ \eqref{polyvector for one loop}:
\begin{align}
\frac{l_i^2}{2}\frac{\partial}{\partial l_i} \longleftrightarrow -\frac{1}{2\alpha_{S'_{i+1}}}\frac{\partial}{\partial \xi_{i0}} \qquad ; \qquad  \frac{\xi_i^2}{2}\frac{\partial}{\partial \xi_i} \longleftrightarrow \frac{1}{2\alpha_i}\frac{\partial}{\partial \xi_i} 
\end{align}
\newcommand{\bble}{\begin{tikzpicture}[scale=0.1]
	\draw (0,0) circle [radius=0.5];
	\draw (0.5,0) -- (1.5,0);
	\draw (-0.5,0) -- (-1.5,0);
	\end{tikzpicture}}
The following is the Corolla differential for the bubble:
\begin{align}
\hat{C}_{\bble} \\
&\hspace{-1.5cm} = \bigg[\eta^{\nu_1\nu_2}\left( \frac{1}{2\alpha_{S'_1}}\,\p_{\xi_{20}}+\frac{1}{2\alpha_{S'_2}}\,\p_{\xi_{10}} \right)^{\mu_1} + \eta^{\nu_2\mu_1}\left(\frac{1}{2\alpha_1}\,\p_{\xi_1}-\frac{1}{2\alpha_{S'_1}}\,\p_{\xi_{20}}\right)^{\nu_1} + \eta^{\mu_1\nu_1}\left( -\frac{1}{2\alpha_{S'_2}}\,\p_{\xi_{10}} - \frac{1}{2\alpha_1}\,\p_{\xi_1} \right)^{\nu_2}\bigg] \nonumber \\
&\hspace{-1.3cm}\times \bigg[\eta^{\mu_2}{}_{\nu_2}\left( -\frac{1}{2\alpha_{S'_1}}\,\p_{\xi_{20}} - \frac{1}{2\alpha_2}\,\p_{\xi_2} \right)_{\nu_1} + \eta_{\nu_1\nu_2}\left( \frac{1}{2\alpha_{S'_2}}\,\p_{\xi_{10}} +\frac{1}{2\alpha_{S'_1}}\,\p_{\xi_{20}} \right)^{\mu_2} + \eta_{\nu_1}{}^{\mu_2}\left(\frac{1}{2\alpha_2}\,\p_{\xi_2} - \frac{1}{2\alpha_{S'_2}}\,\p_{\xi_{10}}\right)_{\nu_2} \bigg] \nonumber \\
&- \left(\frac{1}{2\alpha_{S'_1}}\,\p_{\xi_{20}}\right)^{\mu_1}\left(\frac{1}{2\alpha_{S'_2}}\,\p_{\xi_{10}}\right)^{\mu_2} - \left(\frac{1}{2\alpha_{S'_1}}\,\p_{\xi_{20}}\right)^{\mu_2}\left(\frac{1}{2\alpha_{S'_2}}\,\p_{\xi_{10}}\right)^{\mu_1}
\end{align}
As the sum over all tadpole contributions  cancel, we do not discuss them further. We are then left with merely one top-dimensional cone, and  the one-loop scalar two-point amplitude can be written as, \eqref{integrated one loop planar scalar}:
\begin{align}
A_2{}^{\text{Tr}\phi^3} &= \prod_{i=1,2}\xi_i^2\int \dd^2s_i\int_{\sum t_i\leq 0}\dd^2\vec{t}\,\left(\frac{-\pi}{\sum t_i}\right)^{D/2}\exp\left(- \frac{\left(\alpha_{S_1'}\xi_{20}+ \alpha_{S_2'}\xi_{10}\right)^2}{t_1+t_2} - \alpha_{S_1'}\xi_{20}^2- \alpha_{S_2'}\xi_{10}^2 \right)
\end{align}
Where we have set $p_\Sigma \to 0$ as tadpoles have been eliminated. Now, the YM amplitude is as follows:
\begin{align}
\mathcal{A}_2[1,2]^{\text{YM}} &= \prod_{i=1,2}\xi_i^2\int \dd^2s_i(\text{Reg-Res}) \ \frac{\sum_{\mathcal{T}}\hat{C}_{\mathcal{T}}\prod_{I\in \mathcal{T}}\alpha_I(\vec{t})}{\sum_{\mathcal{T}}\prod_{I\in \mathcal{T}}\alpha_I(\vec{t})} \nonumber \\
&\hspace{3cm}\times\exp\left(- \frac{\left(\alpha_{S_1'}\xi_{20}+ \alpha_{S_2'}\xi_{10}\right)^2}{t_1+t_2} - \alpha_{S_1'}\xi_{20}^2- \alpha_{S_2'}\xi_{10}^2 -\alpha_1 \xi_1^2- \alpha_2\xi_2^2\right)
\end{align}
Note that we have added Schwinger parameters for external particles as well. 

Due to absence of tadpole contribution, we merely have a bubble:
\begin{align}
\mathcal{A}_2[1,2]^{\text{YM}} &= \prod_{i=1,2}\xi_i^2\int \dd^2s_i(\text{Reg-Res}) \ \frac{\alpha_{S_1'}\alpha_{S_2'}}{\sum_{\mathcal{T}}\prod_{I\in \mathcal{T}}\alpha_I(\vec{t})}\ \nonumber \\
&\qquad \quad \hat{C}_{\bble}\, \exp\left(- \frac{\left(\alpha_{S_1'}\xi_{20}+ \alpha_{S_2'}\xi_{10}\right)^2}{t_1+t_2} - \alpha_{S_1'}\xi_{20}^2- \alpha_{S_2'}\xi_{10}^2 -\alpha_1 \xi_1^2- \alpha_2\xi_2^2\right)
\end{align}

Here is the action of relevant derivatives on the integrated exponential \eqref{derivatives}:
\begin{align}
-\frac{1}{2\alpha_{S'_k}}\,\p_{\xi_{k+1,0}} e^{-\sum\hdots} = \left(\frac{\alpha_{S'_1}\,\xi_{20}+\alpha_{S'_2}\,\xi_{10}}{t_1+t_2}+\xi_{k+1,0}\right)e^{-\sum\hdots} = \frac{\alpha_{S'_{k+1}}(\xi_{k,0}-\xi_{k+1,0})}{t_1+t_2} \ e^{-\sum\hdots}
\end{align}
\begin{align}
\hat{C}_{\bble}\,e^{\hdots}&= \bigg[\eta^{\nu_1\nu_2}\left( -\frac{\alpha_{S_2'}(\xi_{10}-\xi_{20})}{t_1+t_2}-\frac{\alpha_{S_1'}(\xi_{20}-\xi_{10})}{t_1+t_2} \right)^{\mu_1} + \eta^{\nu_2\mu_1}\left(-{\xi_1} + \frac{\alpha_{S_2'}(\xi_{10}-\xi_{20})}{t_1+t_2} \right)^{\nu_1} \nonumber \\
&\hspace{9cm}+ \eta^{\mu_1\nu_1}\left( \frac{\alpha_{S_1'}(\xi_{20}-\xi_{10})}{t_1+t_2} + {\xi_1} \right)^{\nu_2}\bigg] \nonumber \\
&\hspace{-0.5cm}\times \bigg[\eta^{\mu_2}{}_{\nu_2}\left( \frac{\alpha_{S_2'}(\xi_{10}-\xi_{20})}{t_1+t_2}+\xi_2  \right)_{\nu_1} + \eta_{\nu_1\nu_2}\left( -\frac{\alpha_{S_1'}(\xi_{20}-\xi_{10})}{t_1+t_2} - \frac{\alpha_{S_2'}(\xi_{10}-\xi_{20})}{t_1+t_2} \right)^{\mu_2} \nonumber \\
&\hspace{9cm}+ \eta_{\nu_1}{}^{\mu_2}\left(-\xi_2 +\frac{\alpha_{S_1'}(\xi_{20}-\xi_{10})}{t_1+t_2} \right)_{\nu_2} \bigg] \nonumber \\
&  +(D-1)\,\frac{1}{t_1+t_2}\,\left(\frac{\alpha_{S_1'}}{\alpha_{S_2'}} + \frac{\alpha_{S_2'}}{\alpha_{S_1'}}-1\right)\eta^{\mu_1\mu_2} \nonumber \\
& \hspace{-1.5cm} - \left(\frac{\alpha_{S_2'}(\xi_{10}-\xi_{20})}{t_1+t_2}\right)^{\mu_1}\left(\frac{\alpha_{S_1'}(\xi_{20}-\xi_{10})}{t_1+t_2}\right)^{\mu_2} - \left(\frac{\alpha_{S_1'}(\xi_{20}-\xi_{10})}{t_1+t_2}\right)^{\mu_1} \left(\frac{\alpha_{S_2'}(\xi_{10}-\xi_{20})}{t_1+t_2}\right)^{\mu_2} + \frac{1}{t_1+t_2}\,\eta^{\mu_1\mu_2}
\end{align}

We can impose momentum conservation after the action of Corolla $\hat{C}$. As advertised earlier in \eqref{spurious dep}, all the loop momenta $\xi_{i0}$ appear in a particular combination such that the dependence on $\ell$ vanishes. Imposing momentum conservation, we have $\xi_{10}-\xi_{20} = \xi_1$, and $\xi_2 = -\xi_1$. Since we are restricted in one cone, let us substitute $\alpha_{S_i'}$ in terms of $t_i$: $\alpha_{S_i'}=-t_{i+1}$. After some simplifications, we obtain:
\begin{align}
\hat{C}_{\bble}\,e^{\hdots} &= \frac{1}{(t_1+t_2)^2}\left[ -\xi_1^{\mu_1}\xi_1^{\mu_2}\big(2t_1^2+12t_1t_2+2t_2^2\big) + \xi_1^2\,\eta^{\mu_1\mu_2}\big(5t_1^2+8t_1t_2+5t_2^2\big) \right]\nonumber\\ &\hspace{1cm} + \frac{\eta^{\mu_1\mu_2}}{t_1+t_2}\left[ 3\frac{t_1}{t_2} + 3\frac{t_2}{t_1}-2 \right] e^{- \left(\alpha_{S_1'}\xi_{20}+ \alpha_{S_2'}\xi_{10}\right)^2/{(t_1+t_2)} - \alpha_{S_1'}\xi_{20}^2- \alpha_{S_2'}\xi_{10}^2 -s_1 \xi_1^2- s_2\xi_2^2 }
\end{align}

(Reg-Res) picks up the particular poles in Schwinger parameters and integrates the rest. For this case, in this particular cone, we have
\begin{align}
\big(\text{Reg-Res}\big) &= \int \dd^2s\left[\left(\int\dd t_1\,\dd t_2 \ \underset{t_1,t_2}{\text{Reg}}\right) + \left(\int \dd t_1\ \underset{t_1}{\text{Reg}}\ \underset{t_2\to 0}{\text{Res}}\right) + \left(\int \dd t_2\ \underset{t_2}{\text{Reg}}\ \underset{t_1\to 0}{\text{Res}}\right) +  \underset{t_1,t_2 \to 0}{\text{Res}} \right]
\end{align}
Thus, we finally get,
\begin{align}
\mathcal{A}_2[1,2]^{\text{YM}} &=  \xi_1^2\xi_2^2\,\big(\text{Reg-Res}\big) \hat{C}_{\bble}\,e^{\hdots} \nonumber \\
&= \int_{-\infty}^0 \dd t_1\,\dd t_2 \bigg[\frac{1}{(t_1+t_2)^2}\bigg[ -\xi_1^{\mu_1}\xi_1^{\mu_2}\big(2t_1^2+12t_1t_2+2t_2^2\big) \nonumber\\
&\hspace{2cm}+ \xi_1^2\,\eta^{\mu_1\mu_2}\big(5t_1^2+8t_1t_2+5t_2^2\big) \bigg] - 2\frac{\eta^{\mu_1\mu_2}}{t_1+t_2}\bigg] + 2\int_{-\infty}^0\dd t\,\eta^{\mu_1\mu_2} \,e^0
\end{align}
It can be checked that the above result equals the parametric representation of two point gluon correlator as derived in \cite{prinz}.

\section{Conclusion and Outlook} 
The essence of the positive geometry program is to analyse the combinatorial and geometric structures that \emph{naturally} arise in the bare kinematic space of scattering data and study if these structures can give birth to S-matrix of local unitary quantum field theories. The naturality referred to above arises from the fact that these structures do not emerge within the world of quantum field theory computations but appear at first in an apparently disconnected domain of quiver representation theory and so called gentle algebras. The simplest example of this ``S-matrix without quantum field theory" is the bi-adjoint scalar amplitudes. Namely, inside the Kinematic space of scattering data, a rich class of combinatorial geometries live whose embedding inside the kinematic space makes no reference to Feynman diagrams or physical principles such as unitarity or locality, \cite{nima1711,baziermatte,pppp1906}.  

Once we locate these positive geometries inside the kinematic space of scalar particles, a large class of tree-level scattering amplitudes arise directly as the canonical forms which are fixed uniquely by the geometries. Several intriguing results have started coming out of this program in last few years. The top-form defined by the ABHY associahedron not only generates amplitude of bi-adjoint scalar theory, but simple linear diffeomorphisms in kinematic space changes the shape of the associahedron in such a way that the resulting geometry (and the corresponding form) generate amplitudes of a large class of scalar theories at tree-level, \cite{mrunmay2206}, including the amplitude of ordinary $\phi^{3}$ theory without color, \cite{mrunmay2304}. 

 ABHY associahedron is hence a  ``universal polytope" that underlies tree-level S matrix of an infinite family of scalar theories which appear to be completely unrelated to each other at the level of Lagrangian or Feynman diagrams.  In fact, more is true. In \cite{mrunmay2104}, it was shown that ``shadows" of the associahedron in certain regions of kinematic space are  lower dimensional accordiohedra which are the geometries underlying scalar theories with $\phi^{p}, p > 4$ interactions. As a result, the effect of integrating out massive modes (at least at tree-level) was same as mapping of positive geometry to a \emph{different} class of lower dimensional positive geometry. Both of these results point us towards the universality of the ABHY associahedron as a geometry that underlies a large class of tree-level amplitudes. Perhaps, the most striking evidence of this universality is the new understanding of the NLSM amplitude as a specific deformation of $\textrm{Tr}(\phi^{3})$ amplitude, \cite{nima2312:hiddenzero}. In fact, this relationship between NLSM and positive geometries underlying $\phi^{3}$ S-matrix extends to all orders in perturbation theory, a result completely unforseen in traditional perturbative analysis. 

S matrix of non-supersymmetric gauge theories have so far  proved resistant against their ``integration" into the positive geometry program. There is a certain irony to it, as   the connection between polytopes and amplitudes was first discovered in the seminal work of Hodges\cite{Hodges}  which showed that tree-level NMHV amplitudes were simply volume of certain polytopes in the kinematic space of 4 dimensional Yang-Mills theory. This was followed by the striking discovery of the amplituhedron and the significant insights it continues to bring into the study of SYM amplitudes, \cite{The_Amplituhedron, negativegeometry, deformed-amplituhedron}. However in the non-supersymmetric world, it is the scalar theories which have found their most natural home in the ``amplituhedron program". A breakthrough in this direction has come in the recent paper by Arkani-Hamed, Cao, Dong, He, Figueiredo, He, where gluon amplitude is realised as a ``deformed" amplitude of so called scaffolding scalars. This is a remarkable result with far reaching implications in which  the relationship  $\textrm{Tr}(\phi^{3})$ amplitude (which arise simply from the combinatorial structures such as headlight functions) and the YM S-matrix ``factors through" stringy $\textrm{Tr}(\phi^{3})$ amplitudes. 

In this paper we have shown that the canonical form defined by positive geometries such as associahedron can be used to generate another class of scalars by contraction of the form with certain multi-vector field. This MVF is a sum over all the vertices of the associahedron and is defined via the mutation rules which also determine the canonical form.  At each vertex of the associahedron, the MVF is fixed by a Corolla polynomial which in turn depends on the triangulation as well as the finer structure that triangulation carries with it via it's representation as bracketed words. With such a structure at hand, ABHY had already shown that color-dressed $\phi^{3}$ amplitude at tree-level is a contraction of this MVF with the d-log form. As we have shown in this paper, even the YM ampitudes emerge through such an algebraic operation.   We then showed how these ideas carry over to the planar one loop integrand of Yang-Mills theory which can also be realised as a contraction of the d log form associated to $\hat{D}_{n}$ polytope with the MVF which is generated by the Corolla polynomial associated to pseudo-triangulations. In fact the seminal results of KSVS along with the curve integral formula for $\textrm{Tr}(\phi^{3})$ amplitudes can be combined rather immediately to define \emph{a} curve integral formula for gluons. Although we explicitly derived such a formula for tree-level and one loop amplitudes, as emphasized in the introduction, the extension to higher loop non-planar amplitudes should be possible. 

As the primary motivation of our work is to obtain an insight into how S-matrix of non super-symmetric gauge theories fit into the positive geometries program, it is important to mention the limitation of our results from that perspective. Is there a ``momentum amplituhedron" for perturbative gluon amplitudes? Our analysis does not answer this question. In fact, our results are premised on the observation that the amplitude of YM theory without maximal supersymmetry are not d-log forms.  Of course, this does not rule out the possibility that the meromorphic forms corresponding to e.g. tree-level YM amplitudes \cite{Song1807} may be defined by a new class of geometries, (See e.g. \cite{mom_amplitu_meets_assoc} where the precise relationship via push-forward map of the d-log form defined by the ABHY associahedron with the d-log form defined by the momentum amplituhedron was established). However, the use of Corolla polynomial led us to a different geometric interpretation of these amplitudes, as scalar functions on the kinematic space. Thus we are far from ``integrating gauge theory S-matrix into the positive geometries program". Nevertheless, our hope is that the combinatorial structure that glues together all the marked triangulations may eventually lead us to geometries which would help us discover a ``momentum amplituhedron" for perturbative non-supersymmetric gluon amplitudes.  

We now summarise several avenues for further investigation. 

\subsection*{A homology for each vertex of $A_{n-3}$.}
As we saw in section \ref{mtri}, the action of the Corolla polynomial maps a triangulation to a set of marked triangulations. As each triangulation is dual to a graph, it is natural to quantify the set of marked triangulations with a (direct) sum over graphs is the generator of graph homology $[\gamma]$ (see section \ref{gaugeinvarianceofmn})\begin{align}
[\gamma]\, :=\, e^{\chi}\, \gamma
\end{align}
Hence Corolla maps the set of vertices of  $A_{n-3}$ to the corresponding \emph{unique} homology class $[\gamma_{n}]\, \vert\, \forall\, \gamma_{n}\, \textrm{with fixed ordering of external vertices.}$. It would be interesting to investigate this structure further in light of recent advances in studies in graph homology, see for example \cite{ward2021massey, batanin2022minimal}. 
\subsection*{Relationship with worldsheet formula and the curve integral formula for Stringy YM amplitude.}
The CHY formula for tree-level amplitude of bi-adjoint scalar theory is simply a pull back of the d-log form from the kinematic space to the so-called worldsheet associahedron, \cite{nima1711}.  Worldsheet associahedron is the real section of moduli space of $n$ punctured Riemann sphere compactified using Plucker co-ordinates. The pull back of the d-log form is through scattering equations which are simply diffeomorphism between world-sheet and ABHY associahedra. It is then natural to investigate if there exists a Corolla differential on the worldsheet and if it can be used to ``spin up" the Parke-Taylor form to obtain the CHY integrand for tree-level YM amplitude.

As discussed previously, the most clear relationship between $\textrm{Tr}(\phi^{3})$ and YM amplitude to all orders in the loop expansion has been discovered in \cite{nima2401:scaffold}.  E.g. the tree-level stringy $\textrm{Tr}(\phi^{3})$ amplitude of $2n$ scalars, under specific deformation of the planar kinematic variables generate stringy YM amplitude once pair of scalars $(2i-1, 2i)$ are ``combined" to form a spin-1 state.\footnote{This operation has been called scaffolding in \cite{nima2401:scaffold}.} It will be extremely interesting to see if there is a representation of the Corolla polynomial which maps the  stringy $\textrm{Tr}(\phi^{3})$ amplitude to stringy gluon amplitude at least at tree-level. This may lead to a direct connection between the ideas advocated in this paper and in \cite{nima2401:scaffold}. 
\subsection*{Corolla for Gravitational amplitude}
 An $n$ point tree-level gravitational amplitude requires  higher order contact terms with degree $4 \leq\ d \leq\ n$. Clearly the underlying structure must far more intricate than the Corolla polynomial whose differential representation generates graph homology in which quartic vertices and marked edges are identified.  An intriguing (but highly speculative) possibility is to ``fuse" two Corolla differentials  using the KLT kernel. Needless to say, this is perhaps one of the major open questions that face the idea of using Graph homology to map $\phi^{3}$ amplitude to a gravitational S-matrix. 

\section*{Acknowledgement}

We would like to thank Pinaki Banerjee, Frédéric Chapoton, 
Chandramouli Chowdhury, Nima-Arkani Hamed, Subramanya Hegde,  Mrunmay Jagadale, Vincent Pilaud, Prashanth Raman,  Ashoke Sen, Aninda Sinha and  Adarsh Sudhakar for insightful discussions over last few years on many aspects of the positive geometry program. We would especially like to thank Pinaki Banerjee and Chandramouli Chowdhury for crucial discussions during the initial phase of this project.  AL thanks Center for High Energy Physics (CHEP) at IISc, International Center for Theoretical Sciences (ICTS), and IIT Kanpur for their hospitality during the course of the project.  

\bibliography{Corolla-draft} 
\bibliographystyle{JHEP}

\end{document}